\documentclass[runningheads]{svmult}

\usepackage{graphicx}  

\font\tenib=cmmib10  \skewchar\tenib='177
\font\sevenib=cmmib7 \skewchar\sevenib='177
\font\fiveib=cmmib5  \skewchar\fiveib='177

\def\half{{\textstyle\frac12}}
\def\overarrow#1{\setbox0\hbox to 0pt{\hss$\scriptstyle\rightarrow$}%
                  #1\raise 1.6ex\box0}
\def\doverarrow#1{\setbox0\hbox to 0pt{\hss$\scriptstyle\rightarrow$}%
                  #1\kern.4ex\raise1.5ex\copy0\raise2.2ex\box0\kern-.4ex}
\def\hoverarrow#1{\setbox0\hbox to 0pt{\hss$\scriptstyle\rightarrow$}%
                 #1\kern.6ex\raise 1.6ex\box0\kern-.2ex}
\def\({\raise.3ex\hbox{(}} \def\){\raise.3ex\hbox{)}}
\def\bnabla{{\overarrow\nabla}}
\def\btheta{{\hoverarrow\theta}}
\def\bbeta{{\hoverarrow\beta}}
\def\bTheta{{\hoverarrow\Theta}}
\def\zl{z_{\rm L}}  \def\zs{z_{\rm S}}
\def\Dl{D_{\rm L}} \def\Ds{D_{\rm S}} \def\Dls{D_{\rm LS}}
\def\thee{\theta_{\rm E}}  \def\Thee{\Theta_{\rm E}}
\def\Sigcrit{\Sigma_{\rm crit}}
\def\M{{\bf M}}
\def\vlos{v_{\rm los}}

\def\pderiv(#1/#2){\partial#1/\partial#2}
\def\secderiv(#1/#2,#3){\partial^2#1/\partial#2\partial#3}

\def\<#1>{\langle#1\rangle}

\catcode`@=11
\renewcommand\subsubsection{\@startsection{subsubsection}{3}{\z@}%
                       {-10\p@ \@plus -3\p@ \@minus -3\p@}%
                       {-0.5em \@plus -0.22em \@minus -0.1em}%
                       {\normalfont\normalsize\bfseries\boldmath}}
\renewcommand\paragraph{\@startsection{paragraph}{4}{\z@}%
                       {-4\p@ \@plus -2\p@ \@minus -2\p@}%
                       {-0.5em \@plus -0.22em \@minus -0.1em}%
                       {\normalfont\normalsize\itshape}}
\catcode`@=12

\begin{document}

\title*{Quasar Lensing}

\toctitle{Quasar Lensing}

\titlerunning{Quasar Lensing}

\author{Frederic Courbin\inst{1, 2}
\and Prasenjit Saha\inst{3}
\and Paul L. Schechter\inst{4}}

\authorrunning{Courbin, Schechter \& Saha}

\institute{Pontificia Universidad Cat\'olica de Chile, Av. Vicu\~na 
Mackenna 4860,\\ Departamento de Astronomia y Astrofisica, Casilla 306,
Santiago 22, Chile
\and       Universit\'e de Li\`ege, Institut d'Astrophysique et de
G\'eophysique,\\ 
All\'ee du 6 ao\^ut, Bat. B5C, Li\`ege 1, Belgium
\and       Astronomy Unit, School of Mathematical Sciences\\
           Queen Mary and Westfield College, London E1 4NS, UK
\and       Massachusetts Institute of Technology, \\
           70 Vassar Street, Cambridge, MA 02139, USA}

\maketitle              

\begin{abstract}
Massive  structures, such  as  galaxies, act  as strong  gravitational
lenses on background sources. When  the background source is a quasar,
several lensed images are  seen, as magnified or de-magnified versions
of the same object.  The detailed study of the image configuration and
the  measurement  of  ``time-delays''  between  the  images  yield  
estimates of the  Hubble parameter $H_0$.  We describe  in  a  
simple way  the phenomenon of strong lensing and  review recent 
progress made in the field, including microlensing by stars 
in the main lensing galaxy.
\end{abstract}


\section{Concepts}

\subsection{The formation of multiple images}

There are several ways of understanding the effect of gravity  
on light in
the context of lensing.  We start with an approach which lends itself
particularly well to pictorial representation.

\begin{figure}[!ht]
\begin{center}
\includegraphics[width=.7\textwidth]{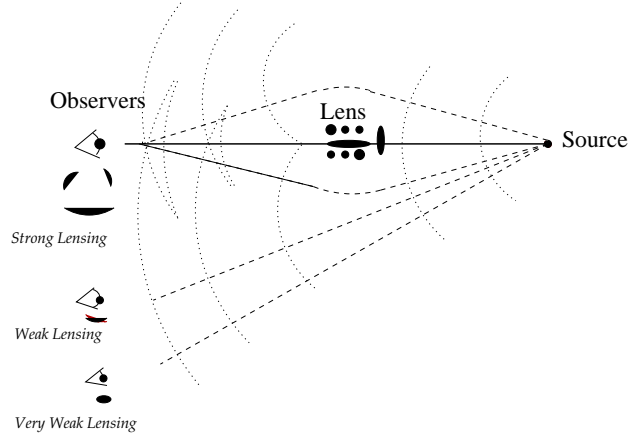}
\end{center}
\caption[]{Schematic illustration \cite{hma98} of the wavefront 
\index{wavefront} and the
different regimes of lensing.  Lensed quasars fall in the strong
lensing regime; the other regimes are important in lensing by clusters
of galaxies.}
\label{fig-wf}
\end{figure}

\subsubsection{Wavefronts} A schematic wavefront \index{wavefront} 
is illustrated in 
Figure \ref{fig-wf}.  Spreading outwards from a point source, the
wavefront is initially spherical.  But as it passes through the
gravitational field of the lens the wavefront gets delayed and bent;
we can interpret this effect as a slowing-down of light by a
gravitational field, usually called the Shapiro \index{time delay!Shapiro}
time
delay \cite{shap64}.  Where the wavefront crosses an observer, they see
an image in the direction normal to the wavefront, and images will be
(de)magnified and/or distorted \index{magnification}
according to how curved the wavefront
is as it crosses the observer.  If the lens is strong enough, the
wavefront can fold in on itself, producing multiple images.  If
moreover the source is variable, different images will show that
variability with time delays proportional to the spacing between these
folds, i.e., the cosmological distance scale.

It is possible to make the above explanation quantitative within the
wavefront picture \cite{Refsd1,kr83}, but for calculations that is
usually not the most convenient route.  Notice that the wavefront
picture has a single source and multiple observers, whereas
astrophysical problems generally involve multiple sources and a single
observer.  So calculations are easier if we use a relative of the
wavefront called the arrival-time \index{arrival time} 
surface \cite{bn86,n90}.

\begin{figure}[!ht]
\begin{center}
\includegraphics[width=.35\textwidth]{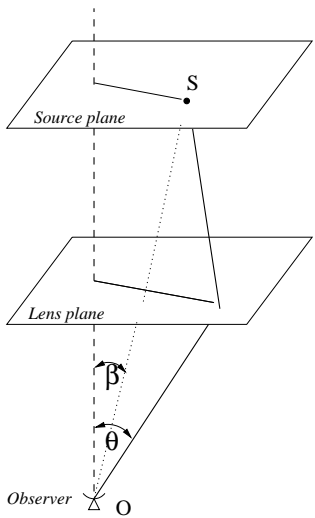}
\end{center}
\caption[]{Illustration of a virtual light ray: $\bbeta$ is the unlensed
sky position of the source, and $\btheta$ is its apparent position.
In the text, we use $\Dl$, $\Ds$, and $\Dls$ for angular diameter
distances \index{distances!angular diameter} 
from observer to lens, observer to source, and lens to
source.}
\label{fig-planes}
\end{figure}

\subsubsection{Arrival times} \index{arrival time}

Consider Figure \ref{fig-planes}: in the usual astrophysical
approximation of small angles and thin lenses, this figure shows a
virtual light ray getting deflected by the lens and reaching the
observer from direction $\btheta$, the source being at angular
position $\bbeta$.  (Vector signs denote 2D angles on the sky.)  The
arrival time is the light travel time---with irrelevant constants
discarded---of such a virtual ray as a function of $\btheta$, with
$\bbeta$ held fixed.  It has two contributions: a `geometrical' part
and a `gravitational' part \cite{bn86}. The geometrical part \index{time
delay!geometric} is simply
the difference between the continuous and dotted paths in
Figure~\ref{fig-planes}, and is given by
\begin{equation}
t_{geom}(\btheta) = \half (1+\zl) {\Dl\Ds\over c\Dls} \(\btheta-\bbeta\)^2,
\label{t-geom}
\end{equation}
where $\zl$ is the lens redshift and the $D$ factors are angular
diameter distances as shown in Figure~\ref{fig-planes}.  The
gravitational part is the Shapiro time delay \index{time delay!Shapiro}
in a gravitational field
from general relativity, and depends on the surface density
$\Sigma\(\btheta\)$ of the lens.  A concise way of writing the
Shapiro delay is
\begin{equation}
t_{Shapiro}(\btheta) = (1+\zl) {8\pi G\over c^3} \, \nabla^{-2} 
\Sigma\(\btheta\).
\label{t-grav}
\end{equation}
Here $\nabla^{-2}$ denotes the inverse of a 2D Laplacian with respect
to $\btheta$,\footnote{By this we mean an operator that solves
Poisson's equation in 2D. Thus, if $\nabla^2 f(\btheta) = g(\btheta)$,
we write $f(\btheta) = \nabla^{-2} g(\btheta)$.  The explicit form of
the inverse Laplacian is as an integral
$$ f(\btheta) = \int \ln|\btheta-\btheta'| \, g(\btheta') \, d^2\btheta' $$
but we will not need it in this article.}
and $\nabla^{-2} \Sigma\(\btheta\)$ is some sort of 2D potential.

Putting equations (\ref{t-geom}) and (\ref{t-grav}) together we have
the arrival time $t(\btheta)$ in full:
\begin{equation}
t(\btheta) = \half (1+\zl) {\Dl\Ds\over c\Dls} \(\btheta-\bbeta\)^2 - 
(1+\zl) {8\pi G\over c^3} \, \nabla^{-2} \Sigma\(\btheta\).
\label{eq-t}
\end{equation}
From Fermat's principle, \index{Fermat's principle}
real light rays take paths that make the
arrival time stationary.  Thus the condition for images is
\begin{equation}
\bnabla t\(\btheta\)=0.
\label{eq-lensprelim}
\end{equation}

Equation (\ref{eq-t}) looks formidable, but it will become much less
so once we introduce some scales.

\subsubsection{Some scales}

Consider a point-mass \index{lens!point mass} 
lens and a point source along the same line of
sight, i.e., $\bbeta=0$ and $\Sigma\(\btheta\)=M\delta\(\btheta\)$.  The
arrival time \index{arrival time} then becomes
\begin{equation}
t\(\btheta\) = \half (1+\zl) {\Dl\Ds\over c\Dls} \, \theta^2 - 
(1+\zl) {4G\over c^3} \, \ln\theta,
\end{equation}
since $\nabla^{-2}\delta\(\btheta\)= \ln\theta/(2\pi)$, and there is a minimum
at $\theta=\thee$ where
\begin{equation}
\thee^2 = {4GM\over c^2}{\Dls\over\Dl\Ds}.
\label{eq-thee}
\end{equation}
This corresponds to a ring image, called an Einstein ring, 
\index{Einstein!ring} \index{Einstein!radius} and $\thee$
is called the Einstein radius.  
If the source is much further than the lens
\begin{equation}
\thee \simeq 0.1\,{\rm arcsec}\times
\left[\hbox{$M$ in $M_\odot$}\over\hbox{$\Dl$ in pc}\right]^\half.
\label{eq-theeapp}
\end{equation}

The combination of a point lens and colinear source is very
improbable, but the Einstein radius is a very useful concept, for two
reasons.  First, even if there is no Einstein ring in a multiple-image
system, the image separation \index{image separation} 
still tend to be of order $\thee$.
Secondly, the Einstein radius also supplies a scale for $\Sigma$, by
the following argument.

From the two-dimensional analog of Gauss's flux law, for any circular
mass distribution $\Sigma(\theta)$, $\bnabla t\(\btheta\)$ will depend
only on the enclosed mass.  So not just a point mass, but any circular
distribution of the mass $M$, will produce an Einstein ring from a
colinear source, {\it provided it fits within $\thee$.}  The condition
of a mass fitting into its own Einstein radius is known as
`compactness'. \index{compactness} 
And because from (\ref{eq-thee}) the area within an
Einstein radius is itself proportional to the mass, compactness is
equivalent to the density exceeding some critical density 
\index{density!critical}.  Working
out the algebra we easily get this critical 
density\footnote{In this article the
units of $\Sigma$ are M$_{\odot}$ arcsec$^{-2}$. 
Some authors prefer M$_{\odot}$ kpc$^{-2}$. This
difference of convention means that different authors' equations may
differ by factors of D$_L$ or D$_L^2$.} to be
\begin{equation}
\Sigcrit = {c^2\over4\pi G} {\Dl\Ds\over\Dls}.
\label{eq-sigcrit}
\end{equation}
From equation \ref{eq-t} we can also define a time scale \index{time scale}
\begin{equation}
T_0 = (1+\zl) {\Dl\Ds\over c\Dls},
\label{eq-T0}
\end{equation}
which is of order the light travel time, or a Hubble time in
cosmological situations.  The interesting time scale in lensing,
however, is not $T_0$, but
\begin{equation}
T_0 \times \<\hbox{image separations}>^2,
\end{equation}
being the scale of arrival-time {\it differences\/} \index{arrival time}
between images.
We will meet the latter presently, in the approximate equation
(\ref{eq-T0app}).

\subsubsection{The arrival-time surface}

Using the scales introduced above, we can render dimensionless the arrival time
(\ref{eq-t}),
\begin{equation}
\tau\(\btheta\) = \half\(\btheta-\bbeta\)^2 - 2\nabla^{-2}\kappa\(\btheta\).
\label{eq-tau}
\end{equation}
Here the scaled arrival time $\tau$ and the scaled surface density 
\index{density!surface} 
$\kappa$ (also called convergence) are both dimensionless. 
\index{convergence} The
last term in equation (\ref{eq-tau}) is called the lens or projected 
potential \index{potential!projected}
\begin{equation}
\psi\(\btheta\) \equiv 2\nabla^{-2}\kappa\(\btheta\).
\label{eq-pot}
\end{equation}
The physical arrival time and density are
\begin{equation}
t\(\btheta\) = \tau\(\btheta\) \times T_0, \quad
\Sigma\(\btheta\) = \kappa\(\btheta\) \times \Sigcrit,
\end{equation}
and the scales are approximately
\begin{equation}
T_0 \simeq h^{-1} \zl(1+\zl) \times 80\,{\rm days\,arcsec^{-2}}.
\label{eq-T0app}
\end{equation}
and
\begin{equation}
\Sigcrit \simeq h^{-1} \zl \times 1.2\cdot10^{11}\,M_\odot\,{\rm arcsec^{-2}},
\label{eq-sigcritapp}
\end{equation}
where $h$ is the Hubble constant \index{Hubble constant}
in units of 100 km/s/Mpc.

The scaled arrival time $\tau\(\btheta\)$ in (\ref{eq-tau}), 
\index{arrival time} visualized
as a surface, is called the arrival-time surface.  Much of lensing
theory is effectively the study of the arrival-time surface and its
derivatives, as we see below.

Note that although the wavefront \index{wavefront} 
and the arrival time surface look
similar and indeed are closely related \cite{n90}, they are not quite
the same thing.  The wavefront is a surface in real space whereas the
arrival time surface is in $\(\btheta,\tau\)$ space and thus a little
more abstract.

\subsubsection{Images and magnification}

The condition for images, from Fermat's principle \index{Fermat's principle}
and following (\ref{eq-lensprelim}) is
\begin{equation}
\bnabla\tau\(\btheta\)=0, \quad {\rm or}\quad
\bbeta=\theta-\bnabla\psi.
\label{eq-lens}
\end{equation}
The latter form is called the lens equation. \index{lens!equation} 
Its interpretation is
that the observer sees an image wherever the arrival-time surface has
a minimum, maximum, or saddle point.  Then consider the second
derivative of $\tau\(\btheta\)$, or curvature \index{curvature} 
of the arrival-time
surface. We have
\begin{equation}
\bnabla\bnabla\tau\(\btheta\) =
{\bf 1}-\bnabla\bnabla\psi\(\btheta\),
\label{eq-ddtau}
\end{equation}
a 2D tensor. (The bold-face ${\bf 1}$ denotes an isotropic
tensor---identity matrix in component notation.)  Meanwhile, taking
the gradient of the lens equation (\ref{eq-lens}) gives
\begin{equation}
\bnabla\bbeta = {\bf 1}-\bnabla\bnabla\psi\(\btheta\).
\label{eq-dlens}
\end{equation}
The curious term $\bnabla\bbeta$ expresses how much source-plane
displacement is need\-ed to produce a given small image displacement;
i.e., the inverse of magnification \index{magnification} 
\footnote{An alternative notation, 
$\partial \bbeta /\partial {\btheta}$,
reminds one of this physical interpretation.}. 
Equation (\ref{eq-dlens}) tells
us that magnification is a 2D tensor, and depends on $\btheta$ but not
$\bbeta$; let us write magnification as $\M$.  Comparing the last two
equations we have
\begin{equation}
\M^{-1} = \bnabla\bnabla\tau\(\btheta\).
\label{eq-mag}
\end{equation}
Equation (\ref{eq-mag}) means that the curvature \index{curvature} 
of the arrival-time
surface is the inverse of the magnification.  Thus, broad low hills
and shallow valleys in the arrival-time surface correspond to highly
magnified images; needle-sharp peaks or troughs correspond to images
demagnified into unobservability. \index{demagnification}

By curvature, \index{curvature} 
we mean a tensor curvature, which depends on directions:
$\M$ and $\M^{-1}$ are symmetric 2D tensors, so their components form
$2\times2$ matrices. In particular, we have
\begin{equation}
\M^{-1} = \pmatrix{
1-\pderiv(^2\psi/\theta_x^2) & -\secderiv(\psi/\theta_x,\theta_y) \cr
 -\secderiv(\psi/\theta_x,\theta_y) & 1-\pderiv(^2\psi/\theta_y^2) \cr
}.
\label{eq-magcomp}
\end{equation}
Comparing equations (\ref{eq-tau}) and (\ref{eq-mag} or
\ref{eq-magcomp}) we see that the trace of $\M^{-1}$ must be
$2(1-\kappa)$. Thus $\kappa$, originally defined as the surface
density in suitable units, also has the interpretation of an isotropic
magnification.  Accordingly, $\kappa$ is known as the convergence. 
\index{convergence}
The traceless part of $\M^{-1}$ is called the shear and its magnitude
is denoted by $\gamma$; \index{shear} 
it changes the shape of an image but not its
size.  In full, we have
\begin{equation}
\M^{-1} = (1-\kappa)\pmatrix{1 &0\cr 0 &1} - \gamma
\pmatrix{\cos2\phi &\sin2\phi \cr \sin2\phi &-\cos2\phi}
\label{eq-Meig}
\end{equation}
where $\phi$ denotes the direction of the shear.  Note that any
symmetric $2\times2$ matrix can be written in the form
(\ref{eq-Meig}).  All we have done here is interpret $\kappa$ and
$\gamma$.

The determinant
\begin{equation}
|\M| = [(1-\kappa)^2-\gamma^2]^{-1}
\label{eq-scalmag}
\end{equation}
defines a scalar magnification, \index{magnification} 
or ratio of image area to source area for
an infinitesimal source.

Surface brightness is conserved by lensing.  Although we will not
prove it here, this is a consequence of the fact that the lens
equation is a gradient map. \index{lens!equation} 
Magnification changes only angular sizes
and shapes on the sky.  Thus a constant surface brightness sheet stays
a constant brightness sheet when lensed.  (Were this not the case, the
microwave background would get wildly lensed by large scale
structure.)  However, an unresolved source will have its brightness
amplified according to (\ref{eq-scalmag}).

\subsubsection{Saddle-point contours, critical curves, caustics}
\index{saddle-points} \index{critical curves} \index{caustics}

The equations (\ref{eq-tau}) for the arrival-time surface, 
\index{arrival time}
(\ref{eq-lens}) for the image positions, and (\ref{eq-mag}) for the
magnification are elegant, but they do not give us much intuition for
the shape of the arrival-time surface, the possible locations of
images, and the likely magnification in real systems that we might
observe.  To gain some intuition, it is very useful to
introduce \cite{bn86} three special curves in the image and source
planes.

\begin{figure}[!ht]
\begin{center}
\includegraphics[width=.5\textwidth]{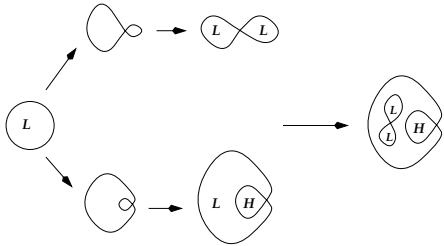}
\end{center}
\caption[]{Multiple images via saddle-point contours in the
arrival-time surface.  Here $L$ marks minima and $H$ marks maxima.}
\label{fig-lemlim}
\end{figure}

Consider the arrival-time surface and contours of constant $\tau$.  In
the absence of lensing $\tau\(\btheta\)$ is a parabola, and the image is
at its minimum, or $\btheta=\bbeta$.  For a small lensing mass, the
shape changes slightly from being a parabola and the minimum moves a
little.  But for large enough mass, a qualitative change occurs, in
that a contour becomes self-crossing.  There are two ways in which a
self-crossing can develop: as a kink on the outside of a contour line,
or a kink on the inside.  These are illustrated in
Figure~\ref{fig-lemlim}.  The outer-kink type is a lemniscate and the
inner-kink type a lima\c con.  With the original contour having
enclosed a minimum, a lemniscate produces another minimum, plus a
saddle-point at the self-crossing, while a lima\c con produces a new
maximum plus a saddle point.  (The previous sentence remains valid if
we interchange the words `maximum' and `minimum'.)  The process of
contour self-crossing can then repeat around any of the new maxima and
minima, producing more and more new images, but always satisfying
\begin{equation}
\hbox{maxima} + \hbox{minima} = \hbox{saddle points} + 1.
\end{equation}
The self-crossing or saddle-point contours form a sort of skeleton for
the multiple-image system.  Lensed quasars characteristically have one
of two configurations: double quasars \index{quasars!double} 
have a lima\c con, while quads \index{quasars!quadruple}
have a lemniscate inside a lima\c con, as in the rightmost part of
Figure \ref{fig-lemlim}.  Both cases have one maximum (marked `{\it
H\/}' in the figure), which will be located at the center of the
lensing galaxy. Since galaxies tend to have sharply-peaked central
densities, the arrival-time \index{arrival time} 
surface at the maximum will be sharply
peaked as well; the corresponding image is highly demagnified and (almost)
always unobservable. \index{demagnification}  
Thus lensed quasars are doubles or quads:
an incipient third or fifth image hides at the center of the lensing
galaxy.

Critical curves \index{critical curves} 
are curves on the image plane where the magnification \index{magnification}
is infinite.  More formally, they are curves where $\M^{-1}$ has a
zero eigenvalue.  From the definition (\ref{eq-mag}) it is clear that
at minima of $\tau$ both eigenvalues of $\M^{-1}$ will be positive, at
maxima both eigenvalues will be negative, and at saddle points one
\index{saddle-points}
eigenvalue will be positive and one negative.  Thus critical curves
separate regions of the image plane that allow minima, saddle points,
and maxima.

If we map critical curves to the source place via the lens equation
(\ref{eq-lens}) we get caustic curves. \index{caustics} 
Caustics separate regions on
the source plane that give rise to different numbers of images.

We discuss examples of saddle-point contours, critical curves, and
caustics in the next section.

\subsection{An illustrative macro-model}\label{sec-macmodel}

We have already met the point lens, \index{lens!point mass} 
which in dimensionless form has
lens potential
\begin{equation}
\psi\(\btheta\)=\thee\ln\theta
\end{equation}
where $\thee$ is effectively a parameter expressing the total mass.
Solving the lens equation, \index{lens!equation} we see that images are at
\begin{equation}
\btheta = \half\left(\beta \pm \sqrt{\beta^2+4\thee^2}\right)\hat\beta,
\end{equation}
where $\hat\beta$ denotes a unit vector in the direction of $\bbeta$.
The scalar magnification \index{magnification} is given by
\begin{equation}
|\M|^{-1} = 1 - {\thee^4\over\theta^4}.
\end{equation}
Another commonly used lens is the isothermal \index{lens!isothermal} 
lens (so called because of its
relation to isothermal spheres in stellar dynamics, and a good zeroth
order model for disk-galaxy halos \index{galaxy!halos} 
and giant ellipticals --- more on this
subject in the modeling section); it has
$\kappa\(\btheta\)=\half\thee/\theta$ and lens potential 
\index{lens!potential}
\begin{equation}
\psi\(\btheta\)=\thee\theta,
\label{eq-isopoten}
\end{equation}
For $\beta<\theta$ there are two images
at
\begin{equation}
\btheta=\bbeta+\thee\hat\beta, \qquad \btheta=\bbeta-\thee\hat\beta
\label{eq-isocims}
\end{equation}
and for $\beta>\theta$ the second of these disappears.  The constant
image-separation \index{image separation} 
in equation (\ref{eq-isocims}) is a peculiar feature
of the isothermal.  The scalar magnification is given by
\begin{equation}
|\M|^{-1} = 1 - {\thee\over\theta}.
\label{eq-isomag}
\end{equation}

Lacking any ellipticity, these lenses by themselves cannot produce
quads. \index{quasars!quadruple} 
But with some added ellipticity, quads and indeed all the main
qualitative features of quasar lenses can be reproduced, as we now
show.

As an example, consider the potential
\begin{equation}
\psi\(\btheta\) = (\theta^2+\epsilon^2)^{\frac12} +
\half\gamma\theta^2\cos(2\phi)
\label{eq-cisshear}
\end{equation}
where $\phi$ is the polar angle and
$\epsilon$ and $\gamma$ are adjustable parameters; $\epsilon$
gives the isothermal a non-singular core, \index{core} 
and $\gamma>0$ contributes
`external shear' which in this case amounts to extra lensing mass
outside the lens in the $y$ direction.  We take $\epsilon=0.1$ and
$\gamma=0.2$, and then examine what happens for different source
positions, through caustics, \index{caustics} critical curves, 
\index{critical curves} and saddle-point \index{saddle-points}
contours.  A similar potential, but with the scale and shear orientation
adjustable, will be used later (cf.~equation \ref{eq-sisshear}) to fit
data on observed systems.

\begin{figure}[!ht]
\begin{center}
\includegraphics[width=1\textwidth]{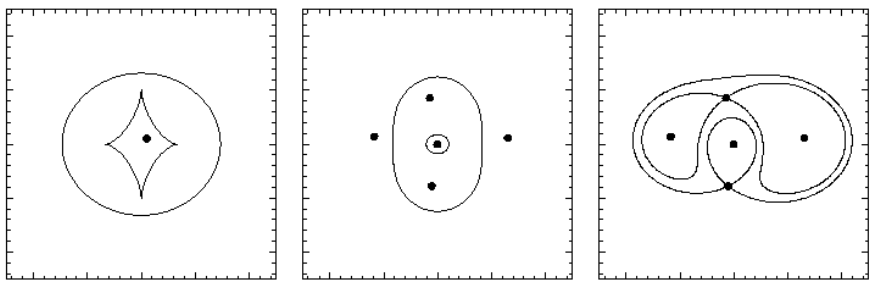}
\end{center}
\caption[]{A central quad: one with source near the center. {\bf Left
panel:}~source positions and caustics; {\bf middle panel:}~image
positions and critical curves; {\bf right panel:}~image positions and
saddle-point contours. In this figure, and in Figures~\ref{fig-long}
to \ref{fig-inc}, the left hand panels (showing the source plane) have
a scale half that of the other panels (showing the image plane).}
\label{fig-cen}
\end{figure}

\begin{figure}[!ht]
\begin{center}
\includegraphics[width=.4\textwidth]{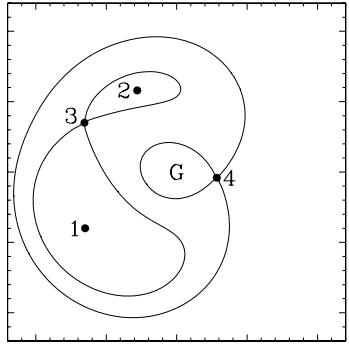}
\end{center}
\caption[]{Saddle point contours in a generic quad. Images 1 and 2 are
minima, 3 and 4 are saddle points; the fifth image would be a maximum
at the galaxy's centre G.}
\label{fig-imorder}
\end{figure}

Figure \ref{fig-cen} shows the situation with the source close to the
center.  The left panel shows what is happening in the source plane,
while the middle and right panel show what is happening in the image
plane.  Several interesting things may be seen.
\begin{itemize}
\item The two caustic curves in the source plane (left panel)
demarcate regions from where a source produces 1, 3, and 5 images.  In
this case the source is well within the inner caustic, and that
results in five images.  The other panels shows these five images,
along with the critical curves (middle panel) or the saddle-point
contours (right panel).  But the image near the center is highly
demagnified, and observationally such a system would be a quad.  Let
us call it a `central quad', to distinguish it from other quads we
will see below.
\item The two critical lines \index{critical curves} 
are maps of the caustics to the image
plane, but the {\it inner\/} caustic \index{caustics} 
maps to the {\it outer\/}
critical line. (Also, the long axes of both of these are aligned with
the potential.) Recall that critical lines are where an eigenvalue of
$\M$ changes sign.  The consequence for this lens is that any image
outside both critical curves is a minimum, any image between the
critical curves is a saddle point, \index{saddle-points} 
and any image inside both critical
curves is a maximum, all irrespective of the source position.  For the
current source position we can verify these statements by comparing
the middle and right panels.
\item The time-ordering of the quad's images is evident from the
saddle-point contours---compare with Figure~\ref{fig-imorder}.
\item The arrival-time contours and the arrangement of the images
appear to be squeezed in the $y$ direction.  Such squeezing is
characteristically along the long axis of the potential, and the
images appear pop out along the short axis of the potential.
\end{itemize}

\begin{figure}[!ht]
\begin{center}
\includegraphics[width=1\textwidth]{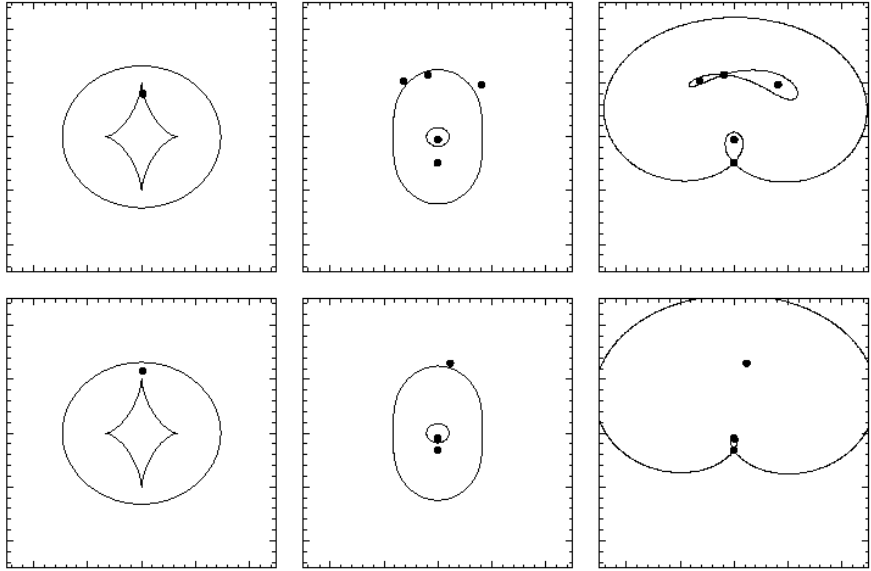}
\end{center}
\caption[]{A long axis quad and double. \index{quasars!double} 
\index{quasars!quadruple} Note how, as the source
crosses the diamond caustic, two images merge on the tangential
critical line and then disappear.}
\label{fig-long}
\end{figure}

Figure \ref{fig-long} shows the situation with the source is displaced
along the long axis of the potential.  As the source nears the inner
caustic curve, two of the images approach the outer critical curve.
We call this configuration a long-axis quad.  Two minima and a saddle
point are fairly close together, displaced in the same direction as
the source, while another saddle point is on the opposite side of the
lens center.  This 3+1 image arrangement reveals the direction of the
source displacement.  Meanwhile, as with the core quad, the
arrangement of the images is squeezed along the direction of the long
axis of the potential.  As the source crosses the inner caustic curve,
a minimum and a saddle point merge on the outer critical curve, and
then disappear.  The system is now a double, which we may call a
long-axis double.

\begin{figure}[!ht]
\begin{center}
\includegraphics[width=1\textwidth]{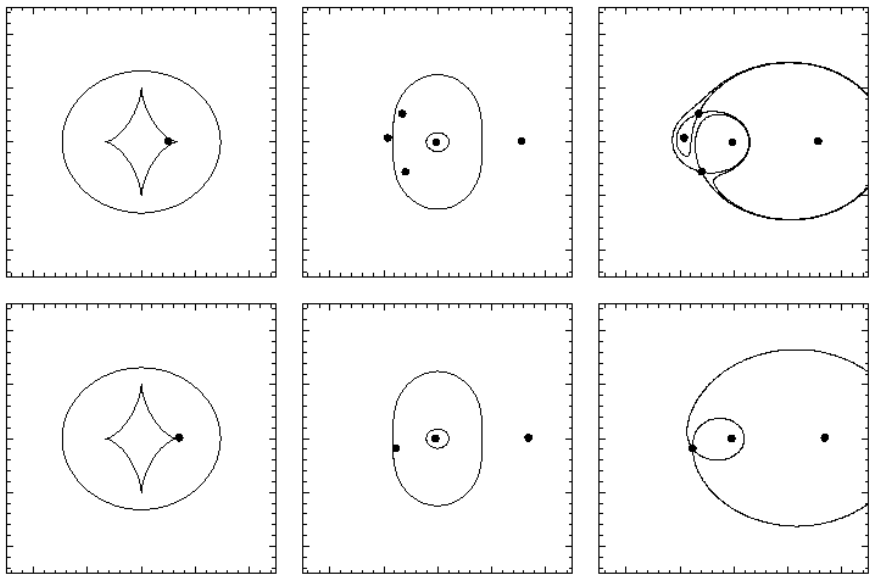}
\end{center}
\caption[]{A short axis quad and double.\hfill}
\label{fig-short}
\end{figure}

Figure \ref{fig-short} has the source displaced along the short axis of
the potential, producing configurations we call a short-axis quad and
a short-axis double.  The morphology of a short-axis quad resembles
that of the long axis quad, but one can tell them apart. First, one of
the four images is far from the others, but this time it is a minimum,
not a saddle point. Secondly, the 1+3 image arrangement indicates
direction of the source displacement, and it is perpendicular to the
long axis of the potential which can be inferred from the squeezing of
the image arrangement.  Moving the source outside the inner caustic
again causes two images to merge, leaving a short-axis double.  The
morphology of a short axis-double is the same as that of a long-axis
double.

\begin{figure}[!ht]
\begin{center}
\includegraphics[width=1\textwidth]{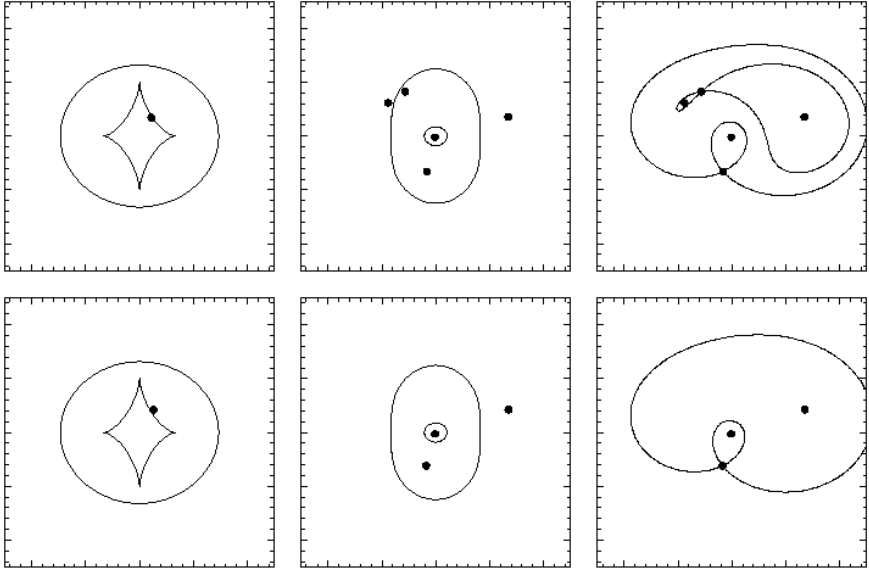}
\end{center}
\caption[]{An inclined quad and double.\hfill}
\label{fig-inc}
\end{figure}

Figure \ref{fig-inc} has the source displaced obliquely to the
potential, producing what we call an inclined quad and an inclined
double.  These are more common than the long and short-axis types, and
easily distinguished because of their asymmetry.

Examining the saddle-point contours in Figures~\ref{fig-cen} to
\ref{fig-inc}, the order of arrival times \index{arrival time} 
of the images is nearly
always evident.  We can summarize image-ordering in quads in the
following simple rules, illustrated in Figure~\ref{fig-imorder}:
(i)~Images 1 and 2 are opposite in Position Angle (PA), 
(ii)~3 and 4 are opposite in
PA, (iii)~1 is the furthest or nearly
the furthest from the lens centre,
(iv)~4 is the furthest or nearly the furthest from the lens centre,
(v)~if there are a nearly merging pair, they are 2 and 3.  For some
cases it is not possible to decide between 1 and 2, but otherwise
there is never an ambiguity.  For doubles, time ordering is trivial:
the image further from the galaxy arrives first.

With a little practice, it is easy to sketch the saddle-point contours
(including image ordering), and from there the critical curves and
caustics, of any quasar lens just from the morphology.

We may summarize the conclusions of this section as follows:
\begin{itemize}
\item From the morphology of a quad, it may be immediately recognized
as one of (i)~central, (ii)~long axis, (iii)~short axis, or
(iv)~inclined; doubles may be recognized as (i)~axis, or
(ii)~inclined, but long and short axis doubles need more information
to distinguish.  The `axis' in each case is of course the axis of the
potential, including any external shear; so morphology already gives
some idea mass distribution.
\item Morphology of quads or doubles also reveals the time-ordering of
images.
\end{itemize}

\subsection{Lenses within lenses: microlensing}
\index{quasars!microlensing}

Stars comprise an appreciable fraction of the mass in lensing
galaxies.  These stars produce small scale fluctuations in lens
potentials which will be seen to have substantial effects on the
magnifications of quasar images. \index{magnification}

\subsubsection{Random star fields}

Suppose that light from a source passes through a screen of $N$ equal
point masses with random positions, and that the Einstein rings of
individual masses are very small compared to the mean spacing between
them.  The delay equation (\ref{eq-t}) \index{time delay!geometric} 
consists of a single geometric
term and a great many Shapiro terms. \index{time delay!Shapiro} 
Each Shapiro term produces three
stationary points: a singular and completely demagnified maximum at
the angular position of the point mass, a significantly demagnified
saddle point close to the point mass and on the side opposite the
source position, and a minimum not far from the unperturbed source
position.  The minimum will be the only non-negligible image. 
\index{demagnification}

\begin{figure}[!ht]
\begin{center}
\includegraphics[height=6cm]{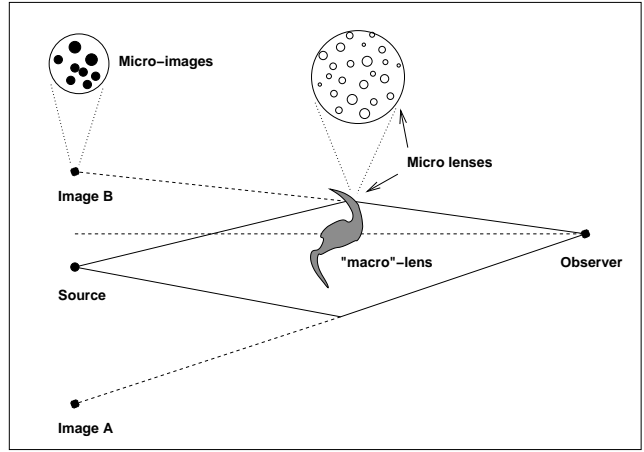}
\end{center}
\caption[]{Schematic representation of microlensing by stars
in a doubly imaged system. In this example, the unresolved ``sea'' of
stars in the main lensing galaxy is responsible for ``microlensing''
of one of the quasar images.}
\label{microlens}
\end{figure} 

Were the same stars to increase in mass without changing position, the
saddle points \index{saddle-points} 
would move further away from the stars, increasing in
brightness.  As their masses continue to increase, close pairs will
create new saddles between them.  For each new ridge, a new valley
will form on the side furthest from the source.  Just after the
formation of this new pair of images, the curvature \index{curvature} 
along the line
connecting them is very small and they are very highly magnified.  As
the masses continue to increase, the images separate and grow fainter,
though the new minimum will never be fainter than the unmagnified
source.

Thus the number of images increases from $N+1$ to $N+3$ to $N+5$ and so
forth. If we lacked the resolution to see the individual images, but
only the combined light, we would find that for the most part the
combined brightness increases steadily, but with bright flashes as new
pairs of images are created.  At any time our star field would have
some average surface density \index{density!surface} 
and an associated dimensionless
convergence, $\kappa$. \index{convergence} 
For an ensemble of such sources placed
randomly behind such a screen, we would expect an average scalar
magnification \index{magnification} 
of $(1-\kappa)^{-2}$ (see equation \ref{eq-scalmag}), but
there would be fluctuations depending upon the accidents of source
position.  Additional images begin to appear (in the absence of
external shear) when $\kappa$ approaches unity.

The general phenomenon of the amplification of unresolved images by
stars (or other point masses) in intervening galaxies is termed
microlensing.  The situation is illustrated schematically in Figure
(\ref{microlens}). The large numbers of highly demagnified saddle
points are not shown.

\subsubsection{Mandatory microlensing}

In the thought experiment of the preceding subsection, additional
positive parity images (minima) and their accompanying saddle points
formed when $\kappa$ approached unity.  The average density interior
to the Einstein ring of an isolated microlens is just the critical
density, \index{density!critical} 
with $\kappa\equiv1$.  The criterion for substantial
microlensing is therefore $\kappa \sim 1$.

Now let us suppose that the galaxy lensing a multiply imaged quasar is
comprised entirely of point masses.  The average surface density interior to
the galaxy's Einstein ring \index{Einstein!ring} 
is exactly the critical density.  Unless
the galaxy is very highly concentrated, the surface density at the
Einstein ring must be a substantial fraction of the critical density
-- one half in the case of an isothermal lens. \index{lens!isothermal} 
The covering factor \index{covering factor}
of the microlenses' Einstein rings \index{Einstein!ring} 
must therefore be a substantial
fraction of unity.  Thus microlensing must be important, {\it if\/} the
galaxy is comprised entirely of point-like objects.

Microlensing will be important only if the Einstein rings of the
particles comprising the galaxy are larger than the projection of the
source onto the sky.  There are two ways in which this might fail to
occur for lensed quasars.  First, the source might be large compared
to the the Einstein rings of the galaxy's stars.  Second, most of the
mass in the galaxy might be in particles with masses very much smaller
than that of a star, as we suspect would be the case for dark matter.
Our understanding of quasar sources and the distribution of dark
matter within galaxies is as yet so limited than we cannot say with
certainty whether microlensing should or should not be expected.  As
we shall see later there is considerable
observational evidence that the conditions for microlensing are met in
at least some lensed quasars.

It should be noted, by contrast, microlensing of sources in the
Magellanic Clouds by stars (and dark objects) in the halo of our Milky
Way is an exceedingly rare event. The covering factor for halo object
Einstein rings \index{Einstein!ring} 
is at most $10^{-6}$.  The largest source of this
difference is the small distance to the Clouds and the correspondingly
large value of $\Sigcrit$. \index{density!critical}

\subsubsection{Static and kinetic microlensing}
\index{microlensing!static}
\index{microlensing!kinetic}

In the above {\em Gedanken\/} experiment neither the source nor stars
were moving.  Imagine a symmetric lens which forms two quasar images
exactly opposite each other.  The images pass through regions of
identical surface density and shear, and would, in the absence of
microlensing, undergo the same magnification.  But since they pass
through different random star fields, they suffer different amounts of
microlensing.  The magnifications predicted from the global galaxy
potential would be only approximate -- one would have to take into
account the local fluctuations.  Static microlensing produces
``errors'' in the predicted fluxes.

Imagine further that the quasar consists of two components, one
smaller than the typical size of stellar Einstein rings and
the other larger.
The smaller component would be microlensed but the larger component
would not. \index{quasars!structure}

The motions of the source and the microlensing pattern add an
additional complication.  Taking the microlens positions to be fixed,
as the source moves the microlensing will change.  To order of
magnitude, the source must move an amount equal to the Einstein radius
of the microlens to produce a substantial change.  If the stars are
moving, they must move an amount comparable to the sizes of their
Einstein radii to produce substantial changes.  The temporal changes
in the brightness of an unresolved source are the result of such
kinetic microlensing.

\begin{figure}[p]
\begin{center}
\includegraphics[height=8cm]{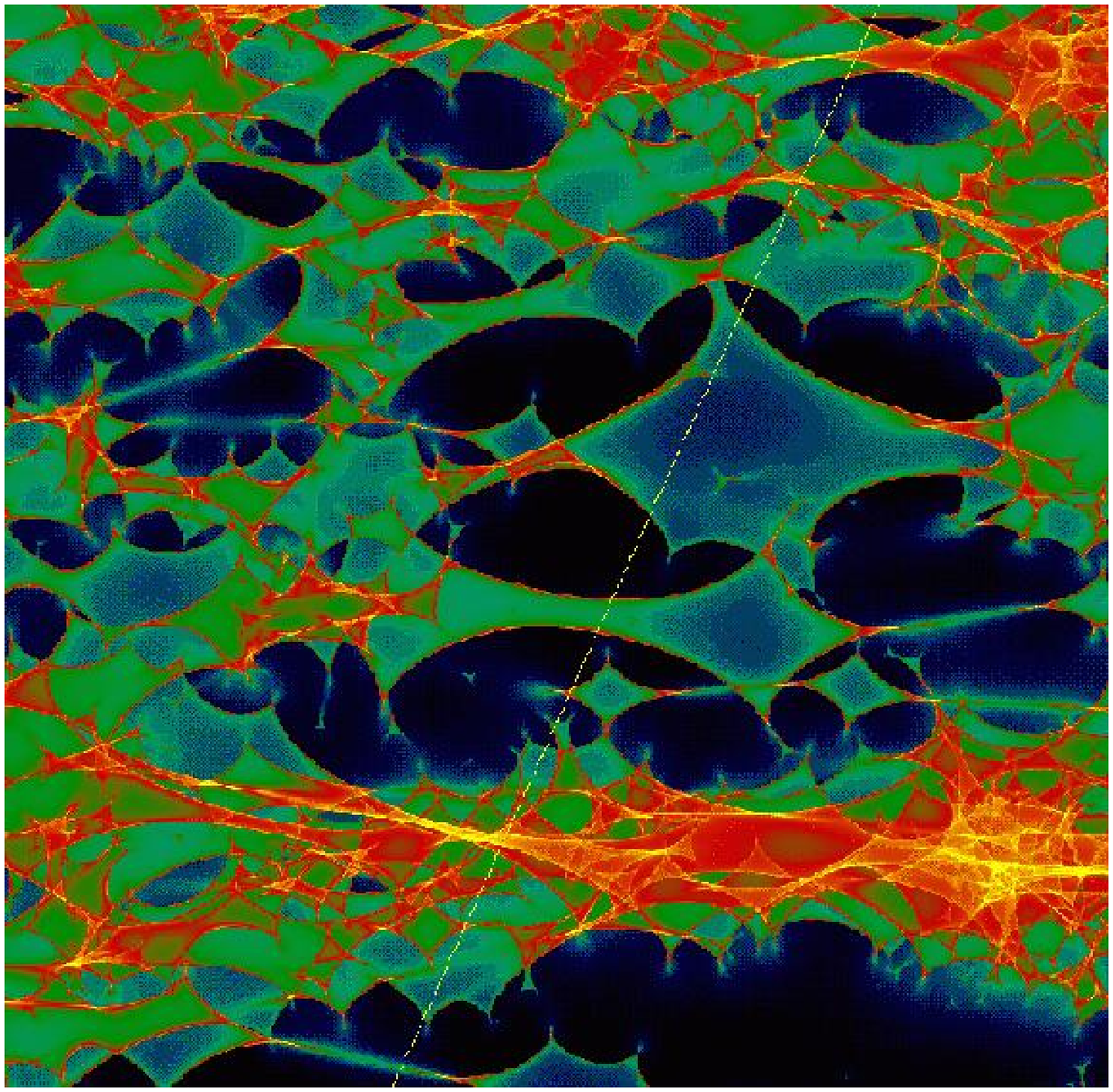}
\includegraphics[height=9cm]{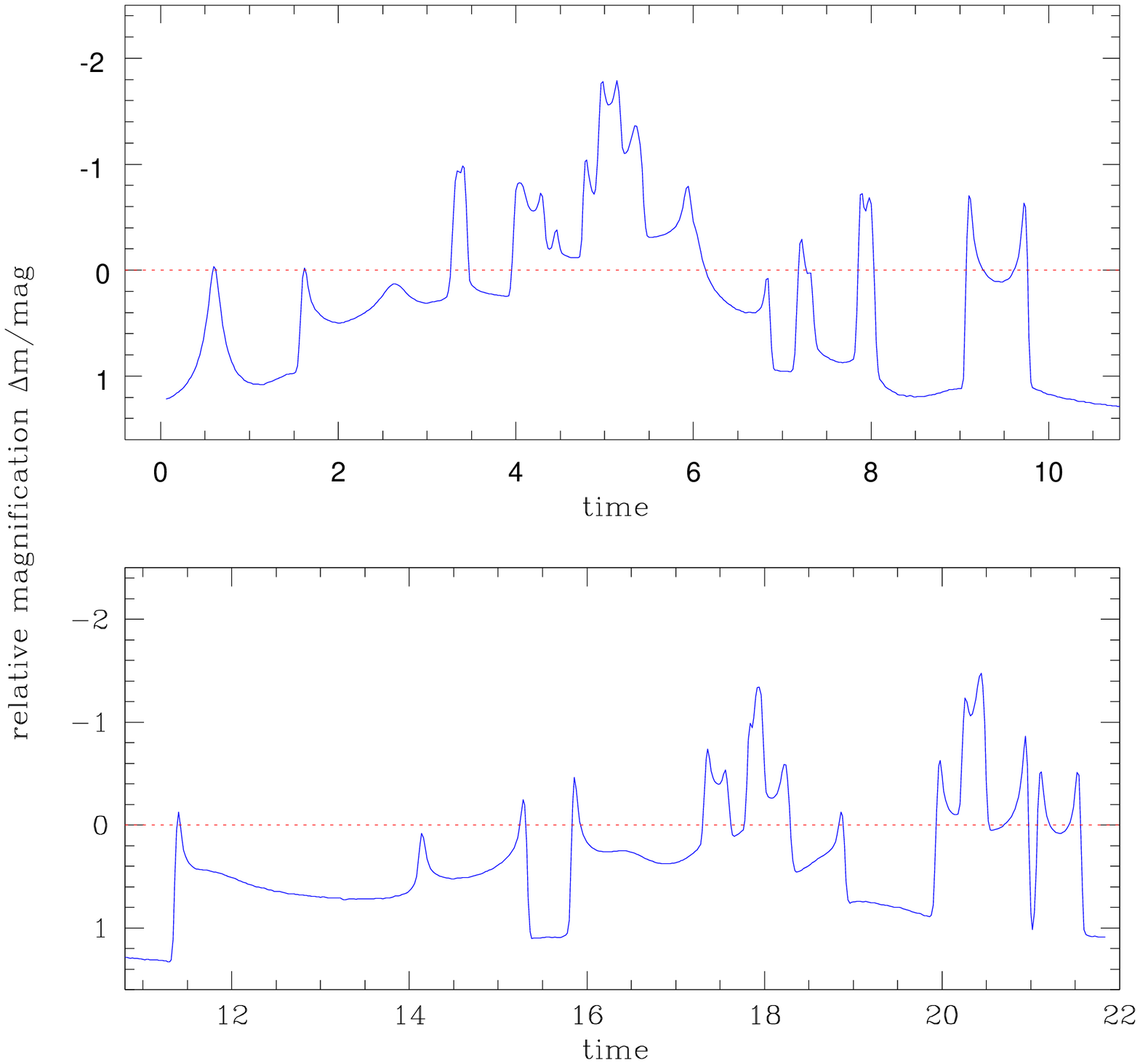}
\end{center}
\caption[]{{\bf Top:} network of micro-caustics in a lensing galaxy. 
The local convergence $\kappa$ is 0.5 and the shear $\gamma$ is 0.6. 
The bright regions correspond to high magnification while the darker ones
show de-magnification.
{\bf Bottom:} predicted light curve \index{light curve} 
when a source crosses the caustics
along the straight line in the top panel.
The time scale is arbitrary (Figure courtesy Joachim Wambsganss).}
\label{joachim}
\end{figure} 

\subsubsection{Microlensing caustics} \index{caustics} 
As described in Section 1,
critical curves \index{critical curves} 
are the locii in the image plane along which pairs of
images merge or are created as one varies the position of a background
source.  The scalar \index{magnification} magnification 
is infinite along the critical
curves.  This property suggests a relatively straightforward
computational scheme for identifying caustics, which are the locii in
the source plane which produce images on the critical curves.  Given a
set of (random) microlens positions, one projects rays back from the
observer uniformly in solid angle.  These land with high (low) surface
density in regions of high (low) magnification and the caustics
readily emerge when one plots a ``spot diagram'' for these rays.  Such
a plot also allows rapid computation of kinetic microlensing
light curves for a moving source -- one simply takes linear cuts
through the source plane spot diagram.  The magnification is
proportional to the local density of spots. Figure \ref{joachim} shows
such a plot, with a predicted light curve when a source crosses 
a network of caustics. \index{caustics network}

\subsubsection{Quantitative microlensing}
\index{microlensing!statistical}

Microlensing  is fundamentally  statistical  in nature.   It has  been
surprisingly resistant  to analytic techniques,  and most quantitative
work has been carried out via simulations.  These have shown
\cite{wam92,lewis96} that fluctuations of a 
magnitude or more are  possible for highly magnified images.  Moreover
saddlepoints behave differently  from minima, with larger fluctuations
for the former than for the latter
\cite{Witt92,schech02}.  Among the few interesting analytic results 
are an exact expression for the magnification probability distribution
at high  magnification \cite{schnei87}, and an expression  for the mean
number  of  positive parity  microimages  (minima)  as  a function  of
$\kappa$.


\subsection{The effect of cosmology}
\index{cosmology}

The main observables in lensing, image positions and magnifications,
are all dimensionless; only time delays \index{time delay} 
are dimensional.  The effect
of cosmology is to set the scale of time delays, and we can think of
it as setting $T_0$, the time scale in equation (\ref{eq-T0}).
Cosmology really enters only through the angular-diameter distances,
so fixing $T_0$ also fixes the other important scale, $\Sigma_{\rm
crit}$. \index{density!critical}

\begin{figure}[!ht]
\begin{center}
\includegraphics[width=.7\textwidth]{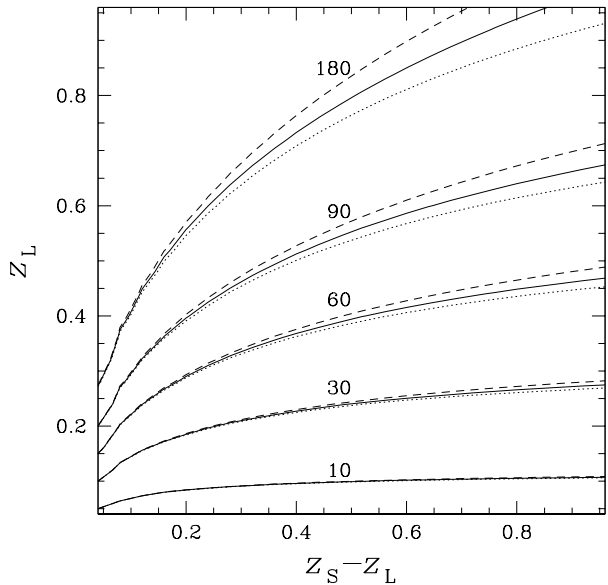}
\end{center}
\caption[]{Contour plots of $T_0$ as a function of $\zs-\zl$ and
$\zl$.  The labels are in units of $h^{-1}\rm days\;arcsec^{-2}$.  The
dashed curves are for $\Omega_0=1,\Omega_\Lambda=0$, the solid curves
are for $\Omega_0=0.3,\Omega_\Lambda=0.7$, and the dotted curves are
for $\Omega_0=0.1,\Omega_\Lambda=0$.}
\label{fig-T0}
\end{figure}

The time scale \index{time scale} has a dependence of the form
\begin{equation}
T_0 = h^{-1} \zl(1+\zl) \times
\<\hbox{weak function of $\zl,\zs,\Omega_0,\Omega_\Lambda$}>
\end{equation}
and is analytic \cite{fukugita92} but messy, so we do not reproduce it
here. Instead we illustrate it in Figure~\ref{fig-T0} for some
cosmologies.  It is worth remarking that
\begin{itemize}
\item $T_0\propto h^{-1}$ exactly;
\item for $\zs\gg\zl$ the approximation (\ref{eq-T0app}) applies.
\item for the same $h$, an Einstein-de-Sitter cosmology gives large
$T_0$, an open cosmology gives small $T_0$, with the currently favored
flat $\Lambda$-cosmology being intermediate; but the differences are
small.
\end{itemize}

The simple dependence on $h$ make it attractive to use time delays to
try and measure $h$.  One can even imagine putting in several time
delays on a sort of Hubble diagram to try and constrain
$\Omega_0,\Omega_\Lambda$.  Both these ideas are due to
Refsdal \cite{Refsd2,Refsd3}.

\subsection{Degeneracies}
\index{degeneracy}

Lensed images correspond to minima, saddle points, \index{saddle-points} 
and maxima of the
arrival-time \index{arrival time} 
surface; the rest of the arrival-time surface is
unobservable.  Thus, lensing observables do not uniquely specify a
lens; another lens that preserves $\tau\(\btheta\)$ and its derivatives
at image positions but changes them elsewhere will produce exactly the
same lensing data.  In this sense, lenses are subject to degeneracies.

An example, which we have already used when deriving the critical
density, is the monopole degeneracy: any circularly symmetric
redistribution of mass inwards of all observed images, and any
circularly symmetric change in mass outside all observed images will
change $\tau$ by at most an irrelevant constant in the image region,
and hence have no effect on lensing observables.  This means in
particular that doubles and quads contain no information about the
monopole part of the interior mass distribution, though they constrain
the total mass enclosed.  So in the example in
Section~\ref{sec-macmodel} our choice of core \index{core} 
\index{core!radius} radius was
irrelevant; it specified the location of the inner critical curve 
\index{critical curves} and
the outer caustic, but those played no part since images and sources
never went near them.

\begin{figure}[!ht]
\begin{center}
\includegraphics[width=.5\textwidth]{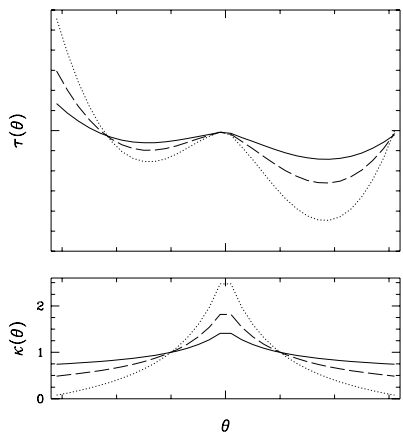}
\end{center}
\caption[]{Illustration \index{degeneracy!mass disk}
of the mass disk degeneracy, showing the
surface density (lower panel) and the arrival time (upper panel) for
three circular lenses. \index{lens!circular}  
The units, except for $\kappa$, are arbitrary.
The arrival time indicates a saddle point (which looks like a local
minimum in this cut), a maximum, and a minimum.  The dashed curves
correspond to a non-singular isothermal lens. 
\index{lens!non-singular isothermal}  Stretching the time
scale amounts to making lens profile steeper (dotted curves) and
shrinking the time scale amounts to making the lens profile shallower
(solid curves).}
\label{fig-mdd}
\end{figure}

In addition to degeneracies of the above type, which all involve
localized changes in the arrival-time surface, there is one special
degeneracy which is particularly serious: the mass disk
degeneracy \cite{fgs85,p86,ss95,ps00}. \index{degeneracy!mass disk} 
In this the $\tau$ scale of the
whole arrival-time surface gets stretched or shrunk.  To derive it we
rewrite equation (\ref{eq-tau}) first discarding a $\half\bbeta^2$
term since it is constant over the arrival-time \index{arrival time} 
surface, and then
using $\nabla^2\btheta^2=4$, to get
\begin{equation}
\tau\(\btheta\) = 2\nabla^{-2}(1-\kappa) - \btheta\cdot\bbeta.
\label{ntau-eq}
\end{equation}
Now the transformation
\begin{equation}
1-\kappa \to s (1-\kappa), \qquad \bbeta \to s\bbeta.
\label{eq-mdd}
\end{equation}
where $s$ is a constant which  just rescales time delays while
keeping the image structure the same; but since the source plane is
rescaled by $s$ all magnifications \index{magnification} 
are scaled by $1/s$, leaving
relative magnifications unchanged.  The effect on the lens is to make
it more like or less like a disk with $\kappa=1$.
Figure~\ref{fig-mdd} illustrates.  Note that in (\ref{eq-mdd}) $s$ can
become arbitrarily small; it can not become arbitrarily large because
otherwise $\kappa$ will become negative somewhere in the image region.
(Negative $\kappa$ {\it outside\/} the image region can always be
avoided by adding an external monopole).

From the modelers point of view, the mass disk degeneracy is a
degeneracy in the central concentration of the lens, or the steepness of
the radial profile, and we will meet this many-headed monster again
in the modeling section.
An easy way of remembering its effect is ``the lens gets
steeper as the universe gets smaller''.  The lensing data stay
exactly the same, and the mass inside an Einstein radius is
unaffected \cite{ps00}, but the sources before lensing get larger and
brighter, and $h$ gets bigger.  Which reminds us that this degeneracy
is particularly inimical to measuring $h$ from lensed quasars, where
it dominates the uncertainty.  In principle it could be broken in
various ways: if the intrinsic brightness of sources were known, or if
sources at very different redshifts were lensed by the same
lens \cite{asw98}, or indeed if $h$ were known from some other method.
But there seem no immediate prospects for any of these.

Another kind of degeneracy is associated with a non-lensing observable
that is often observed in connection with lensing, velocity
dispersion. \index{velocity dispersion} 
Lenses follow an approximate relation between Einstein
radius (or some surrogate for it in non-circular lenses such as the
size of the outer critical curve) and the line-of-sight velocity
dispersion: 
\begin{equation}
\thee \simeq 2'' \times {\<\vlos^2>\over(300\rm\,km\,s^{-1})^2}.
\label{eq-theedisp}
\end{equation}
To see why there should be such a relation, we rewrite the expression
(\ref{eq-thee}) for the Einstein radius of a circular mass
distribution as \index{Einstein!radius}
\begin{equation}
{GM\over\thee\Dl} = {c^2\over 4}{\Ds\over\Dls}\thee.
\label{eq-previrial}
\end{equation}
Now, the left hand side in (\ref{eq-previrial}) will be of order
$\<\vlos^2>$ because of the virial theorem, leading to
(\ref{eq-theedisp}).  The trouble is that the relation
(\ref{eq-theedisp}) cannot be made more precise, because the exact
coefficient that would go into it depends on the mass distribution in
a very complicated way.  In general, more centrally concentrated mass
distributions would give larger velocity dispersions.  On the other
hand, an isothermal sphere \index{isothermal sphere} 
in stellar dynamics gives
$(3\pi/2)\<\vlos^2>$ for the left hand side in (\ref{eq-previrial})
while a barely compact homogeneous sphere gives $5\<\vlos^2>$ --- almost
the same number despite the very different mass profile.

\section{Observations}

\subsection{Historical background}
While the concept of light deflection by massive bodies was already
proposed by Isaac Newton in the 18th century \cite{Newt}, the
astrophysical and cosmological \index{cosmology} 
potential of the phenomenon was, with
notable exceptions, taken seriously only after discovery in of the
first multiply imaged quasar by Walsh, Carswell \& Weymann
\cite{Walsh79}.  The observation of two well separated images of the
same source at $z=1.41$ not only confirmed the existence of what had
previously been seen largely as a theoretical curiosity, but also
established gravitational lensing as a new field of astrophysics.
Indeed, the existence of even a single lensed quasar, lent
considerable hope to the application of Refsdal's method
\cite{Refsd1,Refsd2} for determining the Hubble parameter 
\index{Hubble constant} $H_0$.
Proposed in 1964, the method is based on the measurement of the light
variations in the lensed images of a distant source.  The time lag, or
so-called ``time-delay'' \index{time delay} 
between the arrival times of the luminous
signal from each image of the source to the observer, is directly
related to $H_0$ and to the mass distribution in the lensing object.
Measuring the time-delay therefore provides us, via a mass model for
the lensing galaxy, with an estimate of $H_0$.  Refsdal originally
proposed to apply his method to distant supernovae.  The discovery of
quasars by Schmidt \cite{schmidt63} offered new prospects in using
even more distant light sources.

Measuring time-delays is far from trivial: the angular separations
between the lensed images are usually small, typically 1-2 arcsec, and
not all quasars are willing to show measurable photometric variations.
In addition, characterizing the mass distribution responsible for the
lensing effect, assuming the lensing galaxy is detected at all, was
very challenging at the time of the first discoveries.  CCD detectors,
were only just coming into use.  They were hard to obtain and had small
formats and high read noise.  The uncontrolled thermal environments of 
telescopes produced mediocre seeing, typically larger than 
the angular separations observed in most
presently known objects (see Tables at the end of this chapter). 
The Hubble Space
Telescope (HST) was more than a decade off in the future.  
Despite these difficulties, 
searching for new systems suitable for cosmological investigations became a
major activity in the early eighties.  Based on the
argument that some of the brightest quasars might be magnified
versions of a lower luminosity object (e.g., Sanitt \cite{sanitt71}),
systematic searches for new multiply imaged sources were undertaken
among the apparently brightest quasars.  These searches yielded the
discovery of more doubles, \index{quasars!double}
like UM~673 \cite{surdej87}, but also new image configurations. 
PG~1115+080 \cite{wey80} was thought to be
triple but turned out to be an off-axis quadruple
with higher resolution observations \cite{hege81}. More symmetric
quadruples, \index{quasars!quadruple} such
as the ``cloverleaf'' \cite{magain88}, were also found.  
Almost simultaneously, radio
searches yielded their first results.  As the radio emitting regions
of quasars are larger than the optical ones, lensed radio loud quasars
were often found to be complete Einstein rings: \index{Einstein!ring} 
MG~1131+0416, 
MG~1654+1346, PKS~1830-211 \cite{hewitt88,langston89,subra90}.  The
observation of complete or partial rings offers more constraints than
2 or even 4 point source images and led to the development of more
accurate models \cite{KoNa92}.

During this same period, systematic campaigns were initiated to
measure time delays, \index{time delay} 
much of it concentrated on the first lens
discovered: Q~0957+561.  Early reports \cite{schild86},
\cite{florentin84} gave contradictory results.
Vanderriest et al.
\cite{vander89} and Schild et al.  \cite{schild90} derived a value of
415 days, from ground based optical observations.
Press et al. 
\cite{press91} reanalyzed Vanderriest's data and published a very
different time delay: 536 days, a value supported by the radio
monitoring results obtained at the Very Large Array (VLA)
\cite{roberts91}.  The dictum attributed to Rutherford \cite{bailey67},
``If your experiment needs statistics, you ought to have done 
a better experiment," appears to have been
borne out.  Improved optical \cite{kundic97,oscoz97} and
radio monitorings \cite{Haarsma99} have finally settled the issue.
They reconcile the optical and radio time delays and lead to the value
of  $\Delta t = 417 \pm 3$ days.

The controversy over Q~0957+561 reflects the difficulty
of measuring time delays.  Quasars do not commonly show very sharp
light variations, and their light curves are often corrupted by the
erratic photometric variations induced by microlenses (stars) in the
lensing galaxy (see Chapter 1 and Section 2.3 of the present
Chapter). Photometric monitoring 
over a period considerably longer than the time delay is therefore
necessary.  Temporal sampling must also be sufficiently frequent
to average out short timescale microlensing variations. 
\index{quasars!microlensing} Microlensing may
corrupt quasar light curves but it is of considerable interest for
constraining the 
statistical mass of MACHOs (see Chapter 1) in the lens \cite{wamb00}
and the size of the lensed source \cite{wamb90}. With particularly
good data obtained over a wide wavelength range, it might even be
possible to reconstruct some of the quasar's accretion disk
parameters, such as, size, inclination and details of the
spectral energy distribution of the accretion disk as a function of
distance from the AGN's center \cite{agol99,Wyithe00}.

Progress with CCD detectors, with radio interferometers and
with image processing techniques has made it possible to overcome 
at least some of
the observational limitations on time delay measurements.  The list
of systems with know time delays is rapidly growing, with
optical and radio time delays both available in some cases.
Schechter et al. \cite{schech97} obtained optical light curves for
PG~1115+080 and two time delays between two images and the group of
blended bright images A1+A2. Three time delays have been measured from
radio VLA observation, in the quadruply imaged quasar CLASS B1608+656
\cite{fass99a}. The two bright radio doubles PKS~1830-211 and
B~0218+357 are two other cases with known time delays (e.g.,
\cite{lov98,biggs99}). Note also the lucky case of B~1600+434 
which has both optical \index{time delay}
\cite{Ingunn00} and radio \cite{koop00} time delays
and even overlapping light curves. Many more time delays have
recently been obtained at the Nordic Optical Telescope or at ESO
\cite{burud01,burud01b,hjorth02}.

The level of interest in lensed quasars has followed a more or less
predictable course.  The considerable excitement following what was
effectively the birth of the field in 1979 was followed by
extraordinary growth, as measured by the number of papers published
\cite{mrssurdej} and number of lenses known \cite{kochanekbu}.  
The phenomenon is no longer be so novel, it is entering 
a more mature, and astrophysically and cosmologically, more 
productive stage.  Observational and theoretical advances have
proceeded in parallel, with considerable improvement in
``best'' estimates of $H_0$, in
weighing distant galaxies, and in probing their stellar and
dust content. And as with other areas of astrophysics, there
is an increasing tendency toward large, international teams, 
marking the substantial demands of the enterprise.

\subsection{Observational constraints in quasar lensing}
\index{quasars!observations}

Given the small deflection angles involved in multiply imaged quasars,
high resolution observations are required to measure accurately
the main observables: the position of the quasar images and of the
lens, the time-delay, and the magnification at the position of the
images.
 
\subsubsection{The image configuration and the time delay:} 
For most known lens systems it has become relatively straightforward
nowadays to obtain astrometry of the requisite precision, especially
when HST images are available. However, adequate temporal sampling is
also required as soon as the goal is to measure the time
delay. Photometric monitoring \index{photometric monitoring}
with $0.1''$ resolution would be
possible with HST but one could not realistically expect the large
numbers of orbits necessary.  Until recently, such work was restricted
to radio wavelengths, less affected by weather conditions than optical
ones and providing data with higher resolution on a more regular
basis. Scheduling is also facilitated as one can observe in day time.
But as only 10\% of quasars are radio loud, this restricts the
available sample of lenses.

Recent advances in image processing techniques have extended
the range of ground-based imaging into the subarcsecond regime.
Typically reconstruction and deconvolution techniques can only 
be used with relatively high signal-to-noise observations, 
but in such cases improvements of a factor of two in resolution 
have been possible \cite{magain98}.  

\begin{figure}[!ht]
\begin{center}
\includegraphics[height=3.9cm]{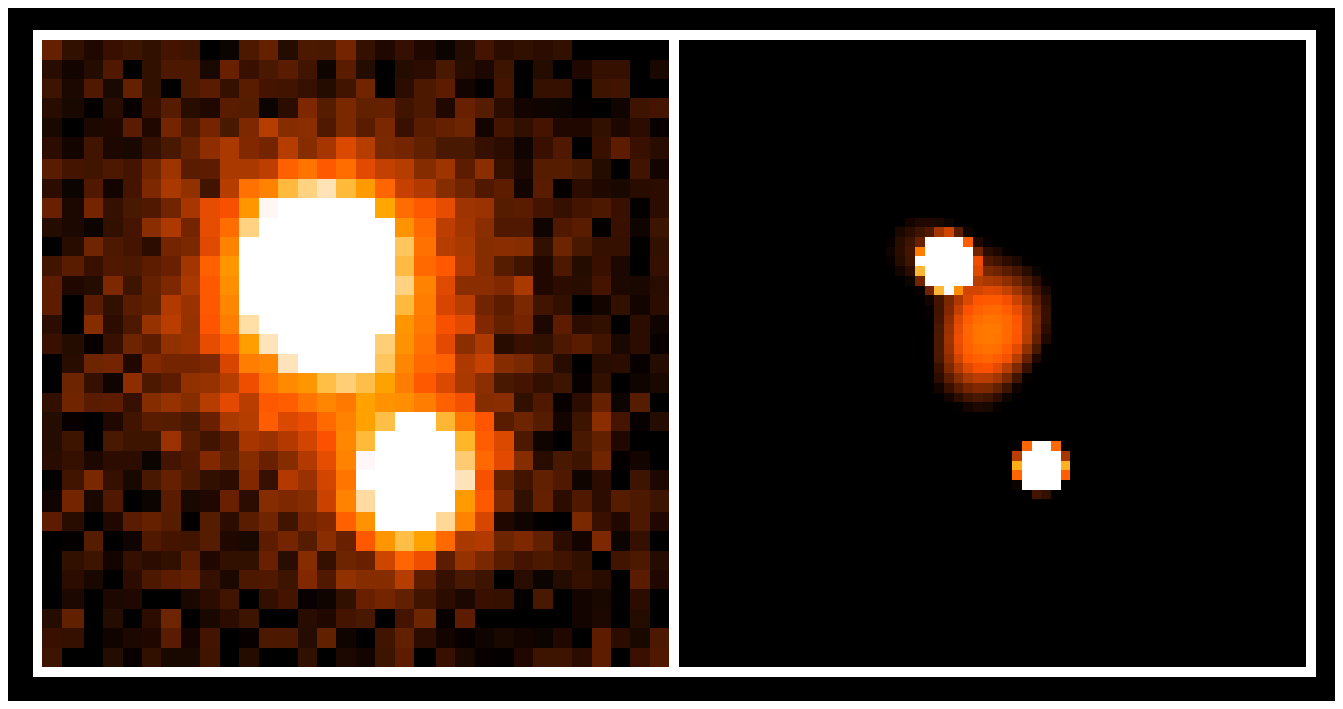}
\leavevmode
\includegraphics[height=3.9cm]{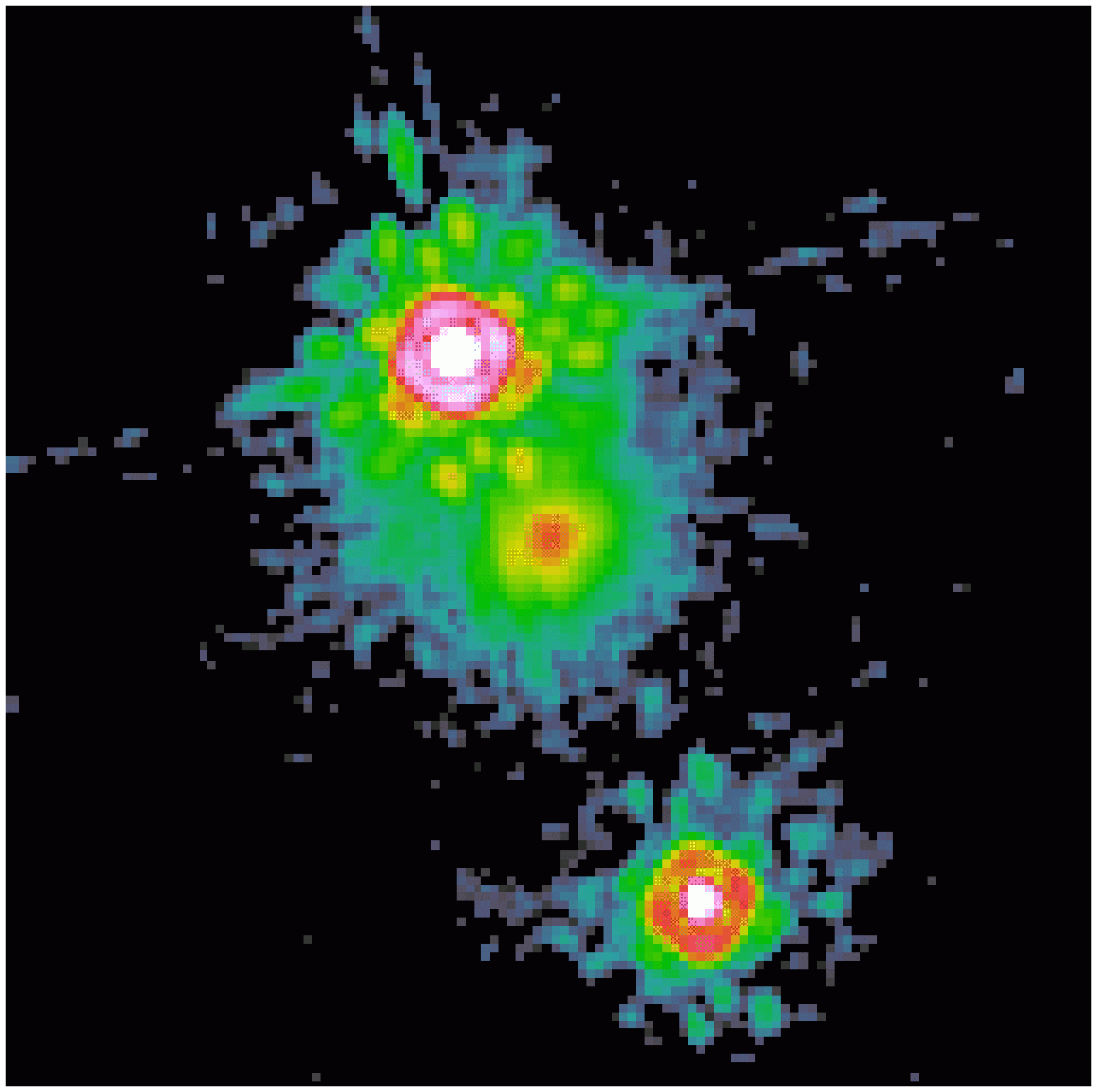}
\end{center}
\caption[]{Two ways of obtaining high resolution images of 
lensed quasars. From left to right: near-IR ground based image of
HE~1104-1805. It has been obtained with the ESO/MPI 2.2m telescope in
the $J$-band under average seeing conditions ($0.7''$). Its
resolution has been improved down to $0.27''$ on the deconvolved
version of the data displayed in the middle panel. The lensing galaxy
is obvious, between the QSO images. Its position and elongated shape
oriented with a PA of about 30 degree are confirmed by the $H$-band
HST/NICMOS image shown on the right (HST image from the CASTLEs
survey).}
\label{he1104}
\end{figure}

Figure~\ref{he1104} shows an
example of high resolution data of the doubly imaged quasar
HE~1104-1805 \cite{wiso93}, as might be obtained either with the HST
(here, in the near-IR) or from post-processed (deconvolved) ground
based images. The data presented in this figure are sufficient to
infer the image and lens positions with an accuracy from a few
milli-arcsec (quasar images) to a few tens of milli-arcsec (lensing
galaxy). In fact, the combination of space observatory data
\cite{lehar2000} and post-processed ground based data now allow for
accurate photometric monitoring in the optical and for detailed
modeling.

The range in properties of lens systems is such that there is no
single factor which consistently limits one's ability to carry out a
determination of $H_0$ through time delay measurement.
It seems that there is no ``golden
lens,'' no ideal case that will give a ``best'' measurement of
$H_0$. In some cases, the error on the time delay dominates (for
example HE~2149-2745 \cite{burud01}), while in other 
systems more symmetric about the center
of the lensing galaxy, the errors introduced by the astrometry of the
quasar images will dominate
\cite{schech97,keeton97,courbin97,impey98}. In still other cases 
the erratic variations of the light curves introduced by microlensing 
events in
the lensing galaxy are the main source of error
\cite{wiso98,Ingunn00}. It therefore seems more reasonable to monitor
as many systems as possible rather than trying to concentrate on a
particular one which might have its own unknown sources of systematic
errors.

\begin{figure}[!ht]
\begin{center}
\includegraphics[height=5.5cm]{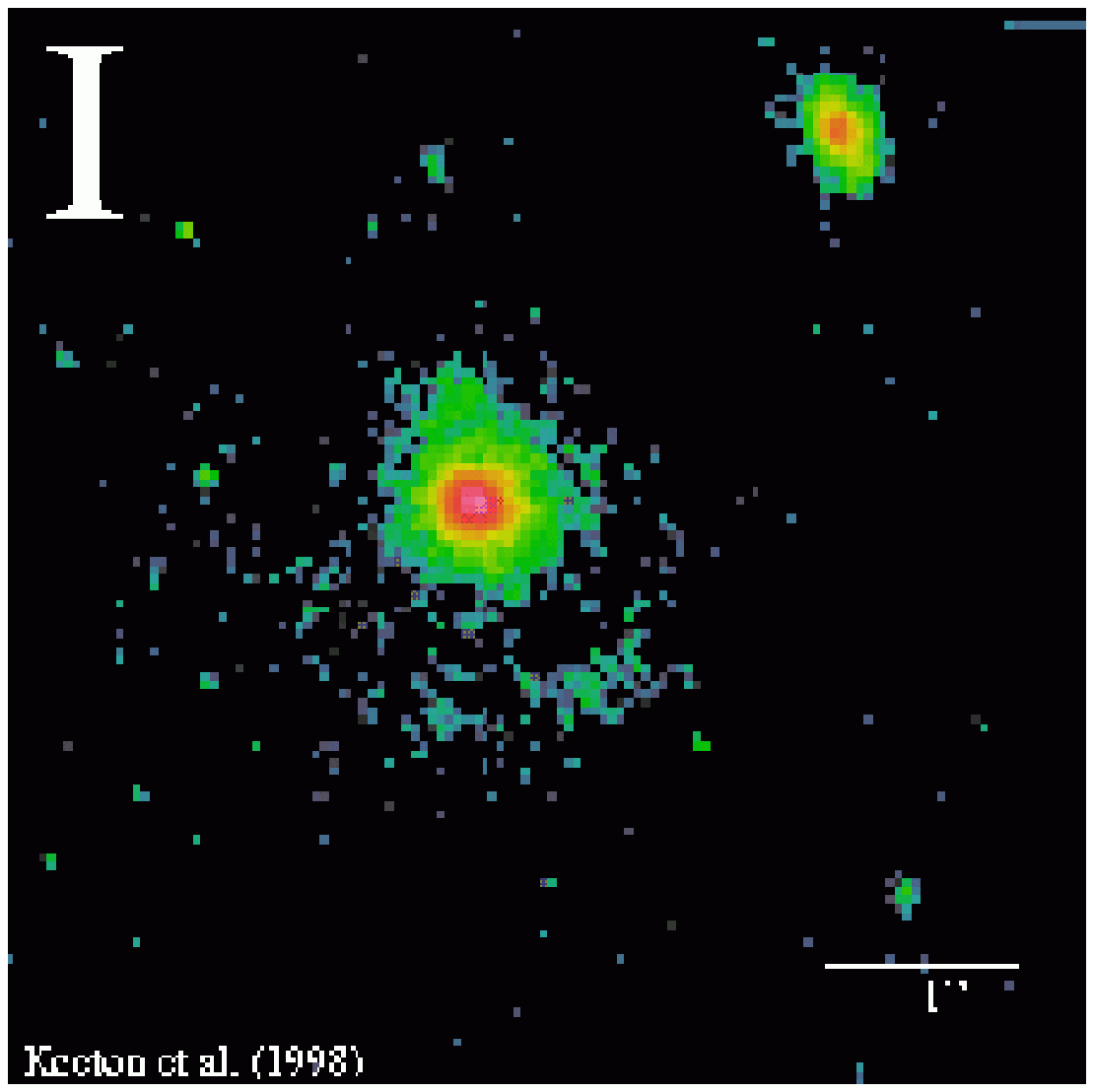}
\leavevmode
\includegraphics[height=5.5cm]{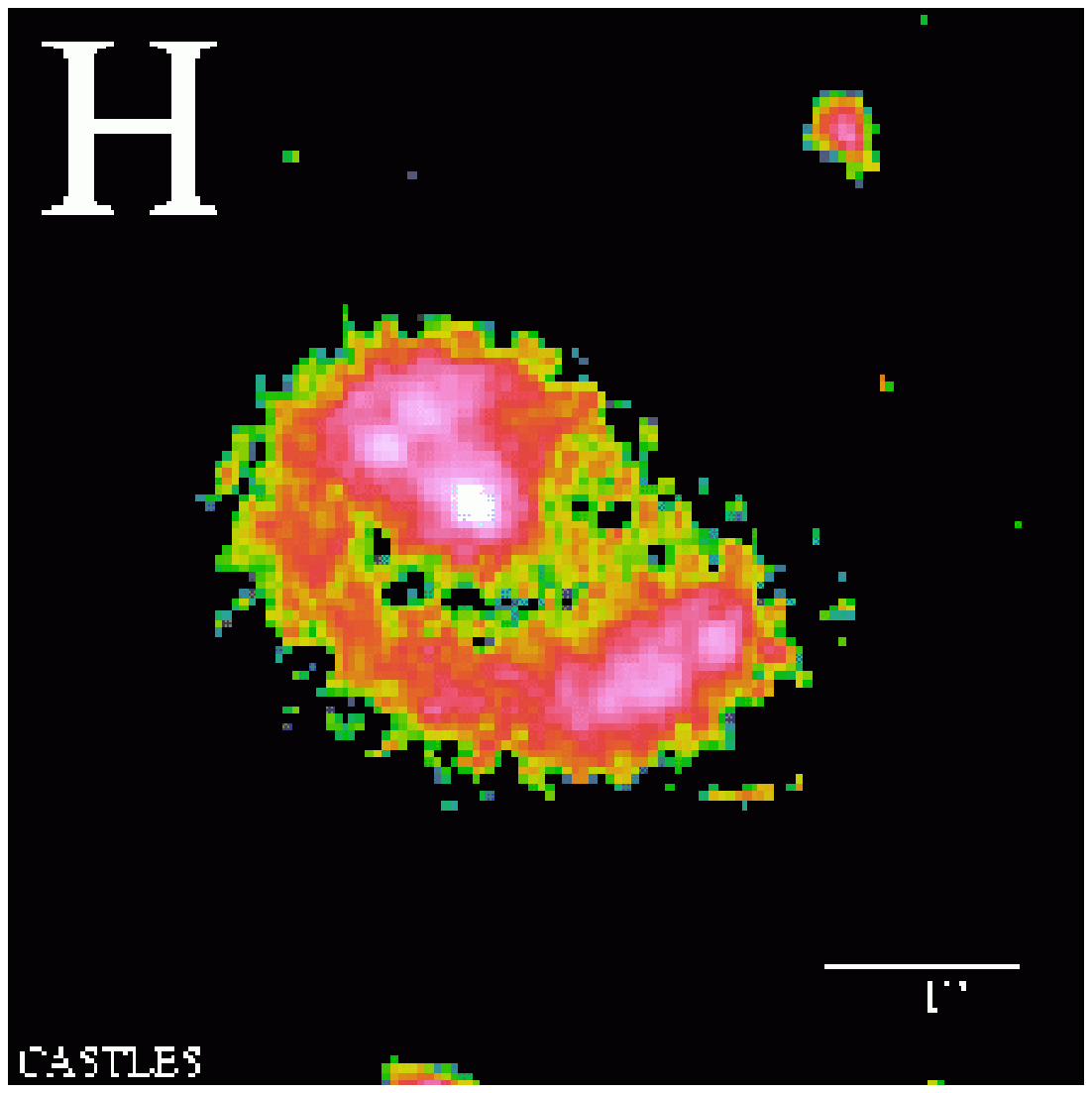}
\end{center}
\caption[]{The lensed radio source MG~1131+0456 is a nice example of system
with the source only visible in the near-IR and longer wavelengths. On
these HST images obtained by the CASTLEs group, only the lens is
visible (left) in the optical $I$-band. On the $H$-band image (right),
the source is seen as a almost full Einstein ring.}
\label{mg1131}
\end{figure} 

\subsubsection{Distances to the source and lens} 
\index{distances}

As seen from equation (\ref{eq-t}), modeling lensed quasars requires
knowledge of the distance $\Dl$ to the lens, and of the distance $\Ds$
to the source.  While the lensing galaxies are not especially faint by
current standards, measuring their redshifts is non-trivial.  In the
optical, the background quasar is often bright and hides the much
fainter lensing galaxy. In some lucky cases, the lensing galaxy shows
emission lines in superposition on the quasar spectrum
\cite{tonry99}, but this is not the rule. To date, no HST
spectrum has been taken of a lensing galaxy, but application of
deconvolution techniques \cite{courbin00} to spectra obtained on
ground based 10m class telescopes have proved useful and have yielded
the measurement of several lens redshifts \cite{lidman2000,burud01}.

There are a number of cases where the lens and source have such
different spectral energy distributions that they must be observed at
very different wavelengths. MG~1131+0456 (see Figure~\ref{mg1131}) and
PKS~1830-211 are examples of systems where the source can be seen
only in the near-IR. In the case of PKS~1830-211, the source's
redshift could be determined only from IR spectroscopy
\cite{lidman99}. Other more extreme cases like MG~1549+3047 show the
lens only in the optical/near-IR and the source only in the radio. As
such systems show no light contamination by the background source,
they allow for a detailed study of the lens. In the case of
MG~1549+3047, the velocity dispersion of lensing galaxy could even be
measured \cite{lehar93,lehar96}. A major drawback however, is that the
redshift of the source remains unknown.

\begin{figure}[!ht]
\begin{center}
\includegraphics[height=9.0cm]{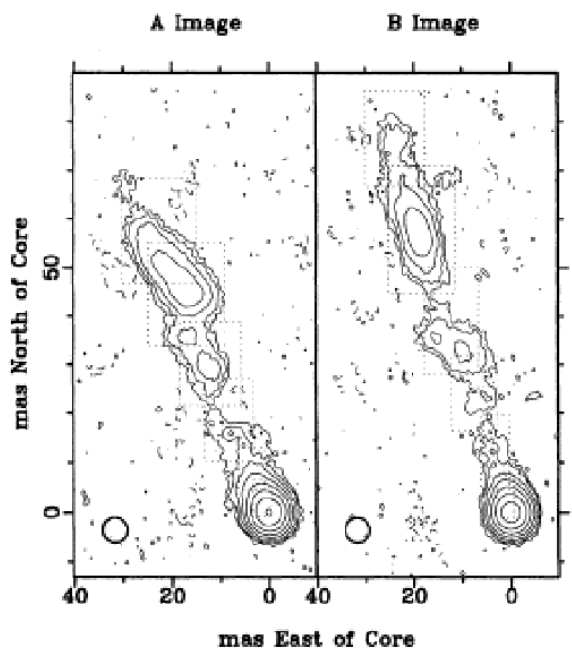}
\end{center}
\caption[]{The double quasar Q~0957+0561 observed in VLBI \cite{campbell95} 
at 6 cm with a resolution of 6 milli-arcsec. The two quasar images show
a very detailed radio jet that is used to place constraints on the
lens model.}
\label{0957vlbi}
\end{figure}
 
\begin{figure}[!ht]
\begin{center}
\includegraphics[height=9.0cm]{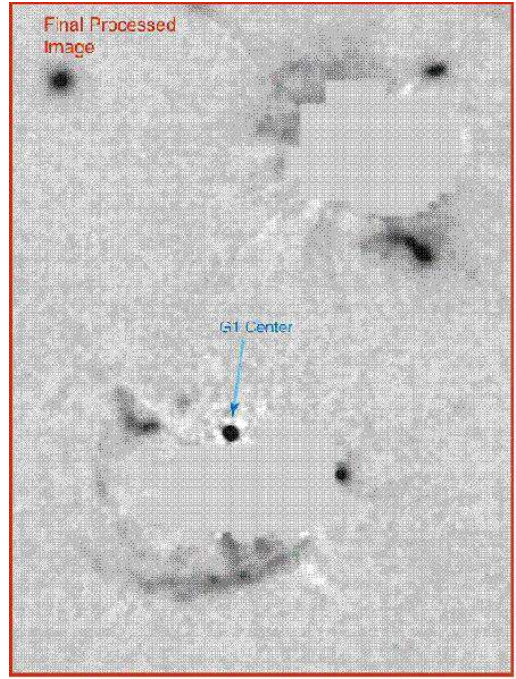}
\end{center}
\caption[]{Using the HST, arcs and arclets are discovered in the field
of Q~0957+0561 \cite{berns97} and help to determine the mass profile
of the lensing galaxy. The quasar components have 
been subtracted on this STIS image, provided to us by Gary Bernstein 
and Phil Fischer, prior to publication. Several arcs are visible as well
as members of a 
foreground galaxy cluster at z$\sim$0.35 \cite{angonin94}. G1 is
the center of the main lensing galaxy which has also been removed 
from the image.}
\label{0957hst}
\end{figure} 

\subsubsection{The quasar host galaxy and background objects}
\index{quasars!host galaxy}

At very high angular resolution, it becomes possible, beyond
measuring the position and brightness of the quasar, to
resolve details in the distorted and amplified quasar host. Observing
the distorted quasar host galaxy brings extra constraints on the
lensing potential, and helps to see distant quasar host galaxies
(up to redshift 4.5) that would have been missed without the lensing
magnification \cite{lehar2000,keeton00}. Figure~\ref{mg1131} shows an
example of a red quasar host galaxy where small details are unveiled, at
the resolution of the HST (about $0.15''$ in the $H$-band). Such
information is of importance as each detail might be identified in the
counter-image and used to place additional constraint on the
reconstruction of the lensing galaxy's mass profile. In the radio,
using Very Long Baseline Interferometry (VLBI) \index{VLBI} 
with resolution on the
order of the milli-arcsec, 
``blobs'' can be seen in the lensed images of the radio jet in the
source. Such observations, producing the spectacular maps shown in
Figure~\ref{0957vlbi} are restricted to very few objects with such
high spatial resolution \cite{campbell95,ros2000,trotter00}.

\begin{figure}[!t]
\begin{center}
\includegraphics[height=10cm]{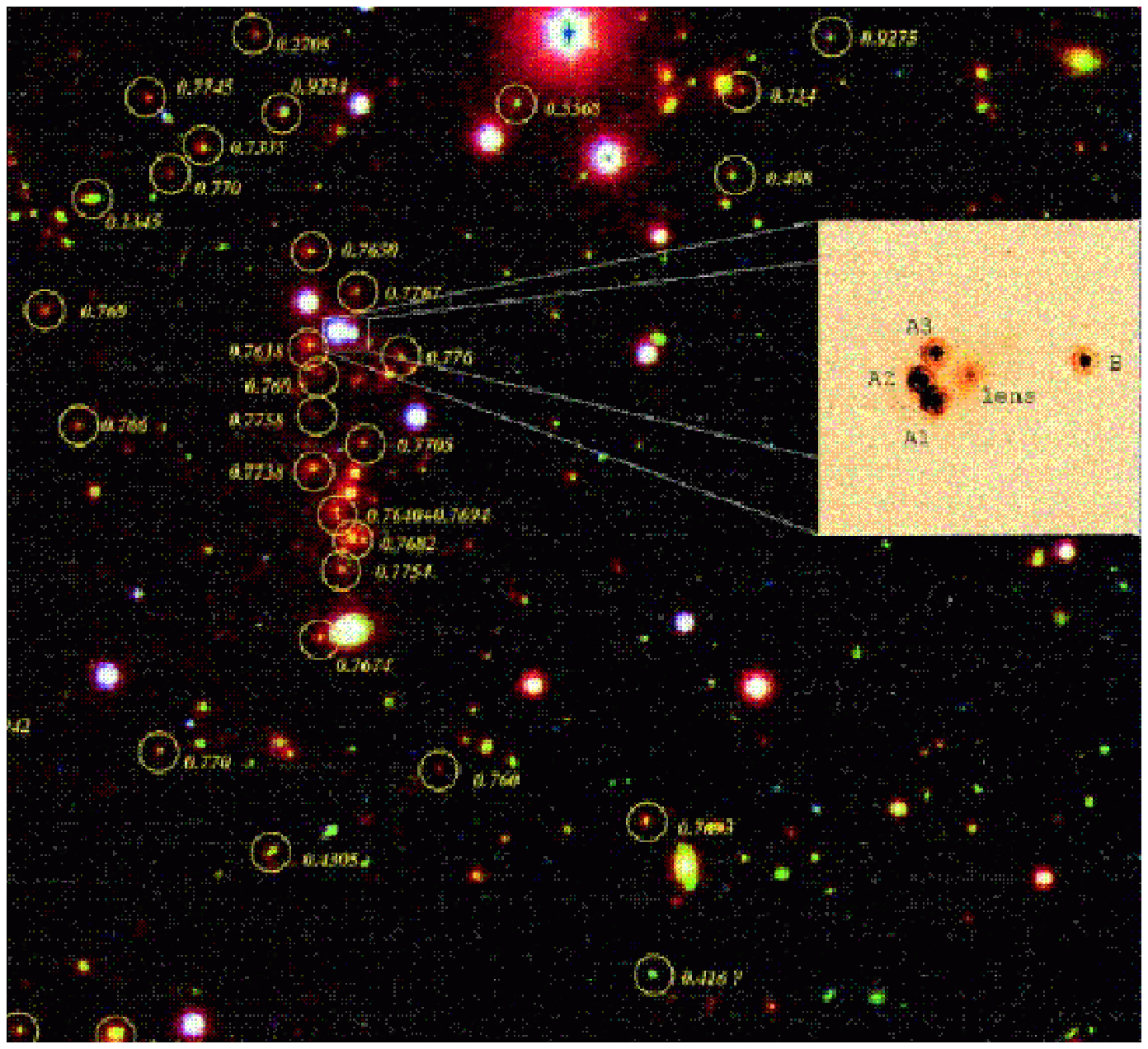}
\end{center}
\caption[]{The quadruply imaged quasar RX~J0911+0551 \cite{bade97} 
at $z=2.8$ and the intervening cluster
at $z=0.77$ \cite{kneib00}
which significantly modifies the overall gravitational potential
responsible for the lensing effect. The field of view is $3.5'
\times 3.5'$.}
\label{rxj0911_jp}
\end{figure} 

Arcs and arclets \index{arcs} \index{arclets} 
are sometimes seen in the immediate vicinity of the
quasar components. These objects, as shown in Figure~\ref{0957hst},
might be companions to the quasar or simply unrelated background
sources \cite{berns97}. As in the case of the quasar host, they probe the
lensing potential, with the further advantage that they do no lie at
the same position as the quasar and therefore probe the lensing
potential in a location otherwise inaccessible.

\subsubsection{Intervening clusters/groups}
\index{galaxy!groups}
\index{galaxy!cluster}

Isolated lenses may be the exception rather than the rule. 
Multiply imaged quasars often lie close to the line of sight to
foreground groups and even clusters of galaxies. A massive galaxy
cluster ($\sigma \sim 600-1000\, {\rm km\,s^{-1}}$), even situated
several tens of arcsecs away from a system, will modify the expected
image position and the time delay, hence also modifying the infered
value for $H_0$. Therefore one has to set constraints not only on the
astrometry and shape of the main lens, but also on additional objects
that may modify the total gravitational potential responsible for a
given image configuration. 

In the case of B~1600+43
\cite{jaunsen97,Ingunn00} the lens is an edge-on spiral 
at  $z=0.41$ \cite{fassnacht98}  with a  lower redshift  spiral  a few
arcsec  South-East  \cite{jaunsen97}.   The  quadruply  imaged  quasar
PG~1115+080 can be  modeled only by taking into  account a small group
\index{galaxy!groups} 
of 4-5 galaxies  at $z=0.31$ (the same as the  lens) about $20''$ away
from the  line of sight  \cite{keeton97}. Two spectacular  examples of
intervening clusters  are RX~J0921+4529 which is situated  in an X-ray
cluster \index{cluster!X-ray} 
at $z=0.32$ \cite{munoz01} and RX~J0911+0551 (see Figure
\ref{rxj0911_jp}). The later is lensed by a galaxy at $z=0.77$, a member
of   an   X-ray  cluster   centered   about   $30''$   to  the   south
\cite{burud98,morgan01b}. The  cluster's velocity dispersion  has been
measured  from   the  redshifts   of  24  members   \cite{kneib00}  as
$836^{+180}_{-200}\, {\rm km\,s^{-1}}$

\subsection{Microlensing of the quasar images}
\index{quasars!microlensing}

As was shown in section 1.3 that microlensing of individual
quasar images is not unexpected, although it depends upon the size of
the source and the constituents of the lensing galaxy.  The angular
scale associated with such microlenses is smaller than that of the
macro lens by $\sqrt(M_{\rm micro}/M_{\rm macro})$.  Stellar mass microlenses
therefore produce microarcsecond splittings, not accessible to
present-day instrumentation.  But the image magnification by
microlenses is nonetheless observable -- flickering of the combined
flux from the unresolved microimages can be detected as the stars are
move randomly in the lensing galaxy.  While the observational evidence
for microlensing is fragmentary, there is enough to indicate that this
phenomenon, predicted immediately after the discovery of Q~0957+561
\cite{chang79}, plays an important role in the lensing of quasars.

The first hints of microlensing were found in the doubly imaged quasar
Q~0957+561 \cite{vander89}.  The difference light curve between the
two components (once corrected for the time delay) showed slow
variations unrelated to the intrinsic variability of the quasar. These
additional variations are thought to be the explanation of 
different time delays measured by different investigators. They
have also been identified as a potentially interesting tool to set
constraints on the stellar content of the lens and on the internal
\index{quasars!structure} 
structure of the lensed quasar on parsec scales (see for example
\cite{refsdal00,pelt98,wamb90}).

\begin{figure}[!ht]
\begin{center}
\includegraphics[height=6cm]{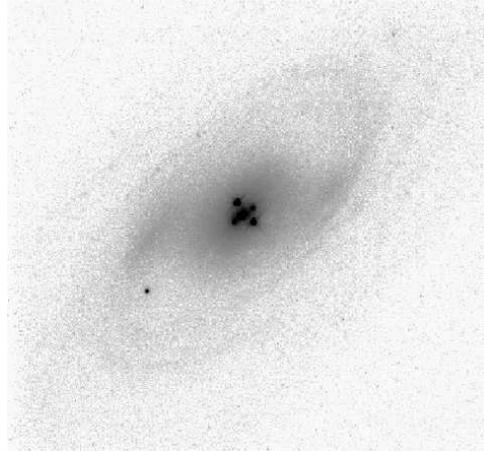}
\end{center}
\caption[]{HST $V$-band image of Q~2237+0305. Four quasar
images at $z=1.69$ are seen about $1''$ away from the nucleus of a
much lower redshift lensing galaxy ($z=0.04$). The high density of
stars in the lens' nucleus and their high projected angular velocity
make of Q~2237+0305 a privileged object for the study of
microlensing.}
\label{HST2237}
\end{figure} 

Q~2237+0305, also known as the ``Einstein cross'' (see Figure
\ref{HST2237}), \index{Einstein!cross} 
was quickly recognized after its discovery
\cite{huchra85} to be particularly susceptible to microlensing.  The
redshift of the lens, $\zl=0.04$, is so low that the apparent angular
velocity of the microlenses, in projection on the plane of the sky, is
much higher than in other systems.  Moreover the Einstein rings 
\index{Einstein!ring} of thes
microlenses have a larger angular diameter, making it more likely that
they are larger than the source.  The quasar is therefore expected to
show frequent and rapid variations, the mean time separating each
microlensing event being approximately the time required for the
microlenses to run across a distance equal to the diameter of their
Einstein ring (see equation
\ref{eq-theeapp}).  Note however that time scales involved can be
significantly different from those calculated with this naive
approach. As can be seen from Figure \ref{joachim} there are regions of high
magnification, \index{magnification} 
in particular close to the cusps of caustics, which are
exceedingly narrow, much smaller than the projection back onto the
source plane of the Einstein ring \cite{wamb90}.  There are also
plateaus, larger than the Einstein ring, over which the magnification
is relatively constant, and usually less than unity.

\begin{figure}[!ht]
\begin{center}
\includegraphics[height=8cm, angle=0]{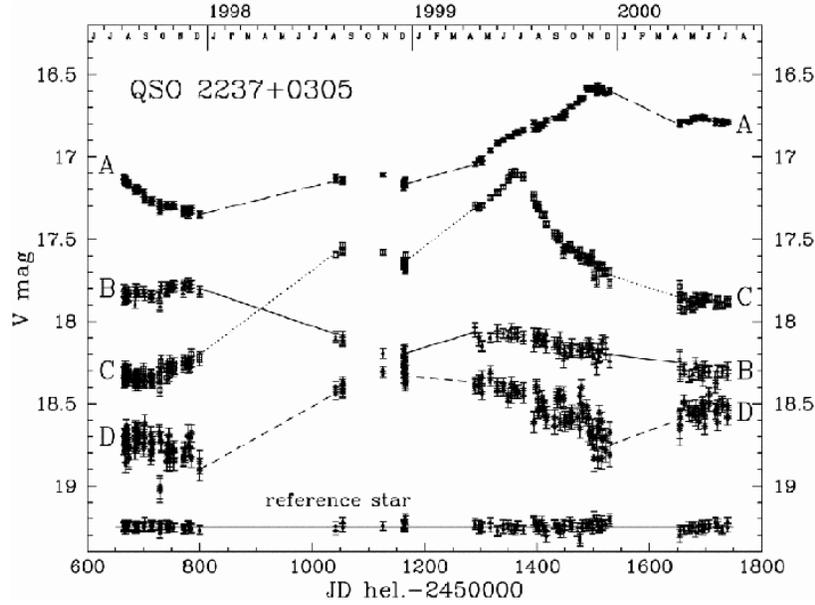}
\end{center}
\caption[]{Optical $V$-band light curves for the four quasar images in
the Einstein cross, Q~2237+0305. \index{Einstein!cross} 
The time-delay in this system is of
the order of a day. The very different behaviour of the four light
curves, with slow variations of the order of a month, strongly support
the idea of microlensing induced variability \cite{wozniak00}.}
\label{Q2237}
\end{figure}

As the optical path to each quasar image intersects the lensing galaxy
at very different locations, microlensing-induced variations in the
light curves \index{light curve} 
of the quasar images are uncorrelated. Intrinsic
variations of the quasar would be seen identical in each image,
separated by the time delay.  Time scales involved for microlensing
events in the Einstein cross \index{Einstein!cross} 
were predicted to be of the order of a few months
\cite{wamb90}, spectacularly confirmed by optical monitoring
\cite{ostensen96,wozniak00}.  This is much longer than the 
time delay of the system (about 1 day), making it easy to discriminate
between intrinsic variations of the source and microlensing events.
Figure \ref{Q2237} illustrates this: erratic variations of the 4 light
curves (especially component C) are seen, with a typical time-scale of
a few months. At the scale of the plot, intrinsic variations of the
quasar would be seen simultaneously in all light curves.

Flickering of quasar light curves is not the only signature of
microlensing. As noted in section 1.3 the Einstein radius of
microlenses is small and may be comparable in size to the inner
regions of quasars.  One may therefore observe differential
magnification of regions of different sizes. As different regions of
quasars are thought to have different colors, this implies {\it
chromatic} magnification. \index{magnification!chromatic} 
There are many instances where the flux
ratios for quasar images are quite different at different wavelengths
\cite{morgan01}.  Static microlensing \index{microlensing!static} 
is often invoked
as a possible explanation.  If one region varies and another does not,
static microlensing might also produce chromatic differences in the
quasar light curves. 

\begin{figure}[!ht]
\begin{center} 
\includegraphics[height=12cm, angle=0]{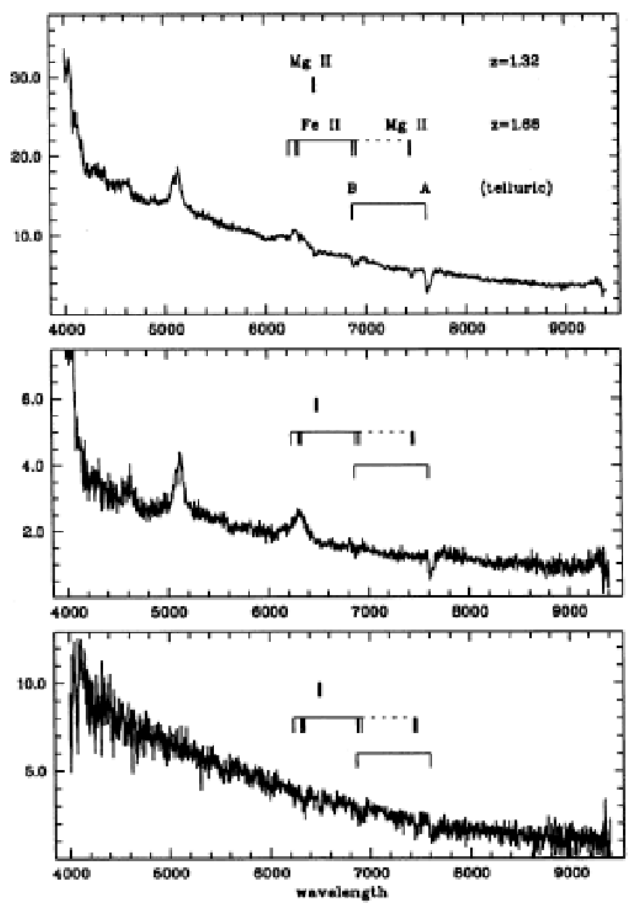} 
\end{center} 
\caption[]{Spectra of the two quasar images in HE~1104-1805 
\cite{wiso93}. The two first panels starting from the top show the
quasar spectra. The bottom panel shows the difference between the spectrum
of the brightest component of the system and a scaled version
of the spectrum of the faint component. A scaling factor of 2.8 is
necessary to subtract the emission lines from the spectrum. The labels
in the figure are related to absorption lines by the lensing galaxy.}  
\label{1104micro}
\end{figure}

More generally, the spectral differences among the regions of a quasar
will involve the presence or absence of emission lines.  One might
therefore expect differential magnification of the emission lines and
continuum.  The deblending of closely separated quasar spectra is not
trivial.  Fortunately, the relatively wide angular separation system,
HE~1104-1805, appears to show the phenomenon \cite{wiso93,wisot99}. In
this double system (see Figure
\ref{he1104}), the spectra of the two components are identical in 
the emission lines but show a different continuum, suggestive of
microlensing. Figure \ref{1104micro} shows the difference spectrum
(bottom panel) between the two quasar images. In order to subtract
properly the emission lines from the spectrum of component A (top
panel), one has to subtract a scaled version of the spectrum of
component B (middle panel). The scaling factor of 2.8 is found to be
stable with time and wavelength, even in the near-IR
\cite{courbin00b}. This suggests that emission lines are unaffected by
microlensing.  The difference continuum is blue and shows photometric
variations.  Part of these variations are intrinsic to the quasar, and
are used to infer the time delay of the system, but additional
flickering can be attributed to microlensing \cite{wisot99}. With
higher signal-to-noise spectra of HE~1104-1805, it has been found that
some emission lines might be affected by microlensing as well
\cite{lidman2000}. For example the red side of the CIII emission line
does not subtract perfectly after subtraction of the B
spectrum. This is also true for HE~2149-2745 \cite{burud01}, but there
are no similar observations so far for other systems.

\begin{figure}[!ht]
\begin{center}
\includegraphics[height=9cm, angle=0]{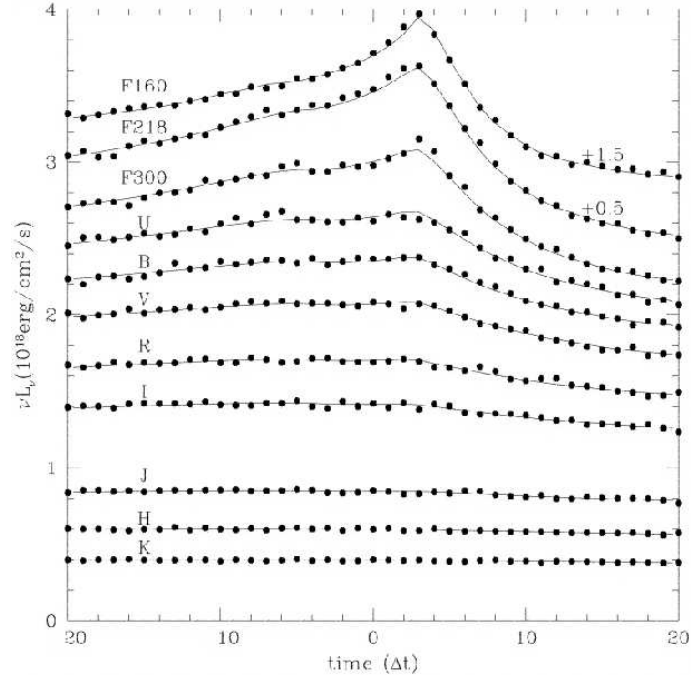}
\end{center}
\caption[]{Expected light curves for the microlensing events 
in a quasar at $z=1.695$ \cite{agol99}. Microlensing
events have larger amplitudes (larger magnification) at short
wavelengths. The time axis is in arbitrary units which depends on the
velocity and redshift of the microlenses.}
\label{agol}
\end{figure} 

In principle, this effect can be used to study the detailed structure
of active galactic nuclei.  For the sake of argument, one adopts a
standard model for of active galactic nuclei \cite{blandford90} where,
for example, the continuum region is much smaller than the broad line
region.  The continuum region itself is composed of an accretion disk
\index{quasars!accretion disk} 
which radiation field is less and less energetic when going from the
center of the accretion disk to its outer parts. As wavelength 
decreases with increasing 
energy, the accretion disk radiates bluer photons in
the center than in the outer parts. Let us now make the realistic
assumption \cite{wamb90} that the mean size of the micro-caustics
corresponds about to the angular size of a quasar accretion
disk. Simple geometric considerations shows that the smaller regions
(compared with the caustics) are entirely magnified, while only a
fraction of the larger regions are amplified. That is, the emission
line region will be less magnified by microlensing than the continuum,
and the outer parts of the accretion disk \index{quasars!accretion disk}
will be less magnified than
the inner parts. In other words, the flickering of the light curves
expected from microlensing should be stronger in the blue than in the
red, and should even be invisible in the emission lines.

This very simple scheme is  a lot more complicated in practice, simply
because one can not map the actual network of micro-caustics
\index{caustics network} present
in  a given  lensed  system: this  would  require a  map  of the  mass
distribution in the  lensing galaxy ! Still, one  can propose a quasar
model and predict the statistical behaviour of the light curves,
\index{microlensing!statistical} as a
function of wavelength. Such a theoretical work has been investigated
\cite{agol99,Wyithe00,wyithe00b}, with the goal to derive the 
relative sizes  of quasar emission  regions. \index{quasars!structure}
Figure \ref{agol} illustrates how  a given distribution of microlenses
preferentially  magnifies  the  innermost  (blue)  regions  of  quasar
accretion disks,  hence producing  light curves with  luminosity peaks
progressively increasing while observing  from the near-IR $K$-band to
the ultraviolet $U$-band \cite{agol99}. 

Unfortunately, the amount of data available so far is too small 
to implement any of the proposed method to probe quasar
structure from microlensing.

\section{Models}

\index{lens!models}
\index{models!parametric}

The small number of observables in lensing means that the 
observational data,
no matter how accurate, can be fit by a huge variety of lens models.
The space of allowed models must be narrowed by the adoption of priors
which reflect our understanding of the relative astrophysical
plausibility of different mass models.  There are two strategies for
doing this.  One is to adopt a parameterized mass distribution, where
the parameters are chosen to include the reasonable and important
variations expected among lensing galaxies.  The other strategy is to
keep the mass map free-form, but impose astrophysical priors as
constraints on it.

We now discuss both these modeling strategies.

\subsection {Parameterized models}

When  building parametrized  models, the  small number  of observables
then demands a small number of
parameters.  Eliminating a parameter (e.g. the octupole moment of the
gravitational potential) means that some aspect of the lensing galaxy is not
being modeled.  A wise choice of parameters models those aspects which
are important for the task at hand.

Fortunately there is a vast literature on the mass distributions and
gravitational potentials of galaxies.  For example (and quite
importantly for the interpreting time delays) we know that galaxies
have mass density profiles \index{lens!mass profile} \index{density!profile}
which vary roughly as $1/r^2$, giving
galaxies flat rotation curves and flat velocity dispersion profiles.
For the sake of discussion we put forward here a ``standard'' model
which incorporates much of what we know about nearby galaxies.

\subsubsection{Some simple models}

We start with the simple monopole potential and describe a number of
additional terms which correspond, at least roughly, to what one might
expect for galaxies in a variety of contexts, adding degrees of
freedom which we have reason to believe nature exploits.

\paragraph{Singular isothermal sphere}
\index{lens!Singular Isothermal Sphere}

The singular isothermal sphere is a cornerstone of galaxy dynamics \cite{bt}.
It gives the flat rotation curves and constant velocity dispersion
profiles characteristic (to a first approximation) of
spiral and elliptical galaxies respectively.
It has a three dimensional potential $\Phi = v_{\rm c}^2 \ln r$, where
the circular velocity $v_{\rm c} = \sqrt 2 \sigma$, with $\sigma$ the one
dimensional velocity dispersion.  Integrating this along the line of
sight and multiplying by $ 2\Dls/(\Dl\Ds c^2)$ gives gives
the lens potential \index{lens!potential}
\begin{equation}
\psi(\btheta)=\thee\theta,
\end{equation}
which is the same as equation (\ref{eq-isopoten}) but now with
with $\thee = 4\pi \sigma^2 \Dls/\Ds c^2 $ (measured in radians)
giving the lens strength.
We recall (cf.~equation \ref{eq-isocims}) that such a lens
produces two colinear images, one on each side of
the lens, with magnifications given by equation (\ref{eq-isomag}).
The infinite second derivative of the potential at the origin gives
infinite demagnification of a central third image.

\paragraph{Power-law monopole}
\index{models!power-law monopole}

The singular isothermal sphere is a special case of the power-law monopole
\begin {equation}
\psi({\btheta})  = {\thee^2 \over (1 + \alpha)} 
\left({\theta\over \thee}\right)^{1+\alpha},
\label{eq-plm}
\end {equation}
where the central concentration index, $\alpha$, measures the deviation
from iso\-thermality.  The normalization has been chosen so that the
strength $\thee$ is again the radius of the Einstein ring.  As the
exponent $\alpha$ approaches $-1$, the potential approaches that of a
point mass. 

\paragraph{Self-similar power law quadrupole}
\index{models!self-similar power law quadrupole}

Since galaxies are not circularly symmetric, there is no reason why
their effective potentials should be.  The flattening of a potential
is dominated by a quadrupole term, which if we use polar coordinates
with ${\theta} = (\theta, \phi)$, varies as $\cos 2 \phi$.  A simple
model for the effective potential which incorporates the
non-negligible quadrupole of galaxies incorporates quadrupole term
with the same radial dependence of the monopole, giving equipotentials
which similar scaled versions of each other,
\begin{equation}
\psi(\btheta) = {\thee^2 \over (1 + \alpha)} 
\left({\theta\over \thee}\right)^{1+\alpha}
[1 + \gamma \cos 2(\phi - \phi_\gamma)] .
\label{eq-ssplq}
\end{equation}
An on-axis source gives 4 images whose distance from the lens center
is approximately equal to the lens strength $\thee$.  Dimensionless
$\gamma$ gives the flattening of that quadrupole and $\phi_\gamma$
gives its orientation.\footnote {Following Kochanek (1992) we (somewhat
confusingly) use the same symbol, $\gamma$, for the flattening as is
used for the shear.  The flattening and the shear are equal at $\theta
= \thee$, but not elsewhere.} The special case $\alpha = 0$ gives a
flattened system with the flat rotation curve and constant velocity
dispersion profile characteristic of isothermals.  While the
equipotentials are self similar for all $\alpha$, the equipotentials
and the equidensity contours are both self-similar only for the
$\alpha = 0$, isothermal case.

\paragraph{Tidal quadrupole (first order tide)}
\index{models!tidal quadruple}

Equipotentials which have the same shape are esthetically appealing
but highly idealized.  In particular galaxy equipotentials will
deviate strongly from self-similarity if the quadrupole is due to a
tide from a neighboring galaxy or cluster of galaxies.  In that case
the quadrupole term shows a $\theta^2$ dependence on distance from the
center of the lens, as was seen in equation (\ref{eq-cisshear}). 
Among others, \cite{kundic97} have noted that
the quadrupoles of many lensed systems appear to be due to tides
rather than to the flattening of the lensing galaxies.  A simple
potential incorporating these features is
\begin{equation}
\psi(\btheta) = {\thee^2 \over (1 + \alpha)} 
\left({\theta\over \thee}\right)^{1+\alpha}
 + \half {\gamma} \theta^2 \cos 2(\phi - \phi_\gamma).
\label{eq-plshear}
\end{equation}
This is much like the self-similar power law quadrupole of equation
(\ref{eq-ssplq}).  While the monopole term is, as in the previous
cases, a power law, the quadrupole term has the $\theta^2$ dependence
characteristic of a tide.  In general we expect a lens to have both a
tidal quadrupole, from neighboring galaxies, and something like the
self-similar quadrupole due to the flattening of the lensing galaxy
itself.

\paragraph{Clusters as mass sheets}
\index{cluster!mass sheet}

Galaxies typically reside in groups and clusters, with considerably
more mass (dark matter) associated with the cluster than with the
individual galaxies.  One must therefore take the gravitational
potential of the associated cluster into account.  The scale of a
cluster is much larger than that of a galaxy, so its surface density
projected onto the galaxy is to first order constant.  A mass sheet of
uniform density produces an effective potential \index{lens!potential}
\begin {equation}
\psi_s({\btheta}) = \half \kappa_s \theta^2 
\end {equation}
where $\kappa_s$ is the dimensionless convergence \index{lens!mass sheet}
associated with the mass sheet. \index{convergence}

As was shown in section 1, differentiating twice one finds that a
superposed mass sheet stretches an image configuration by a constant
factor $1/(1-\kappa_s)$ without changing any of the dimensionless
ratios associated with the image configuration.  A model
that failed to take account of such a mass sheet would predict too
long a differential time delay by the just this same stretch factor.
But there is no way of knowing from image positions or relative
magnifications whether or not such a mass sheet is present.  This
formal degeneracy {\it demands} that one bring to bear ``external''
information regarding the projected density of any such mass sheet.

\paragraph{Clusters and higher order tidal terms}
\index{cluster!mass sheet}

In the above paragraphs we have identified two distinct effects of
clusters of galaxies: they introduce tidal and mass sheet terms into
the effective lensing potential.  There are many lenses for which the
first order tidal terms are so strong (e.g. \cite{kundic97}) 
that higher order terms are likely to be important.  The
simplest way to do this is to drop the tidal term above and to model
the cluster as a isothermal at position $\bTheta$ with effective
potential
\begin{equation}
\psi_c(\btheta) =  \Thee |{\bTheta - \btheta}| .
\end{equation}
This model has three free parameters (replacing the two tidal
parameters, $\gamma$ and $\phi_\gamma$), the lens strength $\Thee = 4 
\pi \sigma^2 \Dls /\Ds c^2$ where $\sigma$ is the the velocity dispersion
of the cluster, and the polar coordinates $(\Theta, \Phi)$.

\def\quarter{{1 \over 4}}
While the cluster potential can be written quite compactly in this
form, it obscures the connection between the cluster properties and
the tidal and mass sheet terms described above.  Taking the lensing
galaxy to be at the origin of our coordinate system, we can expand the
cluster potential in powers of $\theta / \Theta$, where $\Theta$ is the
distance of the cluster from the origin.  Dropping constant terms, we
find

\begin{eqnarray}
\psi_c(\btheta) &=&  
-\Thee \bTheta \cdot \left( {\btheta \over \Theta} \right)
\nonumber \\
&+& \quarter \Thee \Theta \left( {\theta \over \Theta} \right)^2
- \quarter \Thee \Theta \left( {\theta \over \Theta} \right)^2
\cos 2(\phi - \Phi) 
\nonumber \\
&+& \hbox{ terms of order } \left( {\theta \over \Theta} \right)^3
\hbox{ and higher. }
\end{eqnarray}

The first term gives a constant deflection $\Thee$ away from the
cluster, showing that the source position $\bbeta$ may be rather far
from the origin and the lensing galaxy.  The second term is just that
of a mass sheet with $\kappa_s = \half (\Thee / \Theta)$ while the
\index{lens!mass sheet}
third is a tidal term with shear $\gamma = \kappa$.  Noting that the
coefficient of the shear term is negative, we find that the position
angle of the shear, $\phi_\gamma$, as defined in equations
(\ref{eq-ssplq}) and (\ref{eq-plshear}), is at right angles to the
position angle of the cluster, $\Phi$. \index{shear}
 
The equality of the shear and convergence suggests a possible
resolution of the mass sheet degeneracy: measure the shear and infer
the convergence.  We adopt this approach with an obvious caveat.  To
the extent that clusters and groups are {\it not} isothermal, such a
``shear inferred'' mass sheet correction will introduce a systematic
error in a derived Hubble constant. \index{Hubble constant}

The terms of order $(\theta/\Theta)^3$ are useful because they break
the classical tidal degeneracy.  Since $\gamma = \half (\Thee /
\Theta)$, we might produce an equally strong first order tide by
putting an isothermal cluster with twice the Einstein radius at twice
the distance.  Alternatively, we might put the cluster at position
$-\bTheta$ without changing the first order tide.  Keeping the higher
order terms resolves these ambiguities.  But rather than add many
coefficients, it is conceptually simpler and more economical to
replace the two parameters of a first order tide with the three
parameters of a circularly symmetric cluster.  There are several
lenses (e.g. RX J0911+0551, PG 1115+080 and B1422+231) for which
higher order tidal effects have been used to determine the position
and lensing strength of the associated cluster.

\paragraph{Yet more degrees of freedom}

Even in the absence of tides, there is no reason to insist that the
monopole and quadrupole terms of a galaxy potential have the same
dependence on $\theta$, {\it i.e.} that the potential be self-similar.
The self-similar model presented above can readily be extended to
allow for separate $\theta$ exponents, permitting the potential to get
rounder or flatter with increasing $\theta$.  Nor is there any reason,
in principle, why we should limit ourselves to monopole and quadrupole
terms.  Purely elliptical density profiles produce potentials with
higher order multipoles.  Some ellipticals are ``boxy'' while others
are ``disky'' \cite{bender9x}, and these too should have higher order
multipoles.  Power laws like equations (\ref{eq-plm}) and
(\ref{eq-ssplq}) give unphysical mass and density divergences, and
should in principle be cut off at small or large $\theta$ or broken
somewhere in between.  Finally, we might argue that it is naive to
assume that the dark matter in a galaxy is centered on its starlight,
and that we should take the central coordinates of the lensing
potential to be free parameters.

With all these possibilities, it is no surprise that different
investigators modeling the same system come up with different
potentials and derive different values of $H_0$ from the same time
delay.  The number of measurements which constrain the potential is
small, so one cannot allow oneself the luxury of adding extra
parameters just for the sake of insurance.  In introducing new 
parameters the two questions to be kept in mind are the degree to
which they degree to which they affect the deflections and
distortions, which constrain the potential, and the degree to which
they affect the delays, which give the Hubble constant. 

\subsubsection{Useful approximations and rules of thumb}

For the sake of simplicity, suppose that a lens has the power-law
monopole \index{models!power-law monopole} 
potential of equation (\ref{eq-plm}).  Using the lens
equation, we substitute the gradient of the effective potential,
$\pderiv(\psi/\btheta)$, for the deflection, $ {\btheta - \bbeta}$,
in the time delay \index{time delay} 
equation. Under the assumption that two quasar
images, $A$ and $B$ are roughly equidistant from the center of the
potential, we predict a differential time delay
\begin{equation}
\tau_B - \tau_A \approx 
\half T_0 \left({\theta_A}^2 - {\theta_B}^2\right)(1 - \alpha) ,
\label{eq-deltauab}
\end{equation}
where $T_0$ is the time scale defined in equation (\ref{eq-T0})
and $\theta$ is measured in radians.  Had we not assumed circular
symmetry, $\theta_A^2$ and $\theta_B^2$ would each have a coefficient
which differed from unity by a factor of order gamma, usually less
than 10\%.  Equation (\ref{eq-deltauab}) has the important and useful
property that it depends only upon observable quantities, assuming
that the position of the lensing galaxy can be measured.

Several useful lessons can be drawn from equation (\ref{eq-deltauab}).
First, the more distant image leads the closer image (cf.~Figure
\ref{fig-imorder}). Second, if
$\theta_A \approx \theta_B$ high astrometric accuracy is needed in
measuring the position of the lensing galaxy for high precision in the
predicted time delay.
\def\fy{\hbox{$.\!\!^{\rm y}$}}
\def\fd{\hbox{$.\!\!^{\rm d}$}}
Third, the predicted delay scales as the square of the separation.
The differential time delay of Q~0957+561, $1\fy2$, is therefore
atypically long, resulting from its large ($6.1''$) separation and
relative asymmetry.  Fourth, if the lens potential is more sharply
peaked than a singular isothermal sphere, the predicted time delay is
longer.  In particular, a point mass model, with $\alpha = -1$,
predicts a time delay twice as large (yielding a Hubble constant
twice as large for a given observed delay) as the corresponding
singular isothermal, \index{lens!Singular Isothermal Sphere} 
$\alpha = 0$ model. Either $\alpha$
must be measured with high accuracy from the observed image
configuration or we must bring external considerations to bear upon
our models.  In comparing models by different investigators for the
same system, one must pay particular attention to the way in which the
degree of central concentration has been treated.

\subsubsection{Fitting models}
\index{models!fitting}

{\em How and what to fit ?}
On first thought it seems straightforward to adopt a lens potential
and a source, find the predicted images, compare those with the
observed images and adjust the parameters associated with the potential
and the source so as to get better agreement.  On closer examination one
discovers that the lens equation can only rarely be solved in closed
form for image positions.  Worse yet, one finds that small changes in
parameters can cause pairs of predicted images to merge and disappear.
What does one do in a gradient search when a small trial step causes
an image to cease to exist ?  Fortunately robust methods for fitting
data have been developed, some of which are publicly available
\cite{keeton00}.

The fluxes of images can readily be measured to 1\% accuracy, but the
differences between optical and radio flux ratios are of order
10-30\%.  Given the very much greater accuracy of positions, one might
be tempted to dispense with magnifications \index{magnification} 
entirely.  But for double
lenses, even the simplest non-circular models have one too many
parameters to permit fitting using positions alone.  Moreover, fitting
fluxes can help avoid aforementioned disappearing image problem. It is
therefore helpful to use fluxes, but with full awareness of the
associated pitfalls.

Image positions constrain the first derivative of the effective
potential.  Magnifications constrain the second derivative.  In some
systems more than one time delay can be measured.  The first measured
delay goes to solving for the dimensioned combination of angular
diameter distances in equation (\ref{eq-deltauab}) but the ratios of
the second and subsequent delays to the first give dimensionless
constraints on the effective potential itself.
Though not yet incorporated in most parametrized models, such
constraints are are in principle quite powerful.  But a disadvantage
so far is that in practice the uncertainties in all measured delays
for a given system are roughly the same, as measured in days; so while
the fractional uncertainties in the longest delay is typically better
than 10\%, the fractional uncertainties in the shorter delays are
correspondingly greater.

\subsubsection{What constitutes ``good enough''?}

There is little difficulty in finding models for lens systems which
fit \index{models!fitting} 
the observed data perfectly.  The number of constraints is small,
and the number of free parameters is large, and so it should be
possible to find an $N$ parameter model which fits the $N$ available
constraints perfectly.  But that leaves no room for reality testing.
Ideally one hopes to find a model with $<N$ parameters for which the
predicted images agree with the observed images within the measured
uncertainties, giving an acceptable fit to the data.

The words ``unacceptable fit'' have a damning ring which tends to end
discussion.  Were we able to measure the relative positions of the
lensed images to one part in a million, the deflections due to
individual stars within the the lensing galaxy become important.  At
that point we would be unlikely to ever get an acceptable fit from a
macromodel.  But the differences in the time delays induced by such
microlensing are small.

For the purpose of interpreting time delay measurements a less
stringent definition of acceptable may be in order.  Consider the case
of Q~0957+561.  Errors in the positions of 100 milliarcseconds
introduce negligible changes in the time delay predicted by
equation (\ref{eq-deltauab}).  While one can concoct a parameterized 
model for which small differences in the positions produce large
changes in the predicted time delays, these are, with the exception of
the central concentration degeneracy, somewhat artificial.

\subsubsection{The central concentration degeneracy}
\index{degeneracy!central concentration}
\index{degeneracy!mass disk}

The central concentration degeneracy has already surfaced in our
discussion, first theoretically as the mass disk degeneracy,
and then in the approximate rule for
computing time delays, equation (\ref{eq-deltauab}), and it appears
yet again in connection with free-form models.   It has also
surfaced many times in the literature.
 A particularly thorough treatment can be found in \cite{trotter00}, 
though it is
evident as well in other works \cite{bern98,schech97}.
Briefly, it has proven exceedingly difficult to constrain the (radial)
second derivative of the monopole term of the effective potential.
Several factors contribute.  In double systems the associated
parameters are coupled to the quadrupole amplitude.  In quadruple
systems the images all tend to lie at roughly the same distance from
the center of the lens -- otherwise the system wouldn't be quadruple.
The radial displacements of these images depend not only on the
concentration parameter but also on higher order multipoles.  Einstein
rings may be less susceptible to this degeneracy 
because the rings are resolved in the radial direction, though
this is controversial \cite{kkm01,sw01}.

What makes this degeneracy pernicious is its strong influence on the
predicted time delay, increasing them by a factor $(1 - \alpha)$ in
the parameterization of our power-law models (equations (\ref{eq-plm})
and (\ref{eq-ssplq})).  In the face of this, one has two choices: to
search for a ``golden'' lens which doesn't suffer from it or to bring
external constraints to bear.

Golden lenses, at least 24 carat golden lenses, are rare.  MG
J0414+0534 would at first sight seem as good a candidate any, with a
core and 3 VLBI \index{VLBI} features, each quadruply imaged.  But 
\cite{trotter00} conclude ``It is clear that useful information
on the radial profile of MG J0414+0534 is unavailable from this
data.''  Alas even if it were, the object has shown little sign of
variability \cite{moore96,angonin99}.

Measurements of  velocity dispersion \index{velocity dispersion} 
gradient \cite{tonry99} have been made of the lensing galaxies 
in Q~0957+561 and PG~1115+080, 
which in principle constrain the degree of central
concentration of the potential.  This is a particularly difficult
measurement because of the competition from the optical images of the
lensed quasar.  Moreover the effective radius of the lensing galaxy
tends to be considerably smaller than the Einstein ring, making it
difficult to obtain measurements out to the region of interest.  

An alternative approach 
\cite{roman98} is to use what one knows about the potentials of nearby
elliptical galaxies.  They compiled data on the potentials of nearby
elliptical galaxies for which not only velocity dispersions but higher
order moments of the line of sight velocity distribution had been
measured.  Their data show a mean power-law index $\<\alpha> = -0.2$,
with a scatter of roughly 0.2 about that value.  This is somewhat more
centrally concentrated \index{degeneracy!central concentration} 
than for the isothermal index, $\alpha = 0$,
but not nearly so concentrated as the point mass index, $\alpha = -1$.
If the data fail to constrain the power-law exponent, fixing it at its
mean value would introduce errors in the predicted time delays of
roughly 20\%.  But since the observed power-law index is so close to
the isothermal value, $\alpha = 0$, and since the power law index
makes so little difference in the quality of the fit (otherwise it
would be well constrained), one does little harm in fixing the
power-law index at its isothermal value and making a post-hoc
correction to the predicted time delay. \index{time delay}

\subsubsection{A proposed ``standard'' model for lenses}
\index{models!standard}

As the preceding sections make only too clear, predicted time
delays and derived Hubble constants depend sensitively upon how lens
potentials are modeled.  In particular, they are sensitive to the
degree of central concentration of the lens model, \index{degeneracy!central
concentration} which is especially
difficult to constrain using lensed images alone.  Such model
differences have led to widely divergent predicted delays and derived
Hubble constants for what are essentially the same data.

Absent the discovery of a 24 carat lens, one can still make progress
measuring the Hubble constant by accepting that most lenses are
underconstrained and adopting a ``standard'' model for which the
associated systematic errors are well understood and which is
sufficiently simple that it can be applied to a large fraction of the
known lensed systems.

\paragraph{The proposed standard}
\index{models!standard}

In the
belief that it will take us within striking distance of $H_0$,
we propose the following ``standard'' effective potential,
\begin{equation}
\psi(\btheta) = \thee\theta 
+ \half\gamma \theta^2 \cos 2(\phi - \phi_\gamma),
\label{eq-sisshear}
\end{equation}
which is the isothermal variant of the tidal power-law plus quadrupole
of equation (\ref{eq-plshear}).  To the extent that they are
understood, the systematic and random errors associated with this
model are as follows.

As noted above the assumption of isothermality, $\alpha = 0$,
introduces a systematic error, but this can readily be corrected by
multiplying the predicted time delay by the factor $1-\<\alpha>$.  We
choose to fit $\alpha = 0$ because in most cases the availalable data
fail to constrain $\alpha$ any better than this external constraint
and to avoid fussing about second and third generation standards as
the appropriate mean value of $\alpha$ is further refined.

In double systems there are too few constraints to permit
discrimination between the tidal isothermal as in our proposed
standard model and a self-similar isothermal.  Among
quadruple systems tides appear to be more important than the
flattening of the lenses \cite{keeton97}, but then tides may
explain the relatively large number of quadruple systems \cite{kundic97}.

For our proposed standard the differential time delay \index{time delay} 
is given by
\begin{eqnarray}
\tau_B - \tau_A \approx T_0 \times
\{ &{\theta_A}^2& [1+ \gamma \cos 2(\phi_A - \phi_\gamma)] \nonumber \\
  &-{\theta_B}^2& [1+ \gamma \cos 2(\phi_B - \phi_\gamma)] \} . 
\label{eq-deltaustd}
\end{eqnarray}

Had we instead adopted the isothermal variant of the self-similar
power-law potential of equation (\ref{eq-ssplq}), the square bracketed
terms would have reduced to unity as in (\ref{eq-deltauab}).  If we have
made the wrong choice, and if the orientation of the shear, \index{shear}
$\phi_\gamma$, is random, our choice of a tidal quadrupole introduces
a random error but not a systematic one.  If $\gamma$ is small, the
effect is not large.  If $\gamma > 0.1$, the quadrupole term is so
large that an external tide seems the more likely possibility.  So
either we make a small random error or we make the right choice.

If we believe that the shear is largely tidal, it seems reasonable to
assume that the tide is due to an isothermal potential, 
\index{lens!isothermal} and that there
is an associated convergence $\kappa = \gamma$.  The predicted time
delays of equation (\ref{eq-deltaustd}) should 
therefore be multiplied by a factor
$(1-\gamma)$ to account for the ``mass sheet'' \index{lens!mass sheet}
\index{lens!mass sheet} 
associated with the
tidal perturber.  We cannot avoid making a systematic error here, but
we make a larger error in failing to correct for the projected surface
densities associated with tides than in making the correction.  Our
doubly corrected prediction is therefore
\begin{equation}
(\tau_B - \tau_A)_c =
(1 - \<\alpha>) (1 - \gamma)(\tau_B - \tau_A) .
\label{eq-deltaucorr}
\end{equation}

In summary, our standard model is a tidal singular isothermal.  We fit
the model to the available constraints and use equation
(\ref{eq-deltaustd}) to compute the time delay.  We apply a correction
factors of $1 - \<\alpha>$ and $1/(1 - \gamma)$ to the predicted time
delay to account for, respectively, the mean power-law index observed
in ellipticals and the projected surface density associated with
tides.

\paragraph{Application of the proposed standard}
\index{models!standard}

Our standard model is by no means new.  
It is one of two models used by the CASTLES
group to analyze the lens data they have assembled.  We note, however,
that they do not apply the corrections for central concentration
and convergence that we adopt in the previous section.
\index{CASTLES}

The CASTLES model for PG 1115+080 has a shear of 0.12 with a predicted
C-B time delay of 12\fd8 for an $h=1$ EdS universe.  Applying the
corrections of equation (\ref{eq-deltaucorr}) using a mean
concentration, $\<\alpha> = -0.2$, gives a predicted time delay of
$13\fd5$.  Using Barkana's value of $25\fd0$ for the observed delay
gives $h=0.53$.

The CASTLES group has not yet posted a SIS+shear model for RX~J0911+0551, 
but Schechter gives a shear of 0.307 and a predicted time
\index{lens!Singular Isothermal Sphere}
delay of $120\fd5$ between B and (A1+A2+A3).  Again applying equation
(\ref{eq-deltaucorr}) we get a corrected prediction of $100\fd2$.
\cite{burud99} report preliminary delay measurement of $200^{\rm
d}$, giving $h=0.50$ for an EdS universe.

\subsection{Free-form models}

Free-form models \index{models!free-form} 
build a lens as a superposition of a large number of
small components, with minimal assumptions about the form of the full
lens.  They are motivated by three considerations.
\begin{enumerate}
\item The fewness of observables in quasar lensing, and the presence
of degeneracies, \index{degeneracy} means that any one lens reconstructed from
observations is highly non-unique.  One needs some systematic way of
searching through possible lens reconstructions. \index{lens!reconstruction}
\item The high accuracy of observations, despite their fewness, means
that data always show deviations from the parametrized 
\index{models!parametric} models discussed above.
Models with more parameters can fit the
data to observational accuracy, but it is not known what all the
essential parameters are.  Are twisting isodensity contours important ?
Does ellipticity vary significantly with radius ?
\item The most important observational constraints from lensing (being
image positions, tensor magnifications, and time delays) are linear,
which makes it straightforward to fit lenses by superposition.
\end{enumerate}
We will refer to the small components as pixels, but in fact they can
be any kind of components and not necessarily small.  For example,
they may be Fourier or harmonic terms in the mass
profile \cite{trotter00}.  But here we will discuss in detail the case
where the pixels are mass tiles with uniform but adjustable surface
density \cite{sw97,ws00}. \index{density!surface}

Consider a lens made up of $N$ pixels each with mass profile $\kappa_n
f_n\(\btheta\)$.  Here $\kappa_n$ is an adjustable
parameter.\footnote{Hence we deprecate the alternative name
`non-parametric' for this method, favoring `free-form' or
`pixelated'.}  Let $Q_n\(\btheta\)$ be the integral of
$\nabla^{-2}f_n\(\btheta\)$ over the $n$-th pixel.  In other words,
let $\kappa_nQ_n\(\btheta\)$ be the $n$-th pixel's contribution to the
lens potential \index{lens!potential} 
at $\btheta$.  For square tiles or Gaussian tents $Q_n$
is known but messy \cite{sw97,asw98}. For harmonic components or other
eigenfunctions of $\nabla^2$, $Q_n\(\btheta\)$ is simply proportional to
$f_n\(\btheta\)$.  For a pixelated lens the arrival time surface
(\ref{eq-tau}) becomes
\begin{equation}
\tau\(\btheta\) = \half\btheta^2 - \btheta\cdot\bbeta - \sum_n\kappa_n
Q_n\(\btheta\)
\end{equation}
where we have discarded a $\half\bbeta^2$ term from (\ref{eq-tau})
since it is constant over the surface.

We may now implement three kinds of observational constraints.
\begin{enumerate}
\item Image positions: an image observed at $\btheta\!_i$ implies
\begin{equation}
\bnabla\tau\(\btheta\!_i\)=0.
\end{equation}
(We can safely neglect the uncertainty in $\btheta\!_i$, since image
astrometry is typically at the mas level).  A multiple-image system
derives from the same $\bbeta$, but that $\bbeta$ is unknown.  So each
such system introduces $2(\<\hbox{images}>-\<\hbox{sources}>)$
constraints.
\item A time-delay \index{time delay}
measurement between images at $\btheta\!_i$ and
$\btheta\!_j$ implies
\begin{equation}
\tau\(\btheta\!_i\)-\tau\(\btheta\!_j\) = h\,T_0^{-1}
\times \<\hbox{obs delay}>.
\end{equation}
In a quad there may be two or three independent time delays.
\item Tensor magnifications \index{magnification}
are measured from images of a
multiple-component source.  The implied constraints can be included
simply by treating the images of separate components as independent
image systems.  A scalar magnification, or simple flux ratio, cannot
be included in this way; however, flux ratios are sensitive to
microlensing and thus less suitable for constraining macro-models.
\end{enumerate}
All these constraints are linear in the unknowns $\kappa_n\(\btheta\)$
and $\bbeta$.  Schematically, we may write

\bgroup \textfont1=\textfont0
\begin{eqnarray}
& & \pmatrix{Lensing \cr data} = \nonumber \\
& & \pmatrix{A & messy & but & linear & operator\cr
             also & involving & the & same & data }
     \pmatrix{The \cr \hbox{lens's} \cr projected \cr
              mass \cr distribution \cr}
     \nonumber
\end{eqnarray}
\egroup
Note the un-square matrix: there are many more pixels than data, i.e.,
the reconstruction problem is highly underdetermined.

It is easy to find lens profiles formally consistent with
observational constraints as above, but most of them will not look
anything like galaxies.  We now try to exclude the latter with
additional constraints --- in Bayesian terminology we apply a prior.
A reasonable prior is the following.
\begin{itemize}
\item $\kappa_n \geq 0$,
\item $180^\circ$ rotation symmetry (optional),
\item density gradient $\leq 45^\circ$,
\item $\kappa_n \leq 2 \<\hbox{average of neighbors}>$, except for the
central pixel,
\item $\kappa$ steeper than $|\btheta|^{-1/2}$, based on stellar
dynamics evidence that the 3D density in galaxies is steeper than
$r^{-1.5}$.
\end{itemize}
The observational and prior constraints confine allowed lenses to a
convex polyhedron in the space of
$(\kappa_1,\kappa_2,\ldots,\kappa_N)$.  This can now be searched by a
random-walk technique, yielding an ensemble of models.  And then one
can estimate $h$ or whatever from that ensemble \cite{ws00}.

There are three caveats associated with this technique.
\begin{itemize}
\item The results depend on the prior, and the above prior certainly
has too little information.  But it has at least the advantage that
uncertainty estimates will be conservative.
\item Having the mass on tiles means that the models cannot have
central density cusps, contrary to what galaxies are thought to have.
But far from the center, this is not an issue because of the monopole
degeneracy.
\item There is too much pixel-scale structure.  This is not an issue
for $h$, but if one wanted input for microlensing computations then
the pixel-scale structure needs to be smoothed out.
\end{itemize}

\subsubsection{Four well-known systems}

We now describe some new results obtained by one of us (PS) with
L.L.R. Williams, on four lenses.  For each lens, there was an ensemble
of 200 models.  The $\kappa_n$ controlled $\sim500$ mass tiles, plus a
parametrized external shear. \index{shear}

\begin{figure}[!hp]
\begin{center}
\includegraphics[width=.35\textwidth]{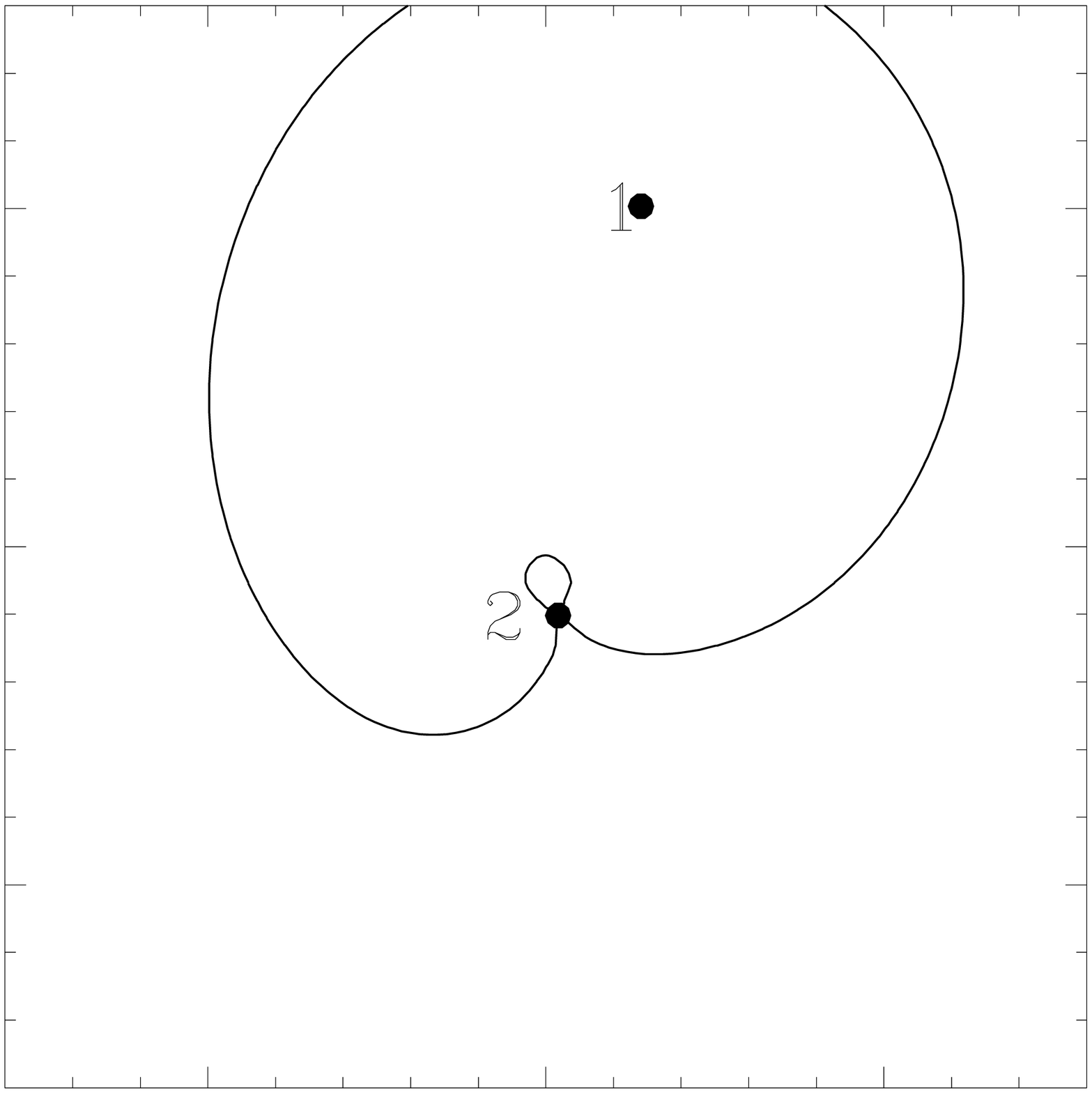}
\hfil
\includegraphics[width=.35\textwidth]{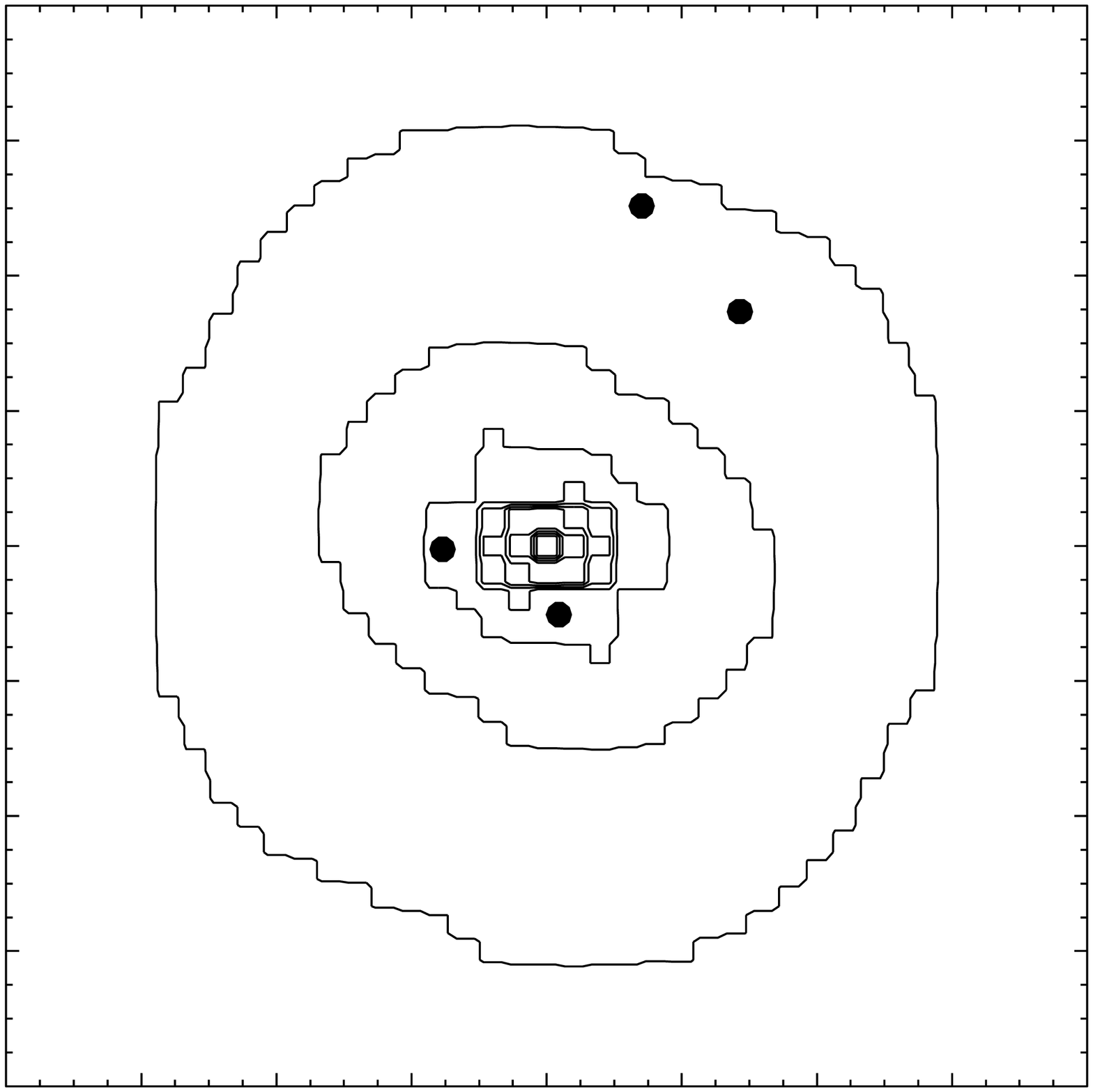}
\end{center}
\begin{center}
\includegraphics[width=.48\textwidth]{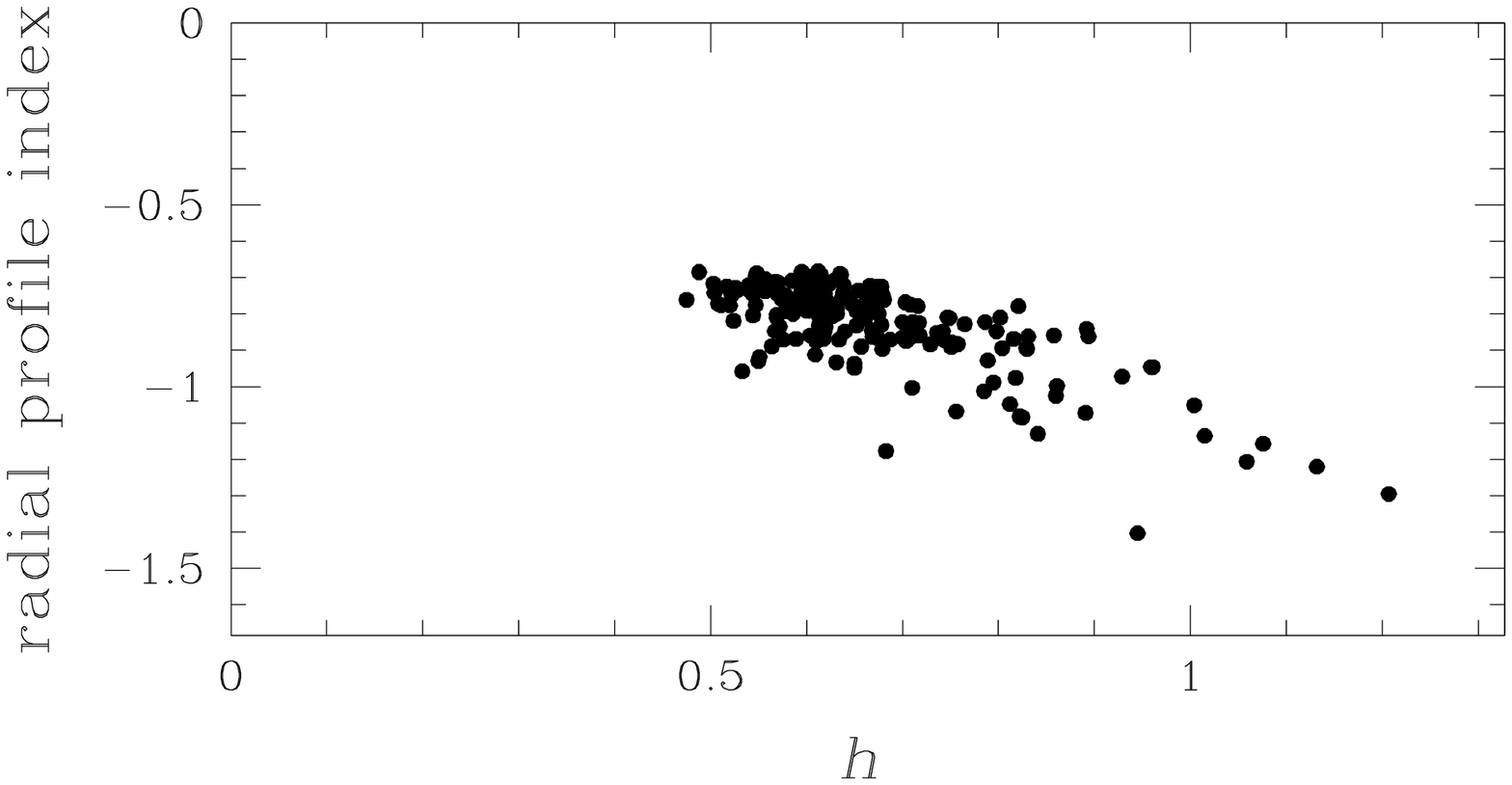}
\hfill
\includegraphics[width=.48\textwidth]{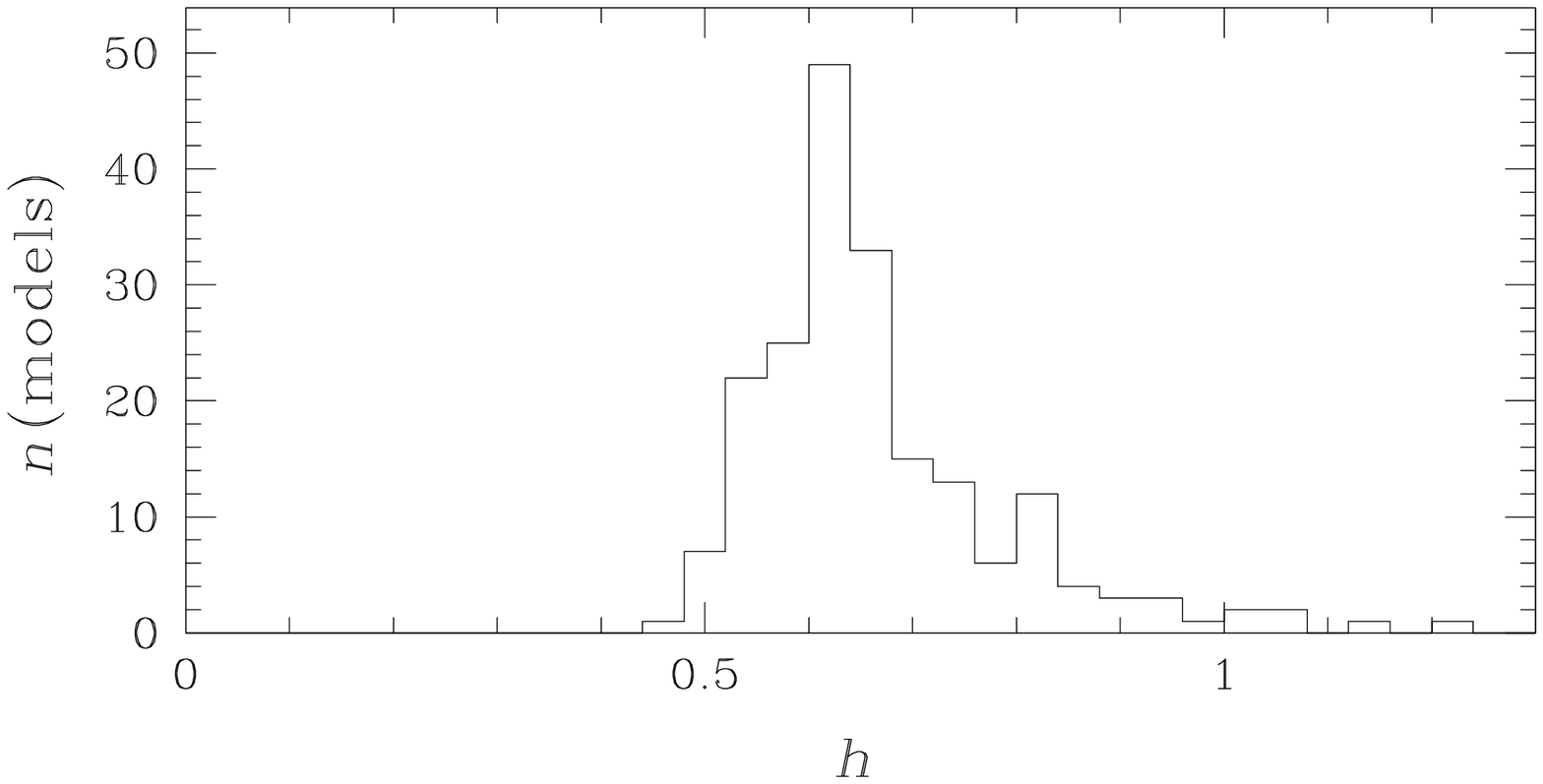}
\end{center}
\caption[]{Models of Q~0957+561. {\bf Upper left:}~schematic image
configuration and saddle-point contours for the quasar.  {\bf Upper
right:}~ensemble-average reconstructed mass map; contours are
$\kappa=\frac13,\frac23,\ldots$ The four images marked are the quasar
double and another double from a knot in the host galaxy.  {\bf Lower
left:}~$h$ against radial index for all 200 models in the
ensemble. {\bf Lower right}:~Histogram of $h$ from the ensemble.}
\label{fig-models0957}
\begin{center}
\includegraphics[width=.35\textwidth]{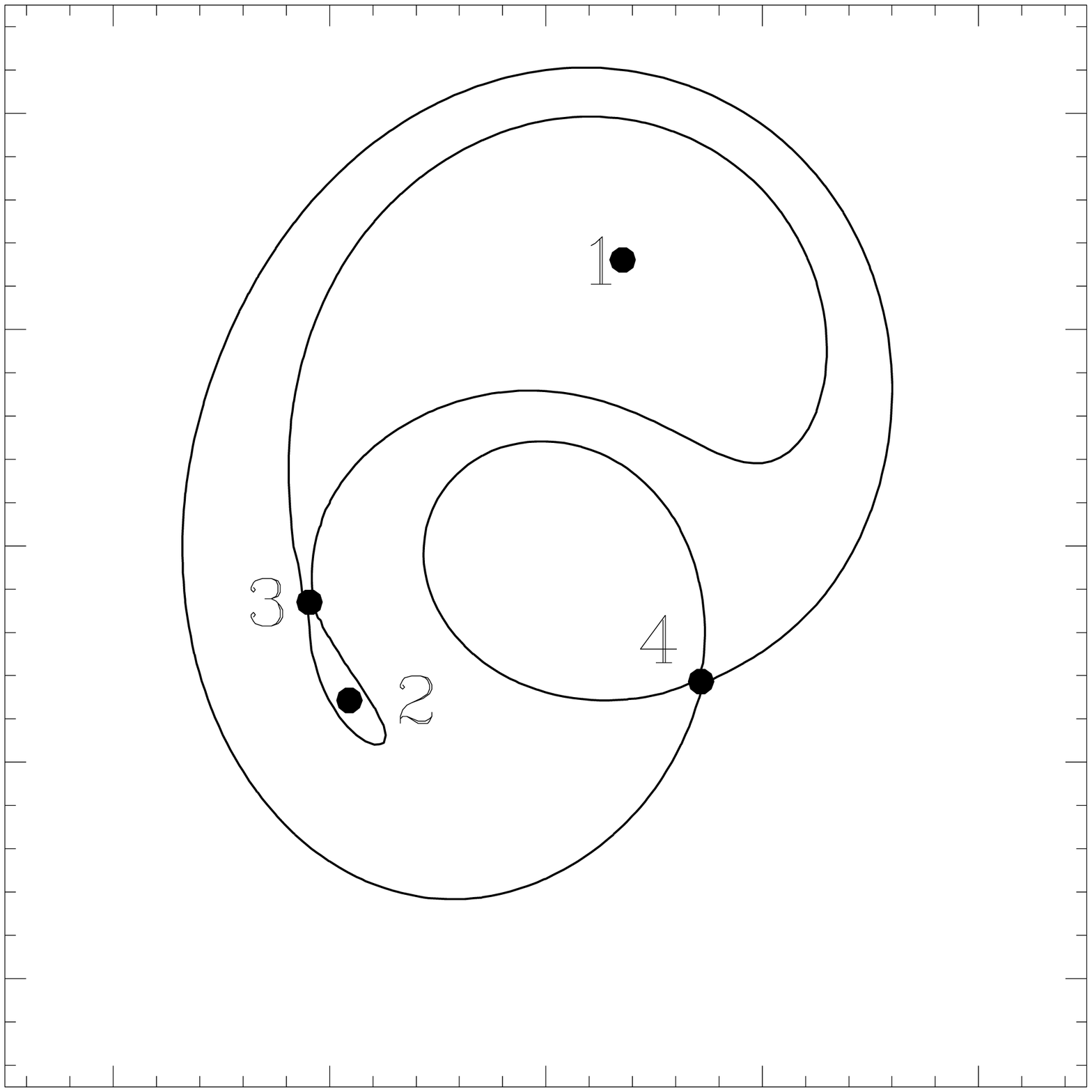}
\hfil
\includegraphics[width=.35\textwidth]{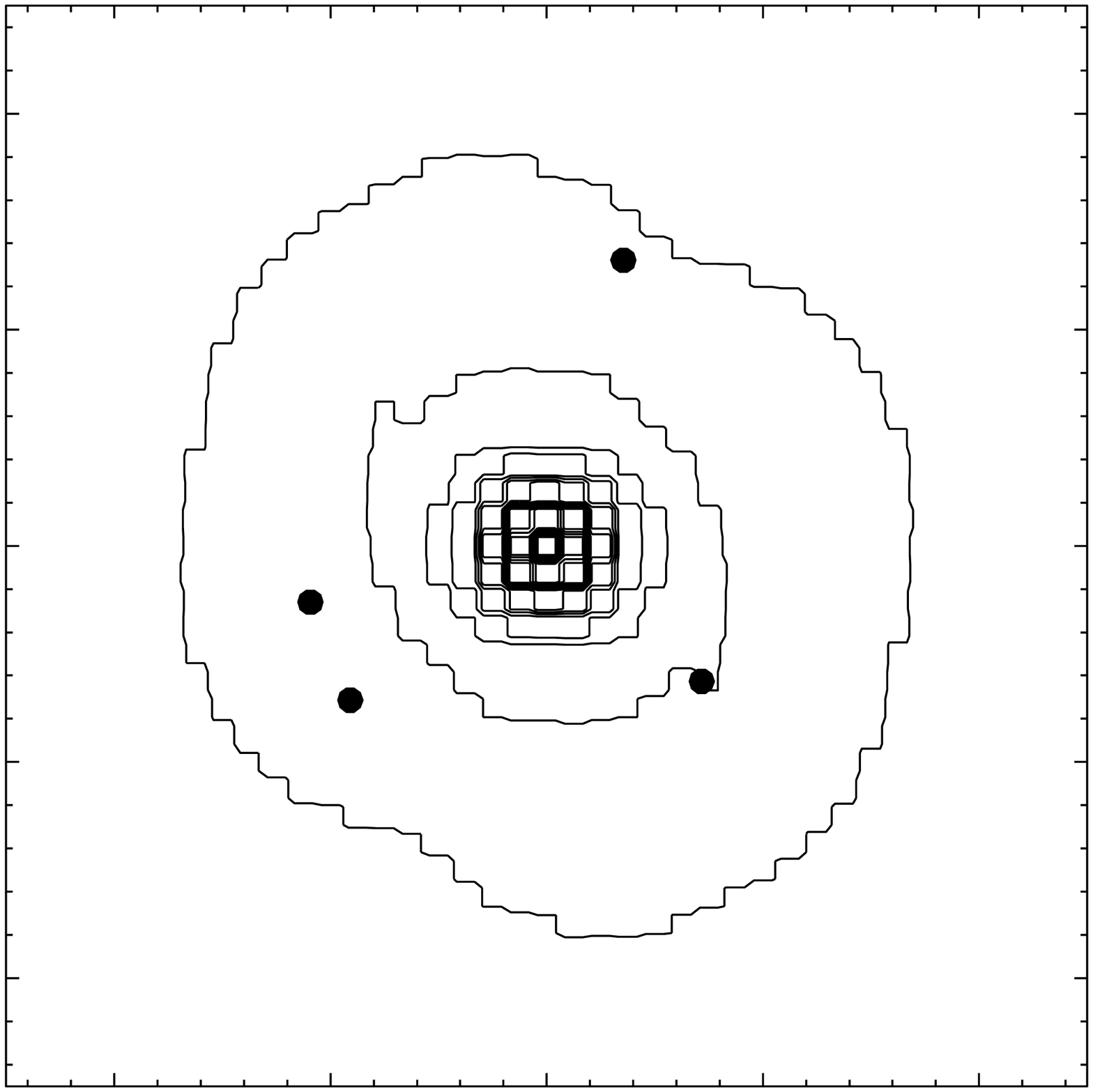}
\end{center}
\begin{center}
\includegraphics[width=.48\textwidth]{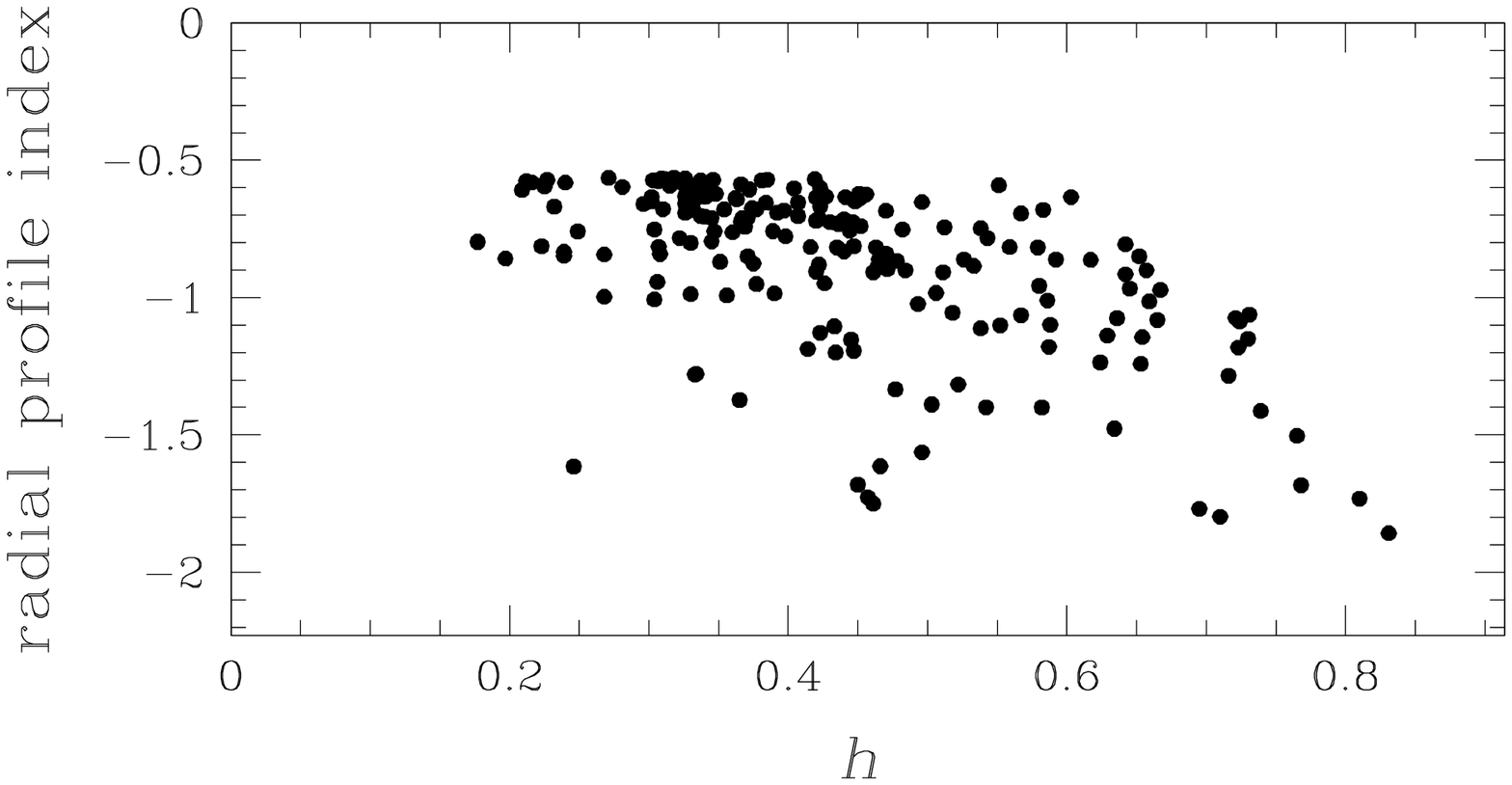}
\hfill
\includegraphics[width=.48\textwidth]{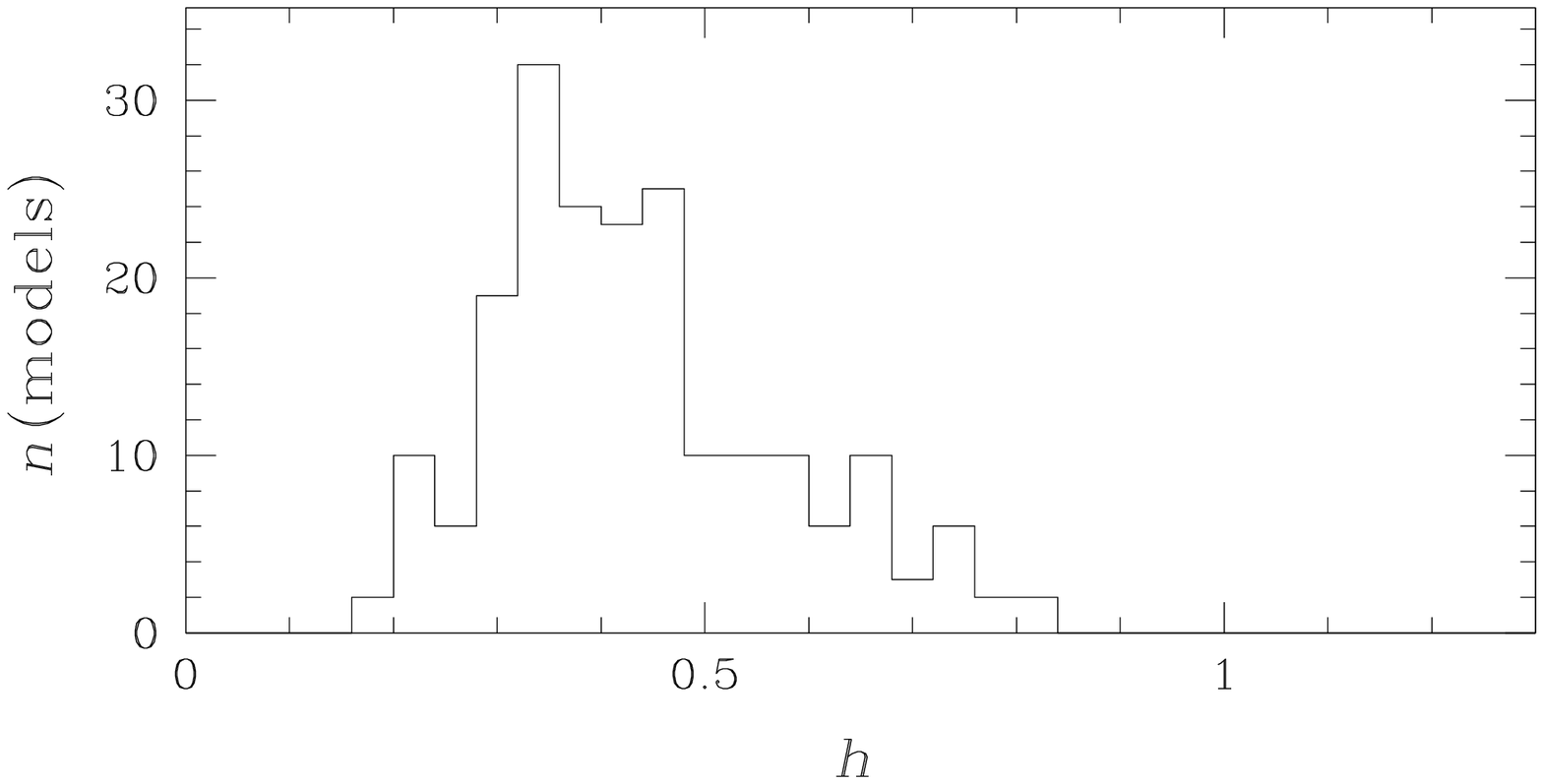}
\end{center}
\caption[]{Models of PG~1115+080. Panels arranged as in Figure
\ref{fig-models0957}.}
\label{fig-models1115}
\end{figure}

\begin{figure}[!hp]
\begin{center}
\includegraphics[width=.35\textwidth]{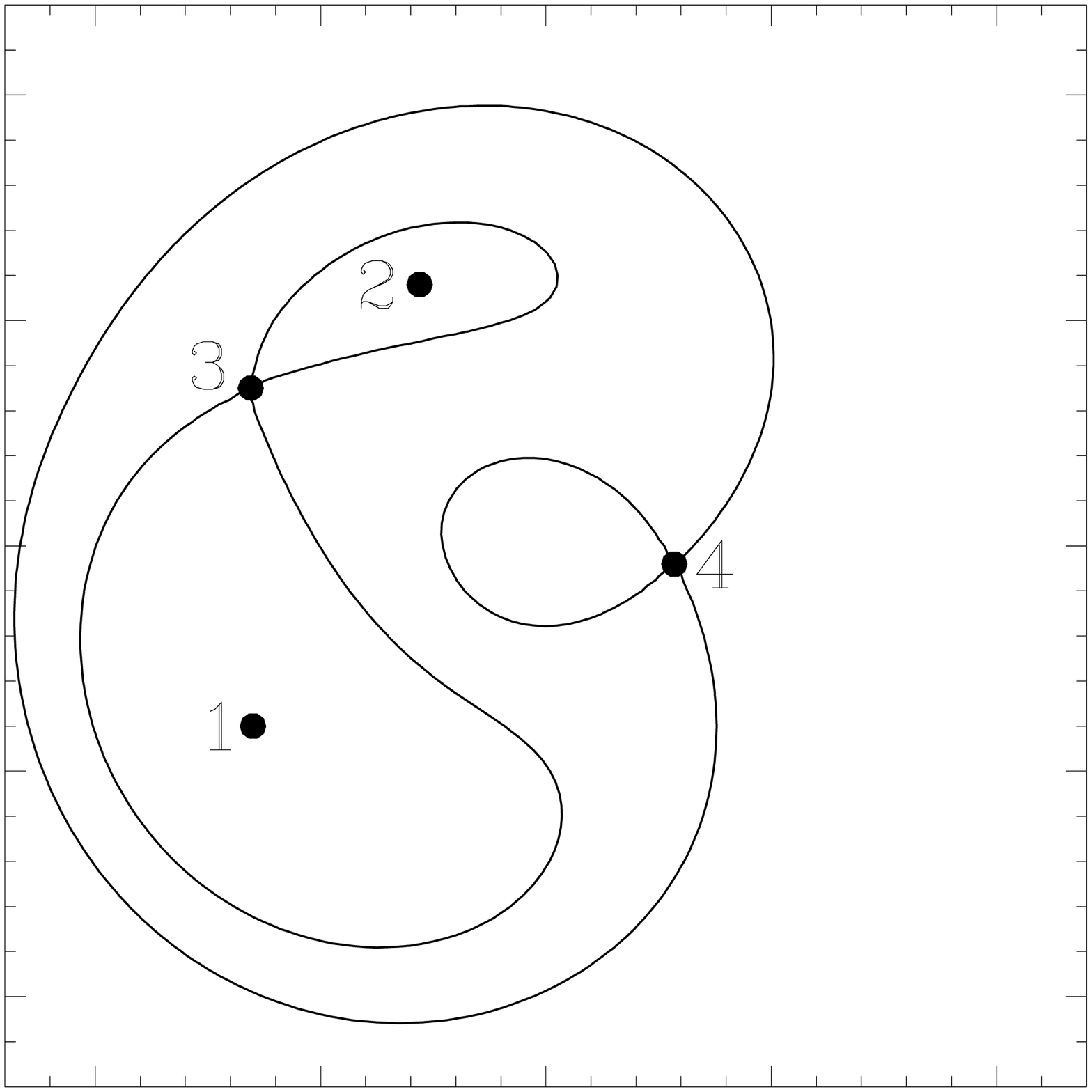}
\hfil
\includegraphics[width=.35\textwidth]{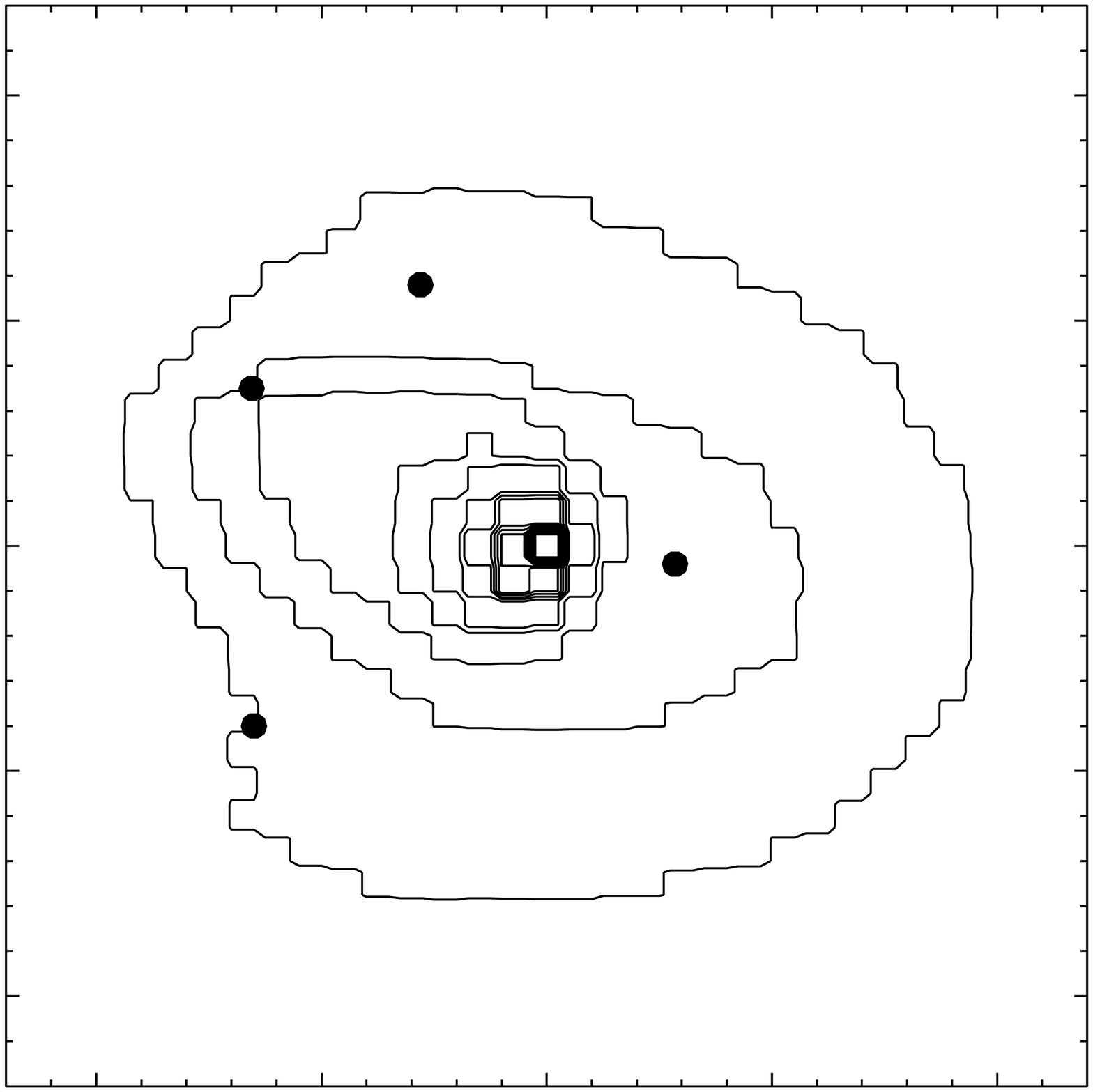}
\end{center}
\begin{center}
\includegraphics[width=.48\textwidth]{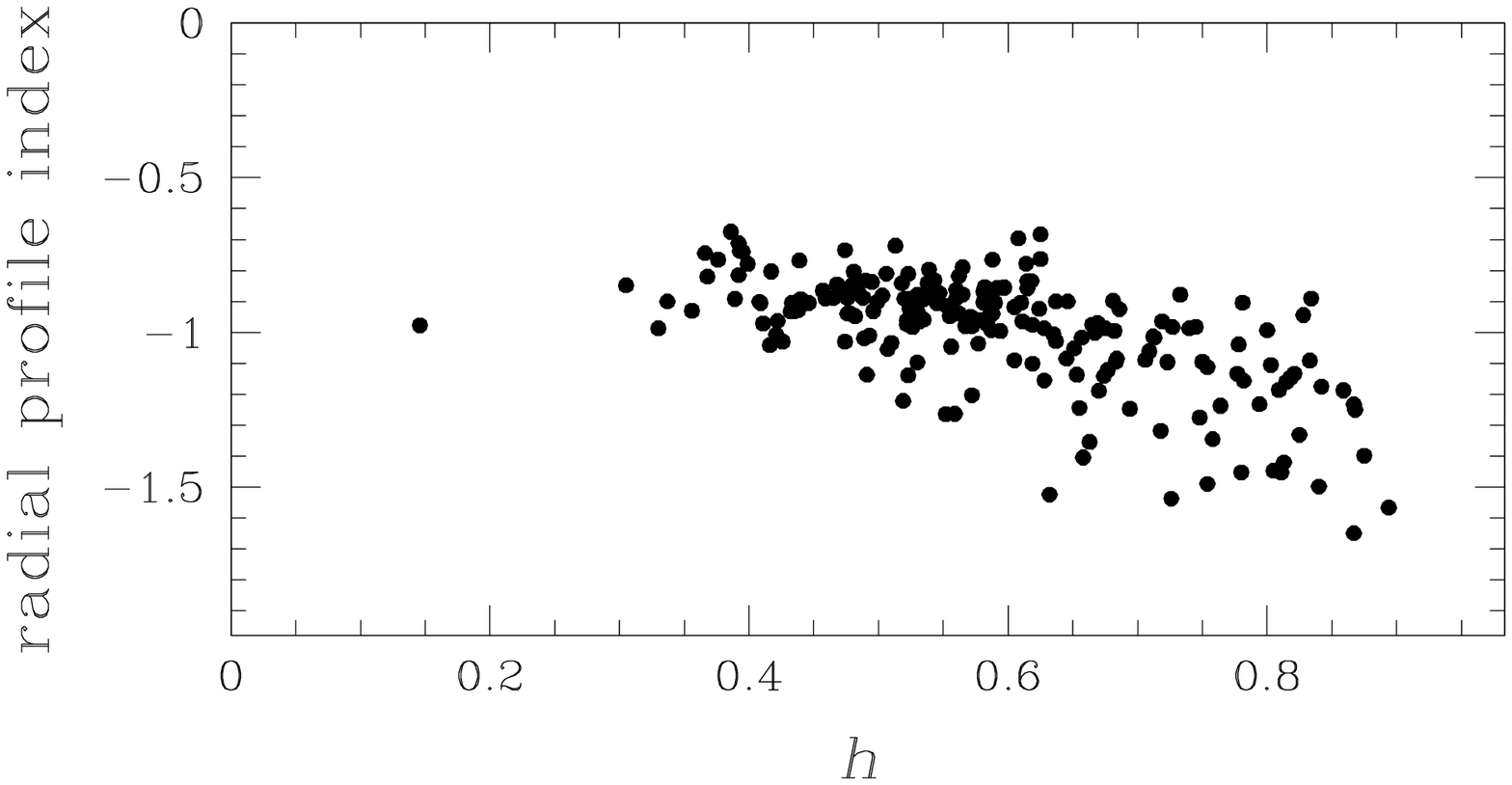}
\hfill
\includegraphics[width=.48\textwidth]{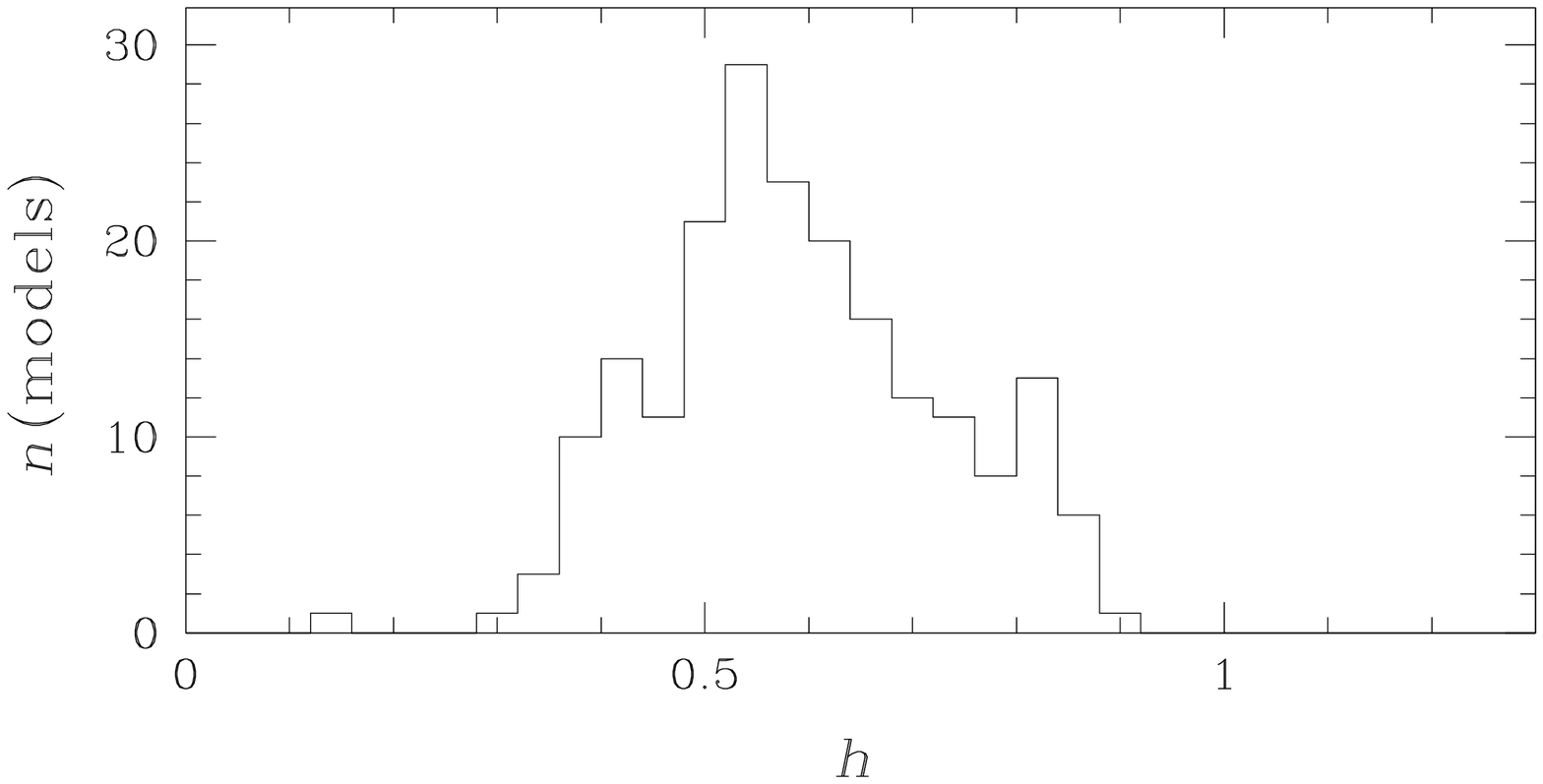}
\end{center}
\caption[]{Models of B~1608+656. Panels arranged as in Figures
\ref{fig-models0957} and \ref{fig-models1115}.}
\label{fig-models1608}
\begin{center}
\includegraphics[width=.35\textwidth]{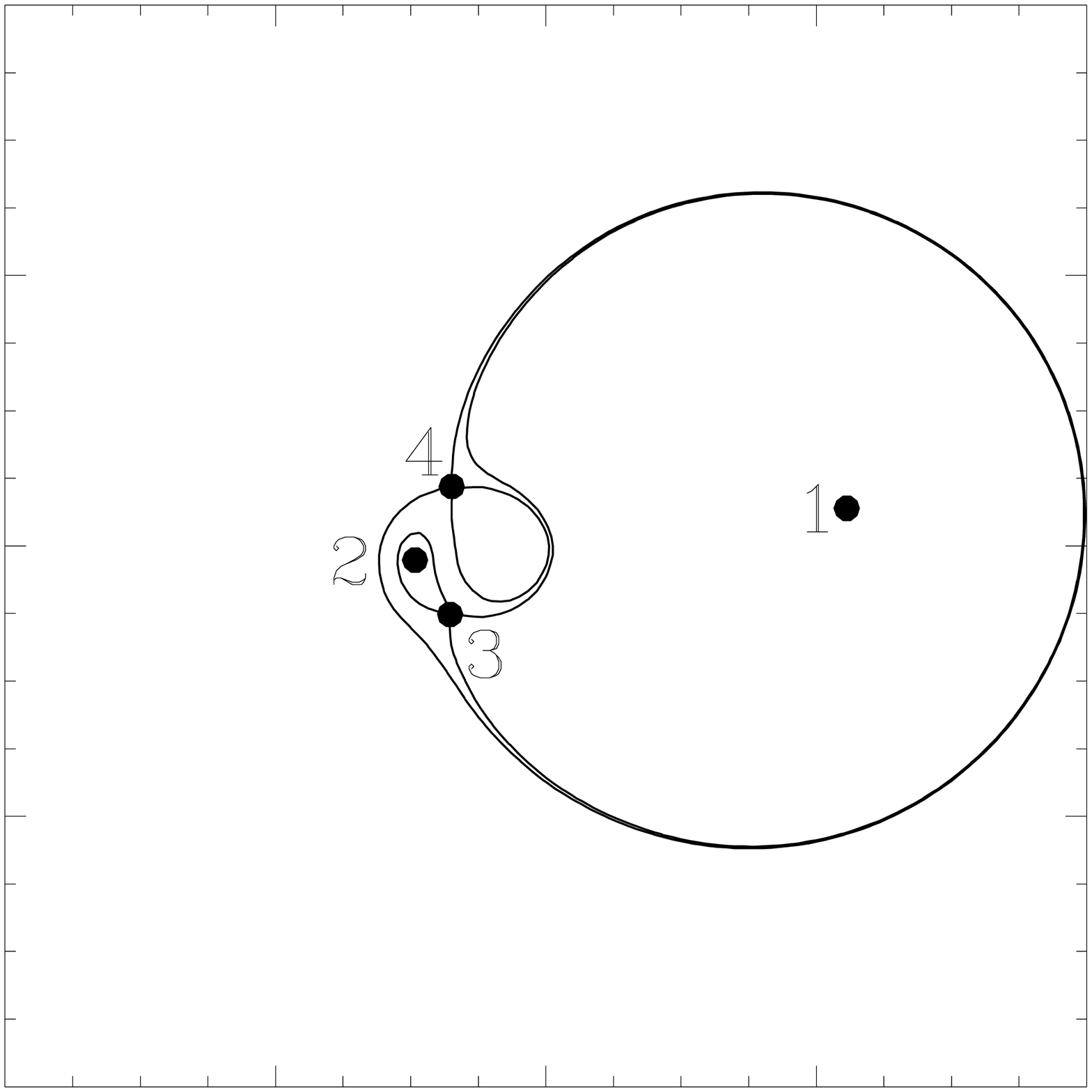}
\hfil
\includegraphics[width=.35\textwidth]{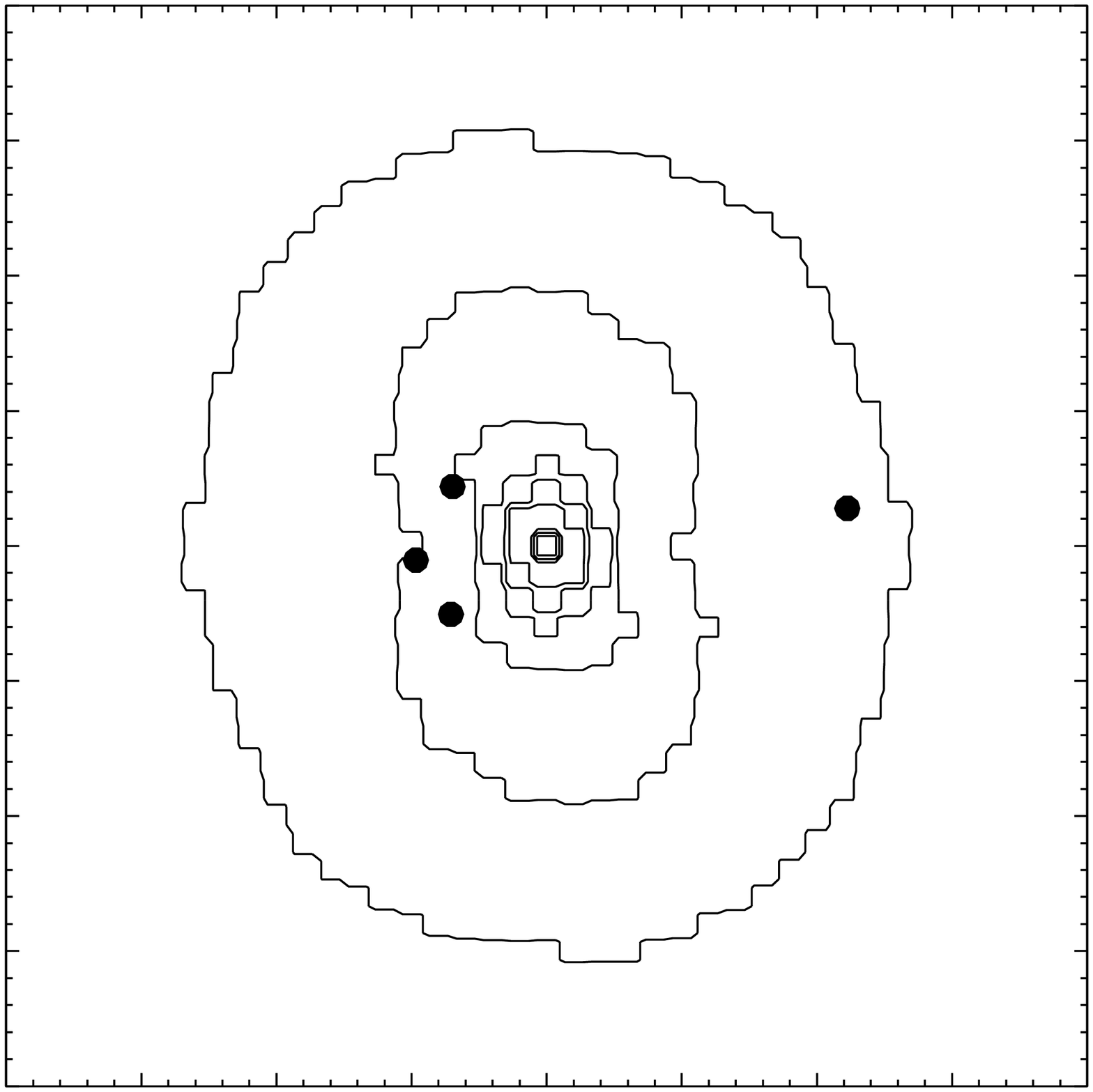}
\end{center}
\begin{center}
\includegraphics[width=.48\textwidth]{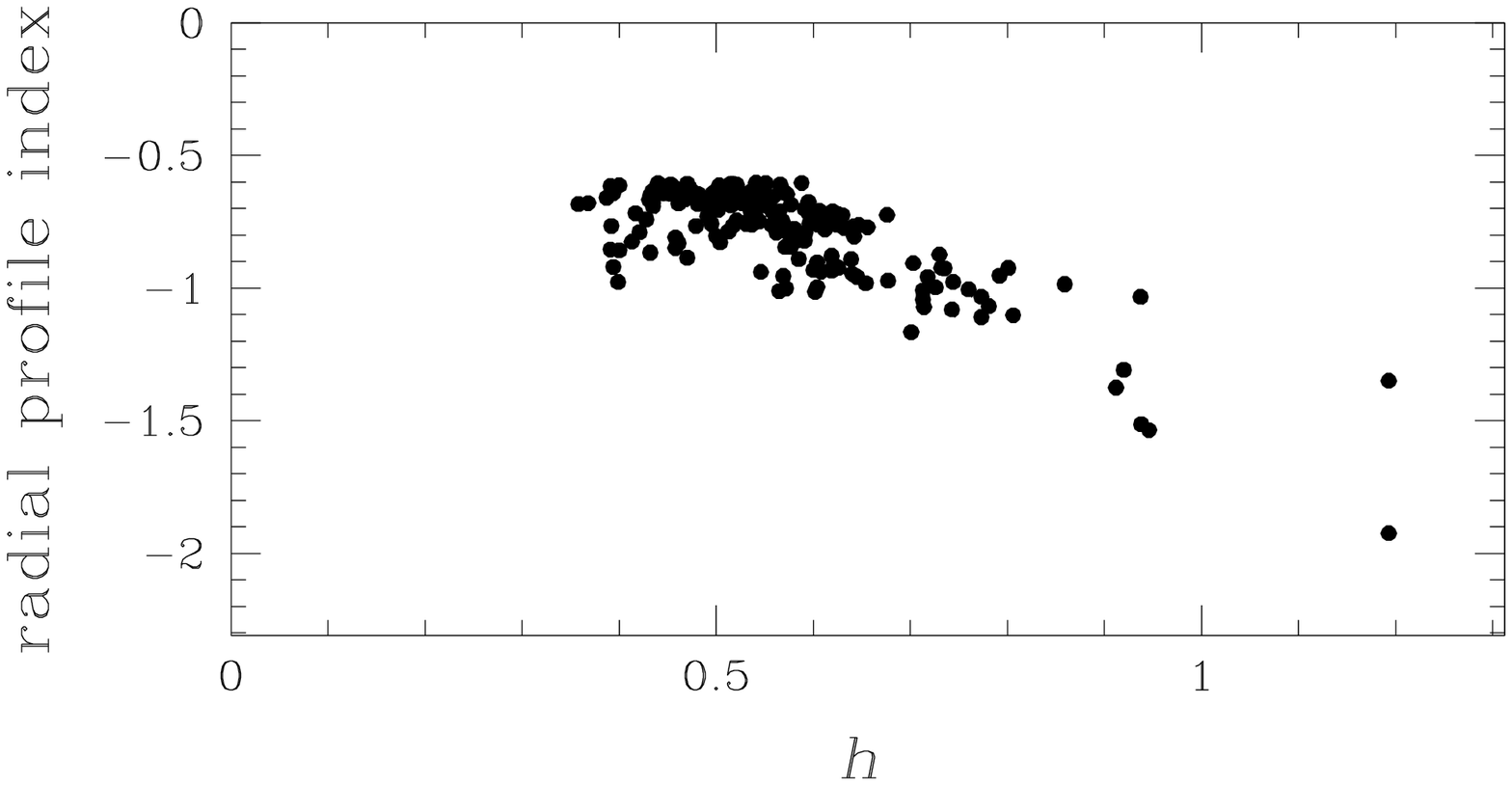}
\hfill
\includegraphics[width=.48\textwidth]{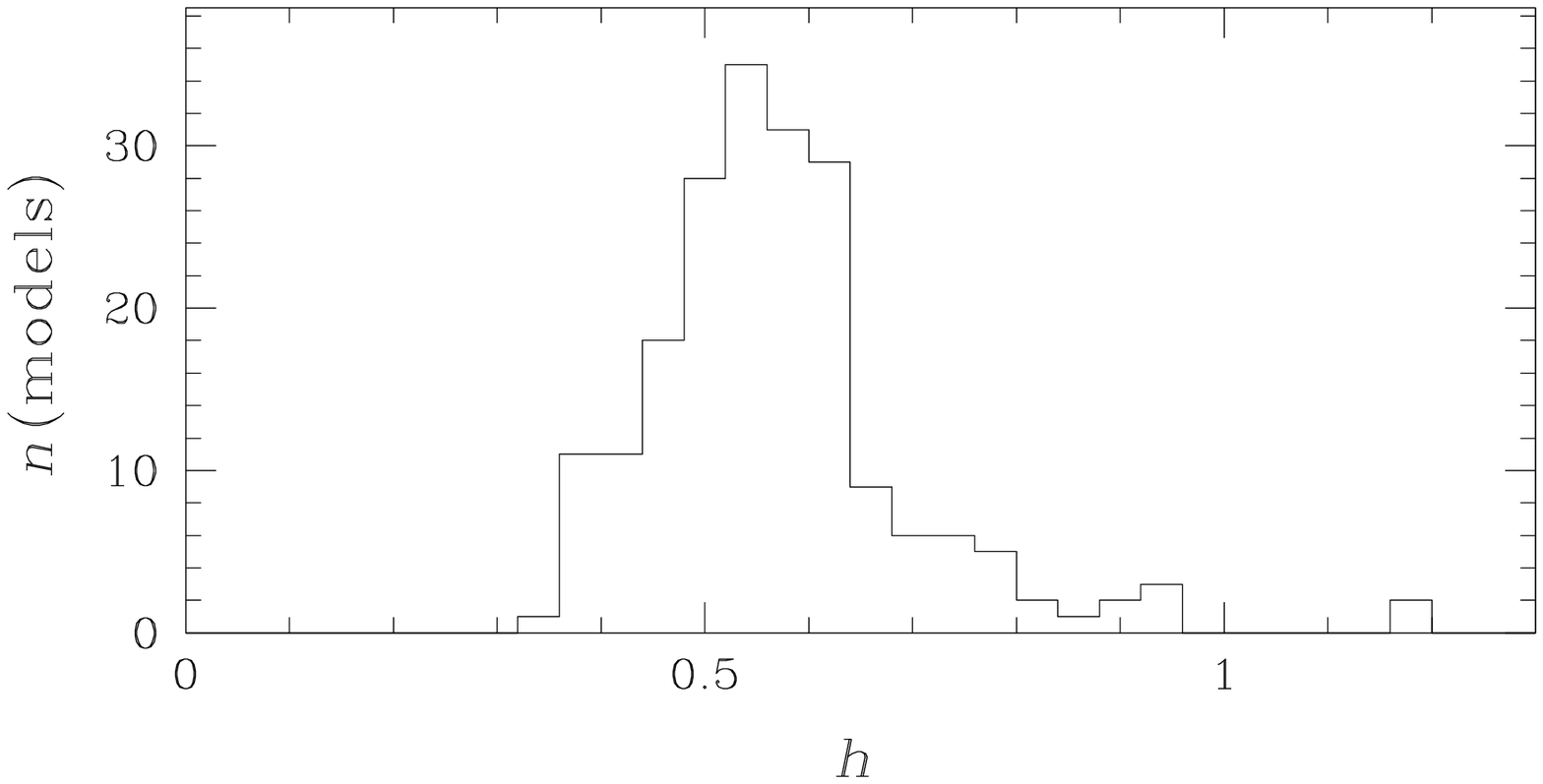}
\end{center}
\caption[]{Models of RX~J0911+055. Panels arranged as in Figures
\ref{fig-models0957}, \ref{fig-models1115}, and \ref{fig-models1608}.}
\label{fig-models0911}
\end{figure}

\paragraph{Q~0957+561}
The reconstructions use the positions, tensor
magnifications \cite{gar94}, \index{magnification} 
and time delays \cite{kundic97} of the
quasar, and another double \index{quasars!double} 
formed by a knot in the quasar's host
galaxy \cite{bern98}. Figure \ref{fig-models0957} shows (i)~the image
configuration and schematic saddle-point \index{saddle-points} 
contours for the quasar,
(ii)~the ensemble-average mass map, (iii)~the $h$ values from each
model in the ensemble plotted against the radial-profile index of that
model between the innermost and outermost images, and (iv)~a histogram
of the $h$ values from the ensemble.  The radial profile index corresponds
roughly to $\alpha-1$ for small values of $\alpha$ as defined in
equation (\ref{eq-plm}).

Two things are very noticeable in Figure~\ref{fig-models0957}.  The
first is the largeness of the uncertainty in $h$; even in this lens
with VLBI structure giving tensor magnifications and a time delay
accurate to 1\%, $h$ values between 0.5 and 1 are all admissible.  The
second noticeable thing is the near-proportionality of $h$ and the
radial index, and it points us to the dominant source of the
uncertainty: changing the radial index is almost equivalent to
applying the mass disk degeneracy transformation, which rescales the
time delays, and hence $h$, while having no effect on image positions
or tensor magnifications.

\paragraph{PG~1115+080}
Here the reconstructions use only image positions of the quasar (an
inclined quad) \index{quasars!quadruple} 
and time delays \cite{schech97,barkana97}.
Figure~\ref{fig-models1115} shows the results, following the same plan
as before.  Again $h$ has a large uncertainty, but is strongly
correlated with the radial profile.  But the distribution of $h$
values is on average lower than for 0957+561.  This promises improved
results if results for several lenses are combined.

\paragraph{B~1608+656}
The reconstructions from this inclined quad use image positions and
time delays \cite{fass99a}. The lensing galaxy in this system appears to
be a binary \cite{bsk00}, so the $180^\circ$-rotation symmetry is not
imposed.  Figure~\ref{fig-models1608} shows the results.  It is
interesting that the mass profile comes out elongated towards the
visible second galaxy, even though the reconstructions had no
information about the light from the lensing galaxies.

\paragraph{RX~J0911+055}
This is a short-axis quad with a preliminary time delay \cite{burud99},
and Figure~\ref{fig-models0911} shows the results. \index{time delay}

\begin{figure}[t]
\begin{center}
\includegraphics[width=.3\textwidth]{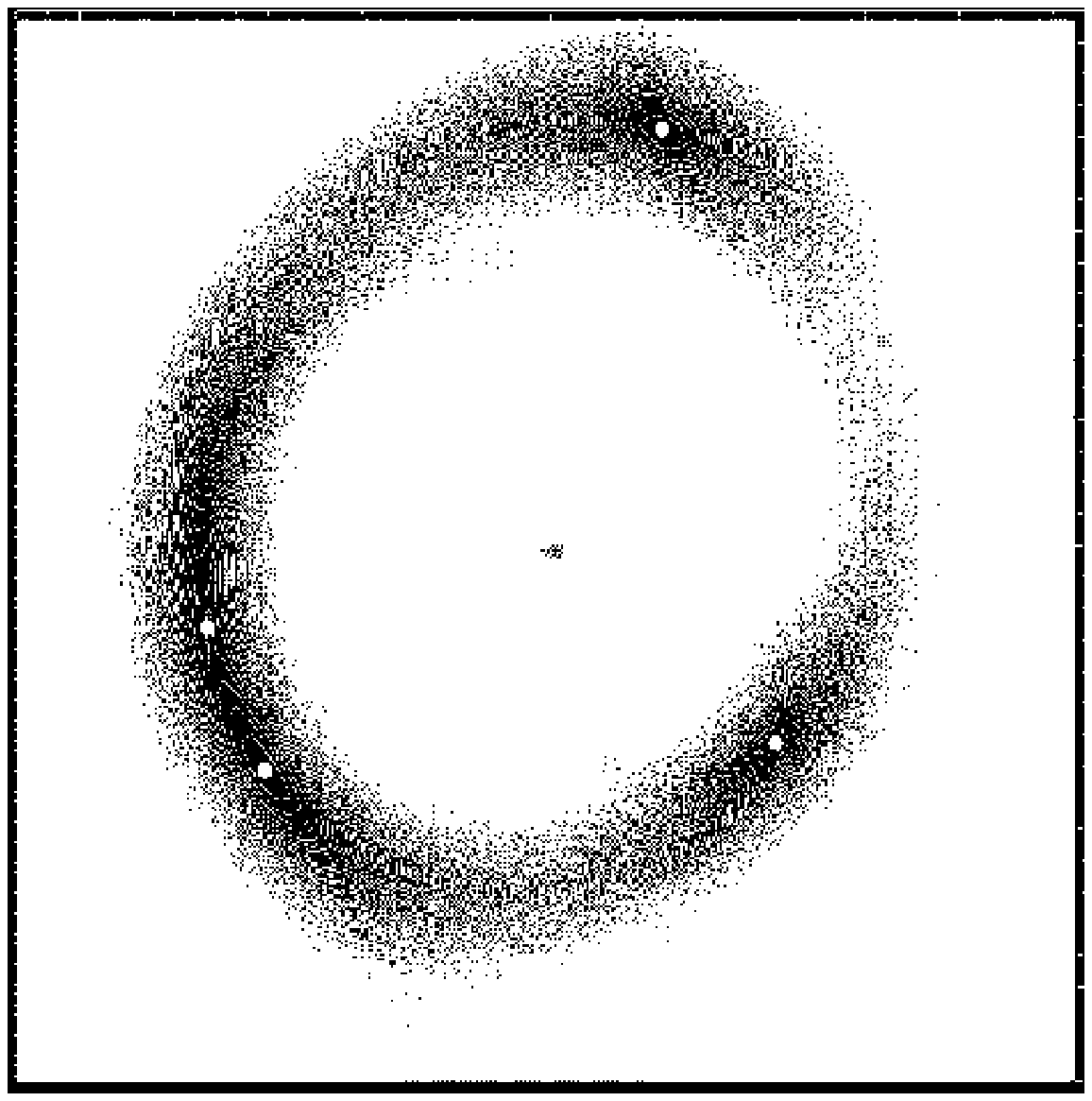}
\includegraphics[width=.3\textwidth]{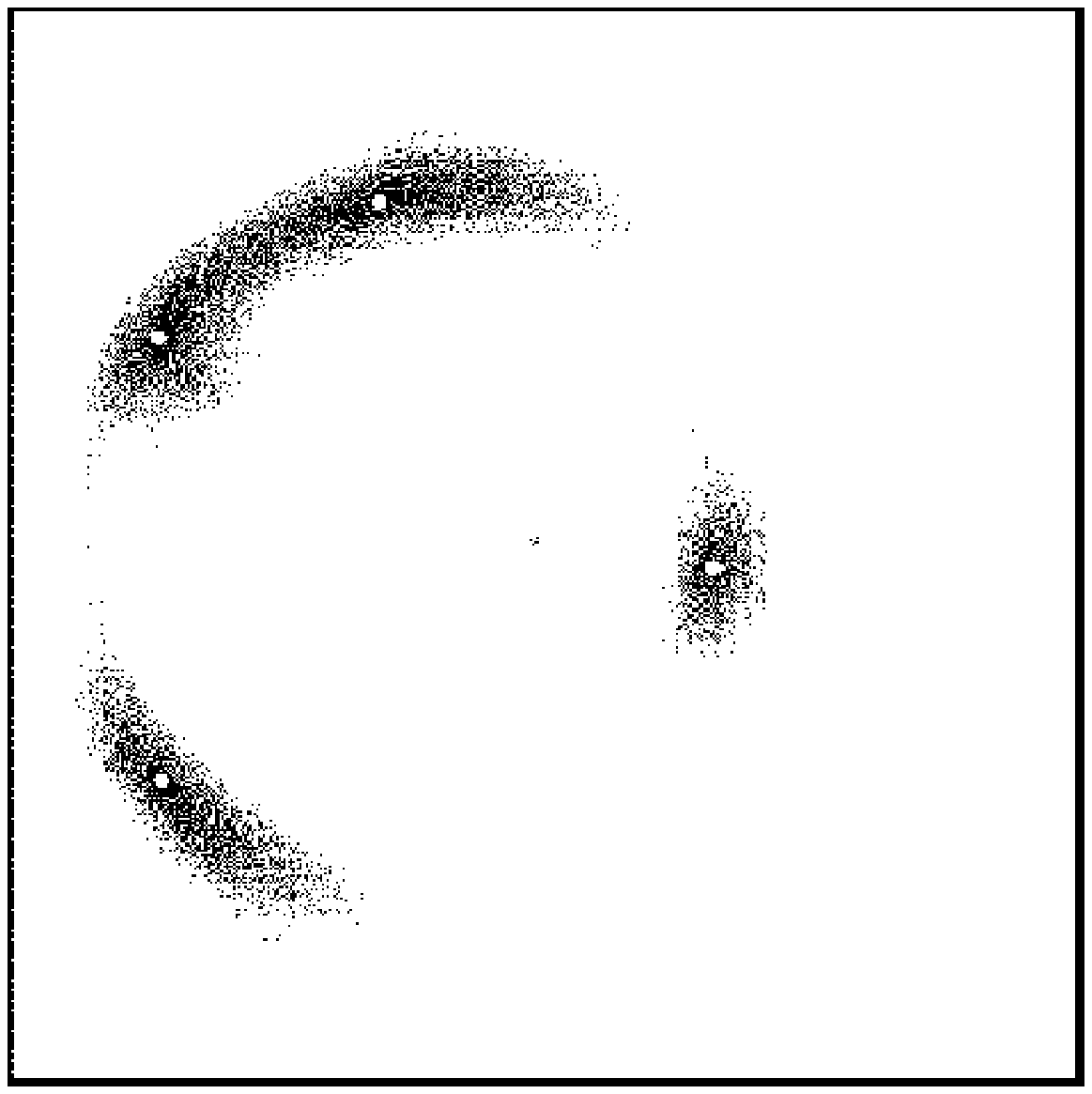}
\includegraphics[width=.3\textwidth]{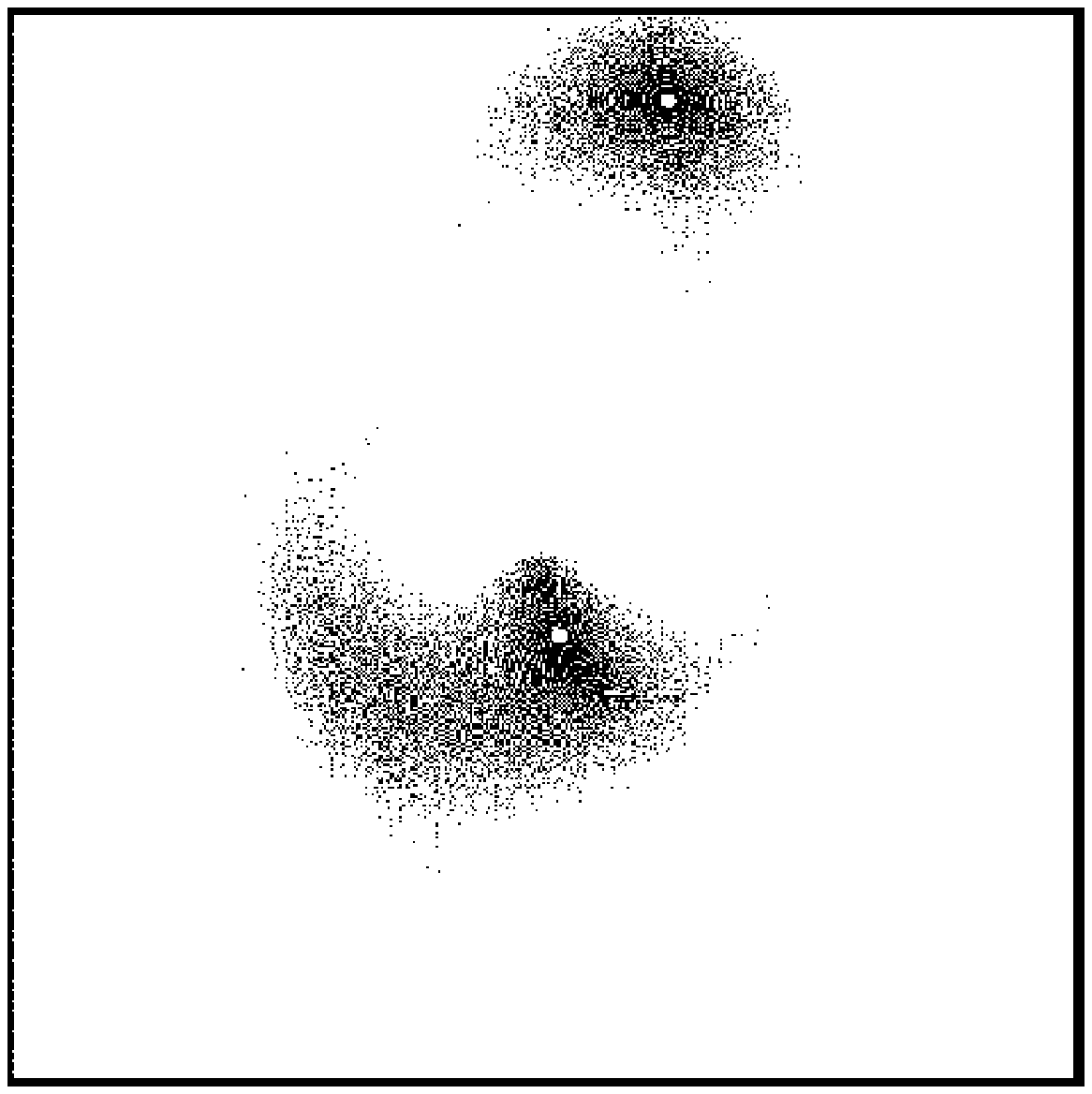}
\end{center}
\caption[]{Ring \index{Einstein!ring} 
and arc \index{arcs} models resulting from plotting arrival-time
\index{arrival time} 
surfaces with dense contours. {\bf Left panel:} PG~1115+080 with
contours 80~min apart; {\bf middle panel:} B~1608+656 with contours 2~hr
apart; {\bf right panel:}~0957+561 with contours 1~day apart.}
\label{fig-rings}
\end{figure}

\subsubsection{Ring and arcs}
\index{Einstein!ring} \index{arcs}

The models described above are designed to fit images of (one or more)
point sources.  But having produced a model, one can check what sort
of image it produces for extended sources.  For a source with a
conical or tent profile for brightness, the image is particularly easy
to produce. We just have to make a dense contour map of the
arrival-time surface for the center of the source and then view this
map from a distance so that the contour lines blur into a
grayscale \cite{ws00,sw01}; the ratio
\begin{eqnarray}
{\hbox{$\tau$-spacing between contours} \over
 \hbox{thickness of contour lines}} \nonumber
\end{eqnarray}
is proportional to the source size.

Figure \ref{fig-rings} shows ring and arc images generated in this way
from the ensemble-average models of PG~1115+080, B~1608+656, and Q~0957+561.
These may be compared with published images of observed
rings \cite{impey98,bsk00,keeton00}.  For PG~1115+080 and B~1608+656 the
model and observed rings overlay extremely well.  (Recall that the
modeling procedure used no ring/arc data.)  For Q~0957+561 the agreement
is not so good: this may indicate simply that the models are less
good, or it may indicate that the observed arc is the image not of the
quasar host \index{quasars!host galaxy} 
galaxy but another galaxy, possibly at different redshift.

\subsubsection{Combined $h$ results}
\index{Hubble constant}

\begin{figure}[!ht]
\begin{center}
\includegraphics[width=.48\textwidth]{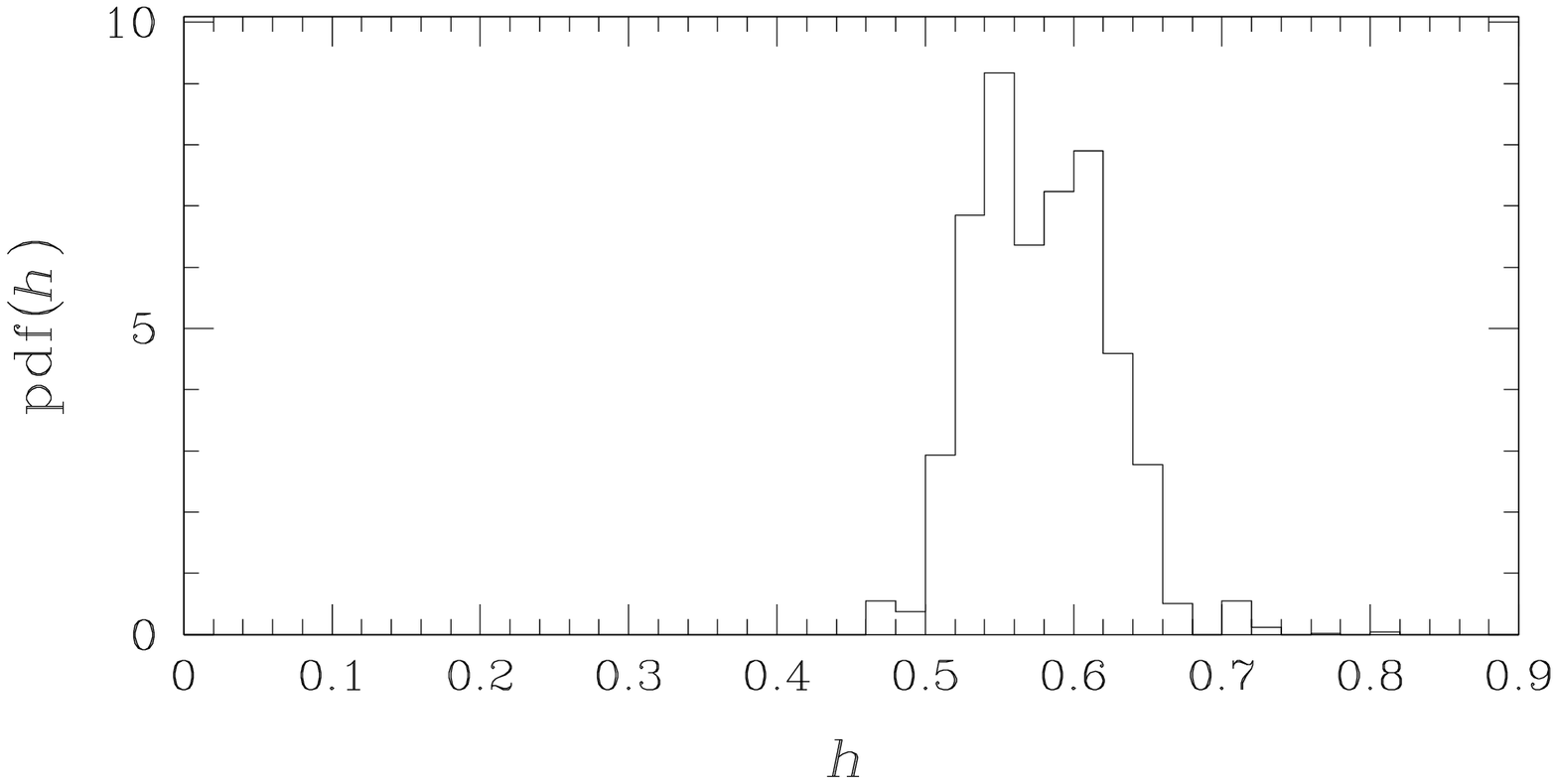}
\includegraphics[width=.48\textwidth]{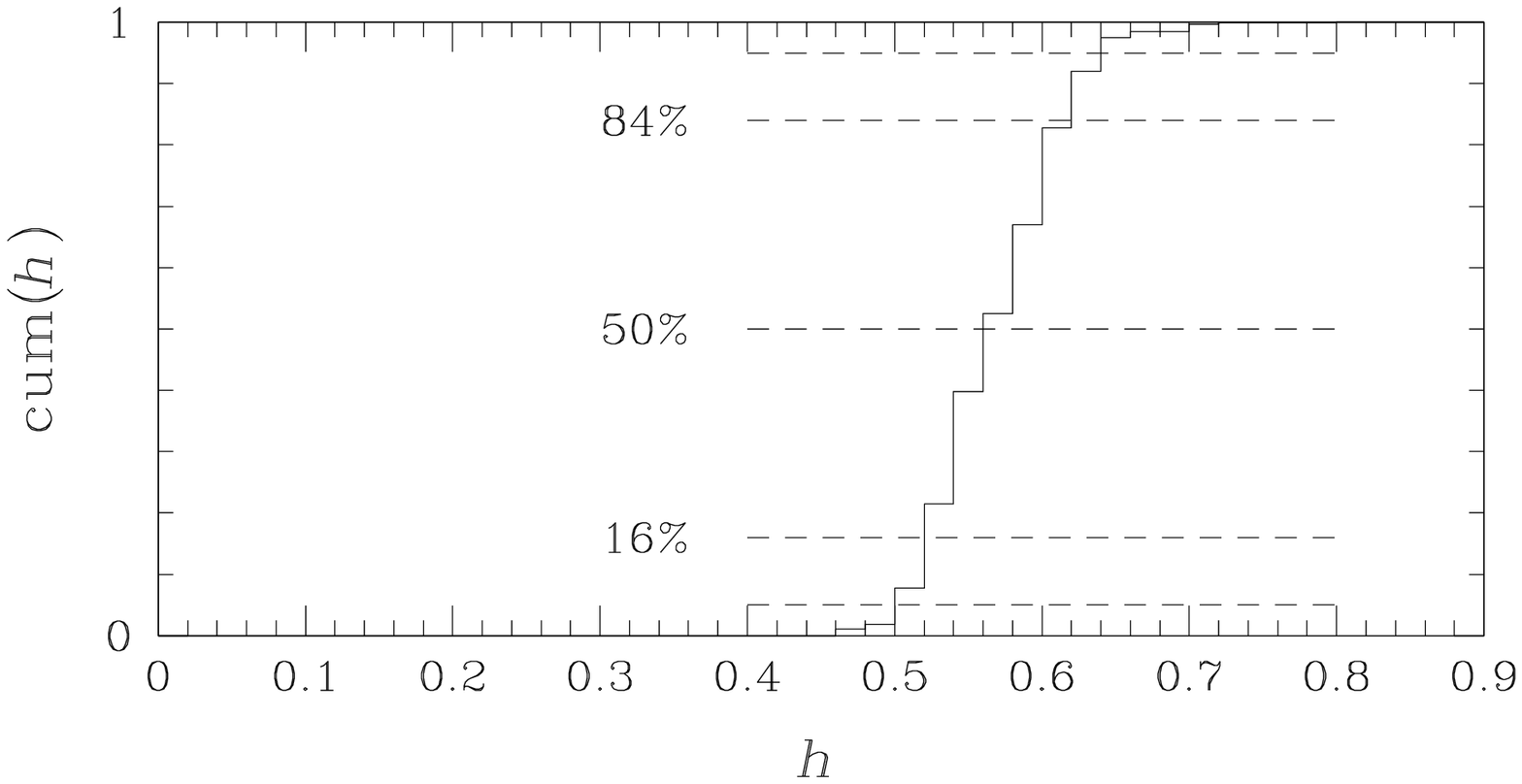}
\end{center}
\caption[]{Combined $h$ results from Q~0957+561, PG~1115+080, B~1608+656, and
RX~J0911+055: histogram on the left and cumulative plot on the right.}
\label{fig-ph}
\end{figure}

Returning to estimates of $h$, in Figure~\ref{fig-ph} we show the
result of combining the $h$ distributions from all four systems above.

The combined result is
\begin{eqnarray}
H_0 &=& 56^{+6}_{-4} \quad \rm(68\%) \nonumber \\
    &=& 56^{+8}_{-6} \quad \rm(90\%) \nonumber 
\end{eqnarray}
The reference cosmology \index{cosmology} 
is Einstein de Sitter; for other cosmologies
the numbers would be 5--10\% higher.

\section{Summary and future prospects}

The  present article concentrates  on some  selected aspects  of quasar
lensing, and in particular on  their use for determining $H_0$. Lensed
quasars have the advantage over other methods that they do not rely on
the knowledge of any standard candle. The disadvantage is that precise
modeling of the  potential well responsible for the  lensing effect is
required.

It has been shown that, with present day instrumentation and efficient
post-processing  techniques,  ``mass  production'' of  time-delays  is
possible,  even  using  2m   class  telescopes  under  average  seeing
conditions  \cite{ingunnthesis}. A typical  precision on  a time-delay
determination is of  the order of 10\%, sometimes  better. However, in
many cases, most of the error on $H_0$ comes from the lens models used
to convert the  time-delay into $H_0$. 
\index{time delay} \index{Hubble constant} The problem  can be overcome in
two  ways:  (1) by  using  any  prior  knowledge available  on  (lens)
galaxies and,  (2) by improving  the observations to  constrain better
the   gravitational   potential  (main   lens   and  any   intervening
cluster/group) in each individual lensed quasar.

The effectiveness  of quasar lensing in producing  a competitive value
for  $H_0$  therefore depends  on  our  knowledge  of the  physics  of
galaxies in general. Gravitational lensing itself should be able to set
suitable constraints on galaxy  mass profiles, \index{mass!profile} 
for example through the
statistical  study of  galaxy halos \index{galaxy!halos} 
using galaxy-galaxy  lensing (see
chapter  on Dark  Matter  Halos). The  development of  two-dimensional
spectrographs used to derive the full velocity field for many galaxies
of  all types will  also yield  important clues  to the  detailed mass
distribution  in galaxies.  Both methods,  direct or  indirect, should
constrain the degree of concentration  of the mass in galaxies and the
extent and shape of dark  matter halos, two quantities which are often
unconstrained  in  present  days  lens  models  and  which  imply  the
exploration  of  huge parameter  spaces  followed  by  a choice  of  a
``best'' model or a best family of models.

Improving the observations  of individual lensed systems is
also important.  Lensing galaxies have  an effect not only  on quasars
but also on galaxies in the  vicinity of quasars.  They should be seen
under the form  of arcs or arclets, \index{arcs} \index{arclets} 
as long  as the angular resolution
and depth are  sufficient.  The Advanced Camera for  Surveys (ACS), 
\index{ACS} on
board of  the HST, \index{HST} 
shall  provide us at  least with depth,  hence with
more  background sources  susceptible to  be  lensed, just  as is  the
quasar. Since we usually observe  only 2 or 4 quasar images, observing
even a few arclets is a  significant constraint for the lens model. In
addition, with the  depth of the ACS, most  lensed quasars should show
their distorted  host galaxy,  and bring even  more constraints  on the
models.  Constraining  lens models  using  many  arclets will  probably
become  an efficient  method with  the launch  of the  Next Generation
Space Telescope. \index{NGST}

Measuring $H_0$  is not the  only application of quasar  lensing. Once
adequate observational constrains are available, or even assuming
$H_0$ can  be measured independently  by other methods, one  shall use
lensing to map the mass  distribution in lensing galaxies and to infer
basic  parameters on  the structure  of quasars,  using  the chromatic
variations due  to microlensing \index{quasars!microlensing} 
events.  Spectrophotometric monitoring
is  the next  obvious observational  step in  the field,  in  order to
enable such applications.

Whether we  will learn  about the mass  distribution in  galaxies once
$H_0$ is measured by other means, or the opposite, will depends on the
speed of the progress made in the fields of the physics of galaxies,
galaxy-galaxy lensing  and on the possible discoveries  of new methods
to infer $H_0$ with a high precision.

\section{Inventory of known systems}
The numerous quasar surveys carried out to date and others still under way
have led to the discovery of many lensed systems.  Consequently, it is
becoming increasingly difficult to keep track of all new lenses discovered.
We try here to provide the reader with a list of all known cases;
we apologize in advance to those who will not see their favorite
system, probably because it has been too recently discovered. Note
also that we list only lensed quasars. There are other cases of
multiply imaged distant galaxies, discovered for example in HST deep
fields (see for example \cite{barkana99}). Basic information such as
coordinates, source and lens redshifts are given, together with the
reference of the discovery paper. When several references are listed,
the first ones corresponds to the discovery paper, and the others to
the time delay \index{time delay} 
measurement, when available. Time delays are given
relative to the leading image. For example, $\Delta$t(BA) means that
image B is the leading image. Note finally that most objects have been
observed or will be observed with the HST, \index{HST} \index{CASTLES} 
either in the context or
individual observing programs or through the CASTLE Survey
whose main results are summarized at {\tt
http://cfa-www.harvard.edu/glensdata/}.


\index{quasars!double}
\begin{table}[h]
\caption{List of confirmed doubles.}
\begin{center}
\renewcommand{\arraystretch}{1.2}
\setlength\tabcolsep{8pt}

\begin{tabular}{llll}
\hline\noalign{\smallskip}
Object          & Coords (2000)    & Redshifts   & Notes \\
\noalign{\smallskip}\hline\hline
Q~0142-100                         & $\alpha$:  ~~01h 45m 16.50s & $z_s$=2.72   & \\
Surdej et al. \cite{surdej87}      & $\delta$: $-$09d 45m 17.00s & $z_l$=0.49   & \\
\hline 
CTQ~414                            & $\alpha$:  ~~01h 58m 41.44s & $z_s$=1.29  & \\
Morgan et al. \cite{morgan99}      & $\delta$: $-$43d 25m 04.20s & $z_l$=?     & \\
\hline 
B~0218+357                         & $\alpha$:  ~~02h 21m 05.48s & $z_s$=0.96  & $\Delta$t(BA) = 10.5\\ 
O'dea et al. \cite{odea92}         & $\delta$:   +35d 56m 13.78s & $z_l$=0.68  & $\pm$ 0.4 days\\
Biggs et al. \cite{biggs99}        &                             &             &               \\ 
\hline 
HE 0512-3329                       & $\alpha$:  ~~05h 14m 10.78s & $z_s$=1.57  & \\ 
Gregg et al. \cite{gregg00}        & $\delta$: $-$33d 26m 22.50s & $z_l$=0.93(?)  & \\
\hline
CLASS~B0739+366                    & $\alpha$:  ~~07h 42m 51.20s & $z_s$=?     & \\   
Rusin et al. \cite(rusin01)        & $\delta$:   +36d 34m 43.70s & $z_l$=?     & \\
\hline 
MG~0751+2716                       & $\alpha$: ~~07h 51m 41.46s  & $z_s$=3.20 & Ring\\
Lehar et al. \cite{lehar97}        & $\delta$: +27d 16m 31.35s  & $z_l$=0.35 & \\ 
\hline
HS~0818+1227                       & $\alpha$:  ~~08h 21m 39.10s & $z_s$=3.12  & \\   
Hagen \& Reimers \cite{hag00}      & $\delta$:   +12d 17m 29.00s & $z_l$=0.39  & \\
\hline
APM~08279+5255                     & $\alpha$:  ~~08h 31m 44.94s & $z_s$=3.87  & \\   
Irwin et al. \cite{irwin98}        & $\delta$:   +52d 45m 17.70s & $z_l$=?     & \\
\hline
SBS~0909+532                       & $\alpha$:  ~~09h 13m 01.05s & $z_s$=1.38  & \\   
Kochanek et al. \cite{kocha97}     & $\delta$:   +52d 59m 28.83s & $z_l$=0.83  & \\
RXJ~0921+4528                      & $\alpha$:  ~~09h 21m 12.81s & $z_s$=1.66  & \\   
                                   & $\delta$:   +45d 29m 04.40s & $z_l$=0.31  & \\   
\hline
FBQ~0951+2635                      & $\alpha$:  ~~09h 51m 22.57s & $z_s$=1.24  & \\   
Schechter et al. \cite{schech98}   & $\delta$:   +26d 35m 14.10s & $z_l$=?     & \\
\hline
BRI~0952-0115                      & $\alpha$:  ~~09h 55m 00.01s & $z_s$=4.5  & \\   
McMahon \& Irwin \cite{macma92}    & $\delta$: $-$01d 30m 05.00s & $z_l$=?    & \\ 
\hline
Q~0957+561                         & $\alpha$:  ~~10h 01m 20.78s & $z_s$=1.41 & $\Delta$t(BA) = 417 \\   
Walsh et al. \cite{Walsh79}        & $\delta$: $+$55d 53m 49.40s & $z_l$=0.36 & $\pm$ 3 days\\
Kundi\'c et al. \cite{kundic97}    &                             &            &             \\ 
\hline
LBQS~1009-0252                     & $\alpha$:  ~~10h 12m 15.71s & $z_s$=2.74 & \\   
Surdej et al. \cite{surdej93}      & $\delta$: $-$03d 07m 02.00s & $z_l$=? & \\
\hline
Q~1017-207                         & $\alpha$:  ~~10h 17m 24.13s & $z_s$=2.55 & \\   
Claeskens et al. \cite{claeskens96}& $\delta$: $-$20d 47m 00.40s & $z_l$=? & \\
\hline
FSC~10214+4724                     & $\alpha$:  ~~10h 24m 37.58s & $z_s$=2.29 & Ring\\   
Graham \& Liu \cite{graham95}      & $\delta$: $+$47d 09m 07.20s & $z_l$=? & \\
\hline
B~1030+074                         & $\alpha$:  ~~10h 33m 34.08s & $z_s$=1.54 & \\   
Xanthopoulos et al. \cite{Xantho98}& $\delta$: $+$07d 11m 25.50s & $z_l$=0.60 & \\
\hline
HE~1104-1805                       & $\alpha$:  ~~11h 06m 33.45s & $z_s$=2.32 & $\Delta$t(AB) = 260\\   
Wisotzki et al. \cite{wiso93}      & $\delta$: $-$18d 21m 24.20s & $z_l$=0.73 & $\pm$ 90 days\\
Wisotzki et al. \cite{wisot99}     &                             &            &         \\
\noalign{\smallskip}

\hline
\noalign{\smallskip}
\end{tabular}
\end{center}
\label{Tab1a}
\end{table}

\begin{table}[!ht]
\caption{List of confirmed doubles (continued)}
\begin{center}
\renewcommand{\arraystretch}{1.2}
\setlength\tabcolsep{8pt}

\begin{tabular}{llll}
\hline\noalign{\smallskip}
Object          & Coords (2000)    & Redshifts   &  Notes \\
\noalign{\smallskip}\hline\hline
B~1127+385                         & $\alpha$:  ~~11h 30m 00.13s & $z_s$=? & \\   
Koopmans et al. \cite{Koopmans99}  & $\delta$: $+$38d 12m 03.10s & $z_l$=? & \\
\hline
MG~1131+0456                       & $\alpha$:  ~~11h 31m 56.48s & $z_s$=?    & \\   
Hewitt et al. \cite{hewitt88}      & $\delta$: $+$04d 55m 49.80s & $z_l$=0.84 & \\
\hline
B~1152+199                         & $\alpha$:  ~~11h 55m 18.37s & $z_s$=1.02  & \\   
Myers et al. \cite{myers99}        & $\delta$: $+$19d 39m 40.39s & $z_l$=0.44  & \\
\hline
Q~1208+1011                        & $\alpha$:  ~~12h 10m 57.16s & $z_s$=3.80 & \\   
Magain et al. \cite{Magain92}      & $\delta$: $+$09d 54m 25.60s & $z_l$=?    & \\
\hline
SBS~1520+530                       & $\alpha$:  ~~15h 21m 44.83s & $z_s$=1.86 & $\Delta$t(BA) = 130\\   
Chavushyan et al. \cite{Chavu97}   & $\delta$: $+$52d 54m 48.60s & $z_l$=0.71 & $\pm$ 6 days\\
Burud et al. \cite{burud01b}       &                             &            & \\ 
\hline
MG~1549+3047                       & $\alpha$:  ~~15h 49m 12.37s & $z_s$=?    & Ring\\   
Lehar et al. \cite{lehar93}        & $\delta$: $+$30d 47m 16.60s & $z_l$=0.11 & \\
\hline
B~1600+434                         & $\alpha$:  ~~16h 01m 40.45s & $z_s$=1.59 & $\Delta$t(BA) = 51\\   
Jackson et al. \cite{jack95}       & $\delta$: $+$43d 16m 47.80s & $z_l$=0.42 & $\pm$ 4 days (radio)\\
Burud et al. \cite{Ingunn00}       &                             &            & $\Delta$t(BA) = 47\\
Koopmans et al. \cite{koop00}      &                             &            & $\pm$ 11 days (radio)\\
\hline
PMN~J1632-0033                     & $\alpha$:  ~~16h 32m 55.98s & $z_s$=3.42 & \\   
Winn et al. \cite{winn02a}          & $\delta$: $-$00d 33m 04.50s & $z_l$=? & \\ 
FBS~1633+3134                      & $\alpha$:  ~~16h 33m 48.99s & $z_s$=1.52 & \\   
Morgan et al. \cite{morgan00}      & $\delta$: $+$31d 34m 11.90s & $z_l$=? & \\
\hline
MG~1654+1346                       & $\alpha$:  ~~16h 54m 41.83s & $z_s$=1.74 & Ring \\   
Langston et al. \cite{langston89}  & $\delta$: $+$13d 46m 22.00s & $z_l$=0.25 & \\
\hline
PKS~1830-211                       & $\alpha$:  ~~18h 33m 39.94s & $z_s$=2.51 & Ring \\   
Subrahmanyan et al. \cite{subra90} & $\delta$: $-$21d 03m 39.70s & $z_l$=0.89 & $\Delta$t(BA) = 26\\
Lovell et al. \cite{lov98}         &                             &            & $\pm$ 8 days\\ 
\hline
PMN~J1838-3427                     & $\alpha$:  ~~18h 38m 28.50s & $z_s$=2.78 & \\
Winn et al. \cite{winn00}          & $\delta$: $-$34d 27m 41.60s & $z_l$=? & \\
\hline
B~1938+666                         & $\alpha$: ~~19h 38m 25.19s  & $z_s$=?    & Full ring \\
Rhoads et al. \cite{rho96}         & $\delta$: +66d 48m 52.20s   & $z_l$=0.88 & \\
\hline
PMN~J2004-1349                     & $\alpha$:  ~~20h 04m 07.02s & $z_s$=? & \\
Winn et al. \cite{winn01}          & $\delta$: $-$13d 49m 31.65s & $z_l$=? & \\ 
\hline
B~2114+022                         & $\alpha$:  ~~21h 16m 50.75s & $z_s$=? & \\
                                   & $\delta$:   +02d 25m 46.90s & $z_l$=0.32/0.59 & \\ 
\hline
HE~2149-2745                       & $\alpha$:  ~~21h 52m 07.44s & $z_s$=2.03 & $\Delta$t(BA) = 103\\
Wisotski et al. \cite{wiso96}      & $\delta$: $-$27d 31m 50.20s & $z_l$=0.49 & $\pm$ 12 days\\  
Burud et al. \cite{burud01}        &                             &            & \\
\hline
B~2319+051                         & $\alpha$:  ~~23h 21m 40.80s & $z_s$=? & \\
Wisotski et al. \cite{wiso96}      & $\delta$:   +05d 27m 36.40s & $z_l$=0.62 & \\  
\noalign{\smallskip}

\hline
\noalign{\smallskip}
\end{tabular}
\end{center}
\label{Tab1b}
\end{table}

\index{quasars!quadruple}

\begin{table}[h]
\caption{List of central quads.}
\begin{center}
\renewcommand{\arraystretch}{1.2}
\setlength\tabcolsep{8pt}

\begin{tabular}{llll}
\hline\noalign{\smallskip}
Object          & Coords (2000)    & Redshifts   &  Notes \\
\noalign{\smallskip}\hline\hline
CLASS~B0128+437                    & $\alpha$: ~~01h 31m 16.26s   & $z_s$=? \\
Phillips et al. \cite{phil00}      & $\delta$:  +43d 58m 18.00s   & $z_l$=? \\ 
\hline
HST~1411+5211                       & $\alpha$: ~~14h 11m 19.60s & $z_s$=2.81   & \\
Fischer et al. \cite{fisch98}       & $\delta$: +52d 11m 29.00s & $z_l$=0.46   & \\
\hline 
H~1413+117                         & $\alpha$: ~~14h 15m 46.40s & $z_s$=2.55 & \\
Magain et al. \cite{magain88}      & $\delta$:  +11d 29m 41.40s & $z_l$=?    & \\
\hline
HST~14176+5226                     & $\alpha$: ~~14h 17m 36.51s & $z_s$=3.4    & \\
Ratnatunga et al. \cite{ratna95}   & $\delta$:  +52d 26m 40.00s  & $z_l$=0.81   &\\
\hline 
B~1555+375                         & $\alpha$: ~~15h 57m 11.93s & $z_s$=?    & \\
Marlow al. \cite{marl99}           & $\delta$:  +37d 21m 35.90s & $z_l$=?    &\\
\hline
Q~2237+0305                        & $\alpha$: ~~22h 40m 30.34s & $z_s$=1.69    & \\
Huchra et al. \cite{huchra85}      & $\delta$:  +03d 21m 28.80s & $z_l$=0.04   &\\
\noalign{\smallskip}

\hline
\noalign{\smallskip}
\end{tabular}
\end{center}
\label{Tab2}
\end{table}


\begin{table}[h]
\caption{List of short axis quads.}
\begin{center}
\renewcommand{\arraystretch}{1.2}
\setlength\tabcolsep{8pt}

\begin{tabular}{llll}
\hline\noalign{\smallskip}
Object          & Coords (2000)    & Redshifts   &  Notes \\
\noalign{\smallskip}\hline\hline
B~1422+231                    & $\alpha$: ~~14h 24m 38.09s  & $z_s$=3.62     & \\
Patnaik et al. \cite{pat92}   & $\delta$:  +22d 56m 00.60s  & $z_l$=0.34     & \\
\noalign{\smallskip}

\hline
\noalign{\smallskip}
\end{tabular}
\end{center}
\label{Tab3}
\end{table}


\begin{table}[h]
\caption{List of long axis quads.}
\begin{center}
\renewcommand{\arraystretch}{1.2}
\setlength\tabcolsep{8pt}

\begin{tabular}{llll}
\hline\noalign{\smallskip}
Object                             & Coords (2000)               & Redshifts      & Notes \\
\noalign{\smallskip}\hline\hline
RXJ 0911.4+0551                  & $\alpha$: ~~09h 11m 27.50s & $z_s$=2.8   & $\Delta$t(BA) = 150\\
Bade et al. \cite{bade97}        & $\delta$:  +05d 50m 52.00s & $z_l$=0.77? & $\pm$ 12 days\\
Hjorth et al. \cite{hjorth02}    &                            &             & \\
\hline
HST~12531-2914                   & $\alpha$: ~~12h 53m 06.70s   & $z_s$=?        & \\
Ratnatunga et al. \cite{ratna95} & $\delta$: $-$29d 14m 30.00s   & $z_l$=?        & \\
\hline 
B~2045+265                       & $\alpha$: ~~20h 47m 20.35s   & $z_s$=1.28        & \\
Fassnacht et al. \cite{fass99b}  & $\delta$:  +26d 44m 01.20s   & $z_l$=0.87        & \\
\noalign{\smallskip}

\hline
\noalign{\smallskip}
\end{tabular}
\end{center}
\label{Tab4}
\end{table}


\begin{table}[h]
\caption{List of inclined quads.}
\begin{center}
\renewcommand{\arraystretch}{1.2}
\setlength\tabcolsep{8pt}

\begin{tabular}{llll}
\hline\noalign{\smallskip}
Object                        & Coords (2000)             & Redshifts  &  Notes \\
\noalign{\smallskip}\hline\hline
0047-2808                     & $\alpha$: ~~00h 49m 41.89s & $z_s$=3.60 &  \\
Warren et al. \cite{wa96}     & $\delta$: $-$27d 52m 25.70s & $z_l$=0.49 & \\ 
\hline
HE~0230-2130                  & $\alpha$: ~~02h 32m 33.10s & $z_s$=2.16 & \\
Wisotzki et al. \cite{wisot99}& $\delta$: $-$21d 17m 26.00s & $z_l$=?    & \\
\hline
MG~0414+0534                  & $\alpha$: ~~04h 14m 37.73s & $z_s$=2.64 & \\
Hewitt et al. \cite{hewitt92} & $\delta$: +05d 34m 44.30s & $z_l$=0.96 & \\
\hline
B~0712+472                    & $\alpha$: ~~07h 16m 03.58s & $z_s$=1.34 & \\
Jackson et al. \cite{jack98}  & $\delta$: +47d 08m 50.00s & $z_l$=0.41 & \\
\hline
PG~1115+080                   & $\alpha$: ~~11h 18m 17.00s & $z_s$=1.72 & $\Delta$t(AB) = 11.7\\ 
Weymann et al. \cite{wey80}   & $\delta$: +07d 45m 57.70s & $z_l$=0.31 &  $\pm$ 1.2 days \\
Schechter et al \cite{schech98}&                           &            &  $\Delta$t(CB) = 25.0\\
                              &                           &            &  $\pm$ 1.6 days\\
\hline 
B~1608+656                    & $\alpha$: ~~16h 09m 13.96s & $z_s$=1.39 & $\Delta$t(BA) = 31\\ 
Myers et al.   \cite{myers95} & $\delta$: +65d 32m 29.00s & $z_l$=0.63 &  $\pm$ 7 days\\
Fassnacht et al. \cite{fass99a}&                            &            & $\Delta$t(BC) = 36\\ 
                              &                            &            & $\pm$ 7 days\\ 
                              &                            &            & $\Delta$t(BD) = 76\\ 
                              &                            &            & $\pm$ 10 days\\ 
\hline
MG~2016+112                   & $\alpha$: ~~20h 19m 18.15s & $z_s$=3.27 & \\ 
Lawrence et al. \cite{law84}  & $\delta$: +11d 27m  08.30s & $z_l$=1.01 & \\
\noalign{\smallskip}

\hline
\noalign{\smallskip}
\end{tabular}
\end{center}
\label{Tab5}
\end{table}


\begin{table}[h]
\caption{List of systems with more than four images.}
\begin{center}
\renewcommand{\arraystretch}{1.2}
\setlength\tabcolsep{8pt}

\begin{tabular}{llll}
\hline\noalign{\smallskip}
Object                        & Coords (2000)             & Redshifts  &  Notes \\
\noalign{\smallskip}\hline\hline
B~1359+154                    & $\alpha$: ~~14h 01m 35.55s  & $z_s$=3.24 & 6 images\\
Myers et al \cite{myers99}    & $\delta$:  +15d 13m 25.60s  & $z_l$=?    & \\ 
\hline
B~1933+507                   & $\alpha$: ~~19h 34m 30.95s  & $z_s$=2.63  & 10 images\\
Sykes et al. \cite{sykes98}  & $\delta$:  +50d 25m 23.60s  & $z_l$=0.76   & \\
\hline
\noalign{\smallskip}

\noalign{\smallskip}
\end{tabular}
\end{center}
\label{Tab6}
\end{table}

\end{document}